\documentclass[twocolumn]{aastex631}

\usepackage{graphicx}
\usepackage{graphbox}
\usepackage{floatrow}
\usepackage{float}
\usepackage{chngcntr}
\usepackage{amsmath}
\usepackage{amssymb}
\usepackage{mathtools}
\usepackage{threeparttable}          
\usepackage{hyperref}
\hypersetup{linkcolor=blue,citecolor=blue,filecolor=cyan,urlcolor=black}
\usepackage[toc,page]{appendix}

\newcommand{\cm}{\ifmmode {\rm cm}\else cm\fi}                              
\newcommand{\cmcube}{\ifmmode {\rm cm^3}\else cm$^3$\fi}                    
\newcommand{\m}{\ifmmode {\rm m}\else m\fi}                                 
\newcommand{\mcube}{\ifmmode {\rm m^3}\else m^3\fi}                         
\newcommand{\km}{\ifmmode {\rm km}\else km\fi}                              
\newcommand{\pc}{\ifmmode {\rm pc}\else pc\fi}                              
\newcommand{\kpc}{\ifmmode {\rm kpc}\else kpc\fi}                           
\newcommand{\Mpc}{\ifmmode {\rm Mpc}\else Mpc\fi}                           
\newcommand{\ly}{\ifmmode {\rm ly}\else ly\fi}                              
\newcommand{\AU}{\ifmmode {{\rm AU}}\else AU \fi}                           
\renewcommand{\deg}{\ifmmode {^\circ}\else $^\circ$ \fi}                         
\newcommand{\s}{\ifmmode {\rm s}\else s\fi}                                       
\newcommand{\Hz}{\ifmmode {\rm Hz}\else Hz\fi}                                    
\newcommand{\yr}{\ifmmode {\rm yr}\else y\fi}                                     
\newcommand{\kms}{\ifmmode{\rm km\;s^{-1}}\else km~s$^{-1}$ \fi}                  
\newcommand{\cms}{\ifmmode{\rm cm\;s^{-1}}\else cm~s$^{-1}$ \fi}                  
\newcommand{\Mpcyr}{\ifmmode {\rm Mpc\,yr^{-1}}\else Mpc\,yr$^{-1}$\fi}           
\newcommand{\K}{\ifmmode {\rm K}\else K\fi}                                       
\newcommand{\Teff}{\ifmmode {T_{\rm eff}}\else $T_{\rm eff}$\fi}                  
\newcommand{\Tex}{\ifmmode {T_{\rm ex}}\else $T_{\rm ex}$\fi}                     
\newcommand{\TK}{\ifmmode {T_{\rm K}}\else $T_{\rm K}$\fi}                        
\newcommand{\Tb}{\ifmmode {T_{\rm b}}\else $T_{\rm b}$\fi}                        %
\newcommand{\Ts}{\ifmmode {T_{\rm s}}\else $T_{\rm s}$\fi}                        
\newcommand{\erg}{\ifmmode {\rm erg}\else erg\fi}                                 
\newcommand{\dyn}{\ifmmode {\rm dyn}\else dyn\fi}                                 
\newcommand{\eV}{\ifmmode {\rm eV}\else eV\fi}                                    
\newcommand{\MeV}{\ifmmode {\rm MeV}\else MeV\fi}                                 
\newcommand{\GeV}{\ifmmode {\rm GeV}\else eV\fi}                                  
\newcommand{\Jy}{\ifmmode {\rm Jy}\else Jy\fi}                                     
\newcommand{\muJy}{\ifmmode $\mu${\rm Jy}\else $\mu$Jy\fi}                         
\newcommand{\MJy}{\ifmmode {\rm MJy}\else MJy\fi}                                  
\newcommand{\solar}{\ifmmode _{\mathord\odot}\else $_{\mathord\odot}$\fi}          
\newcommand{\Rsun}{\ifmmode {R_\odot}\else R$_\odot$ \fi}                          
\newcommand{\Msun}{\ifmmode {M}_{\mathord\odot}\else $M_{\mathord\odot}$\fi}       
\newcommand{\ster}{\ifmmode {\rm ster}\else ster\fi}                               
\newcommand{\muG}{\ifmmode {\mu \rm{G}} \else {$\mu$G}\fi}                         
\newcommand{\kB}{\ifmmode {k_B}\else $k_B$ \fi}                                    
\newcommand{\mH}{\ifmmode {m_{\rm H}}\else $m_{\rm H}$ \fi}                        
\newcommand{\nH}{\ifmmode {n_{\rm H}}\else $n_{\rm H}$ \fi}                        
\newcommand{\NH}{\ifmmode {N_{\rm H}}\else $N_{\rm H}$ \fi}                        
\newcommand{\NHI}{\ifmmode {N_{\rm H\,I}}\else $N_{\rm H\,I}$ \fi}                     
\newcommand{\muH}{\ifmmode {\mu_{\rm H}}\else $\mu_{\rm H}$ \fi}                   %
\newcommand{\radmsq}{\ifmmode {\rm rad\,m^{-2}}\else $\rm rad\,m^{-2}$ \fi}        

\newcommand{\vLSR}{\ifmmode {v_{\rm LSR}}\else $v_{\rm LSR}$\fi}         
\newcommand{\vlsr}{\ifmmode {v_{\rm lsr}}\else $v_{\rm lsr}$\fi}         
\newcommand{\vt}{\ifmmode {v_t}\else $v_t$\fi}                           
\newcommand{\cs}{\ifmmode {c_s}\else $c_s$\fi}                           

\newcommand{\JH}{\ifmmode {J\!-\!H}\else $J\!-\!H$ \fi}                 
\newcommand{\HK}{\ifmmode {H\!-\!K_S}\else $H\!-\!K_S$ \fi}             
\newcommand{\AV}{\ifmmode {A_{\rm V}}\else $A_{\rm V}$ \fi}             
\newcommand{\AK}{\ifmmode {A_{\rm K_S}}\else $A_{\rm K_S}$ \fi}         
\newcommand{\BV}{\ifmmode {B\!-\!V}\else $B\!-\!V$ \fi}                 
\newcommand{\EBV}{\ifmmode {E(B\!-\!V)}\else $E(B\!-\!V)$ \fi}          

\newcommand{\Pvec}{\ifmmode {\boldsymbol{P}}\else $\boldsymbol{P}$ \fi} 
\newcommand{\polgrad}{\ifmmode {|\boldsymbol{\nabla}\boldsymbol{P}|}\else $|\boldsymbol{\nabla}\boldsymbol{P}|$ \fi}                                  
\newcommand{\polgradmax}{\ifmmode {|\boldsymbol{\nabla}\boldsymbol{P}|_{\rm max}}\else $|\boldsymbol{\nabla}\boldsymbol{P}|_{\rm max}$ \fi}           
\newcommand{\polgradrad}{\ifmmode {|\boldsymbol{\nabla}\boldsymbol{P}|_{\rm rad,max}}\else $|\boldsymbol{\nabla}\boldsymbol{P}|_{\rm rad,max}$ \fi}   
\newcommand{\polgradtan}{\ifmmode {|\boldsymbol{\nabla}\boldsymbol{P}|_{\rm tan,max}}\else $|\boldsymbol{\nabla}\boldsymbol{P}|_{\rm tan,max}$ \fi}   
\newcommand{\polgradarg}{\ifmmode {{\rm arg}(\boldsymbol{\nabla}\boldsymbol{P})}\else ${\rm arg}(\boldsymbol{\nabla}\boldsymbol{P})$ \fi}             
\newcommand{\polgradmaxarg}{\ifmmode {{\rm arg}(\boldsymbol{\nabla}\boldsymbol{P}_{\rm max})}\else ${\rm arg}(\boldsymbol{\nabla}\boldsymbol{P}_{\rm max})$ \fi} 
\newcommand{\gradchi}{\ifmmode {|\boldsymbol{\nabla}\chi|_{1.4}}\else $|\boldsymbol{\nabla}\chi|_{1.4}$ \fi} 
\newcommand{\phimax}{\ifmmode {\phi_{\rm max}}\else $\phi_{\rm max}$ \fi}                        

\newcommand{\halpha}{\ifmmode {{\rm H}\alpha}\else H$\alpha$ \fi}         
\newcommand{\hbeta}{\ifmmode {{\rm H}\beta}\else H$\beta$ \fi}            
\newcommand{\hgamma}{\ifmmode {{\rm H}\gamma}\else H$\gamma$ \fi}         
\newcommand{\HI}{H\,{\textsc{i}}}                                         
\newcommand{\HII}{H\,{\textsc{ii}}}                                       

\newcommand{\B}{\ifmmode {\boldsymbol{B}}\else $\boldsymbol{B}$ \fi}                                
\newcommand{\Bparallel}{\ifmmode {\boldsymbol{B}_\parallel}\else $\boldsymbol{B_\parallel}$ \fi}    
\newcommand{\Bperp}{\ifmmode {\boldsymbol{B}_\parallel}\else $\boldsymbol{B}_\perp$ \fi}            

\newcommand{\rmtools}{\tt RMtools}
\newcommand{\rmclean}{\tt RM-clean}
\newcommand{\Planck}{Planck}
\newcommand{\ROHSA}{\tt ROHSA}
\newcommand{\xspec}{\tt Xspec}
\newcommand{\webspec}{\tt WebSpec}

\newcommand{\del}{\ifmmode {\nabla} \else $\nabla$\fi}               %
\newcommand{\calc}{\ifmmode {{\cal C}} \else ${{\cal C}}$\fi}        %
\newcommand{\calf}{\ifmmode {{\cal F}} \else ${{\cal F}}$\fi}        %
\newcommand{\calg}{\ifmmode {{\cal G}} \else ${{\cal G}}$\fi}        %
\newcommand{\calh}{\ifmmode {{\cal H}} \else ${{\cal H}}$\fi}        %
\newcommand{\call}{\ifmmode {{\cal L}} \else ${{\cal L}}$\fi}        %
\newcommand{\calm}{\ifmmode {{\cal M}} \else ${{\cal M}}$\fi}        %
\newcommand{\caln}{\ifmmode {{\cal N}} \else ${{\cal N}}$\fi}        %
\newcommand{\calo}{\ifmmode {{\cal O}} \else ${{\cal O}}$\fi}        %
\newcommand{\calp}{\ifmmode {{\cal P}} \else ${{\cal P}}$\fi}        %
\newcommand{\calq}{\ifmmode {{\cal Q}} \else ${{\cal Q}}$\fi}        %
\newcommand{\calr}{\ifmmode {{\cal R}} \else ${{\cal R}}$\fi}        %
\newcommand{\cals}{\ifmmode {{\cal S}} \else ${{\cal S}}$\fi}        %
\newcommand{\calt}{\ifmmode {{\cal T}} \else ${{\cal T}}$\fi}        %
\newcommand{\calv}{\ifmmode {{\cal V}} \else ${{\cal V}}$\fi}        %
\newcommand{\calw}{\ifmmode {{\cal W}} \else ${{\cal W}}$\fi}        %

\defcitealias{Green2019}{Bayestar19}

\received{June 18, 2021}
\revised{December 1, 2021}
\accepted{December 2, 2021}
\submitjournal{ApJ}

\shorttitle{Multi-Phase and Magnetized Filamentary ISM}
\shortauthors{Campbell et al.}

\begin{document}


\title{A Comparison of Multi-Phase Magnetic Field Tracers in a High-Galactic Latitude Region of the Filamentary Interstellar Medium}

\correspondingauthor{J. L. Campbell}
\email{campbell@astro.utoronto.ca}

\author[0000-0002-2511-5256]{J. L. Campbell}
\affiliation{David A. Dunlap Department of Astronomy \& Astrophysics, University of Toronto, 50 St. George St., Toronto, ON, M5S 3H4, Canada}
\affiliation{Dunlap Institute for Astronomy and Astrophysics, University of Toronto, 50 St. George St., Toronto, ON, M5S 3H4, Canada}

\author[0000-0002-7633-3376]{S. E. Clark}
\affiliation{Department of Physics, Stanford University, Stanford, California 94305, USA}
\affiliation{Kavli Institute for Particle Astrophysics \& Cosmology, P. O. Box 2450, Stanford University, Stanford, CA 94305, USA}

\author[0000-0002-3382-9558]{B. M. Gaensler}
\affiliation{Dunlap Institute for Astronomy and Astrophysics, University of Toronto, 50 St. George St., Toronto, ON, M5S 3H4, Canada}
\affil{David A. Dunlap Department of Astronomy \& Astrophysics, University of Toronto, 50 St. George St., Toronto, ON, M5S 3H4, Canada}

\author[0000-0002-5501-232X]{A. Marchal}
\affiliation{Canadian Institute for Theoretical Astrophysics, University of Toronto, 60 St. George St., Toronto, ON, Canada, M5S 3H8}

\author[0000-0002-7641-9946]{C. L. Van Eck}
\affiliation{Dunlap Institute for Astronomy and Astrophysics, University of Toronto, 50 St. George St., Toronto, ON, M5S 3H4, Canada}


\author[0000-0002-5146-2163]{A. A. Deshpande}
\affiliation{Inter-University Centre for Astronomy and Astrophysics, Pune 411007, India}
\affiliation{Indian Institute of Technology, Kanpur 208016, India}

\author{S. J. George}
\affiliation{School of Physics \& Astronomy, University of Birmingham, UK, B15 2TT}

\author[0000-0002-1495-760X]{S. J. Gibson}
\affiliation{Department of Physics \& Astronomy, Western Kentucky University, 1906 College Heights Blvd., KY 42101, USA}

\author[0000-0003-4631-1528]{R. Ricci}
\affiliation{Istituto Nazionale di Ricerche Metrologiche, strada delle cacce 91, ZIP code 10135 Torino, Italy}

\author[0000-0003-2623-2064]{J. M. Stil}
\affiliation{Department of Physics and Astronomy, The University of Calgary, 2500 University Drive NW, Calgary AB T2N 1N4, Canada}

\author[0000-0001-9885-0676]{A. R. Taylor}
\affiliation{Inter-University Institute for Data Intensive Astronomy, and Department of Astronomy, University of Cape Town}
\affiliation{Department of Physics and Astronomy, University of the Western Cape}


\begin{abstract}

Understanding how the Galactic magnetic field threads the multi-phase interstellar medium (ISM) remains a considerable challenge, as different magnetic field tracers probe dissimilar phases and field components. We search for evidence of a common magnetic field shared between the ionized and neutral ISM by comparing 1.4 GHz radio continuum polarization and {\HI} line emission from the Galactic Arecibo L-Band Feed Array Continuum Transit Survey (GALFACTS) and Galactic Arecibo L-Band Feed Array {\HI} (GALFA-{\HI}) survey, respectively. We compute the polarization gradient of the continuum emission and search for associations with diffuse/translucent {\HI} structures. The polarization gradient is sensitive to changes in the integrated product of the thermal electron density and line-of-sight field strength ($B_\parallel$) in warm ionized gas, while narrow {\HI} structures highlight the plane-of-sky field orientation in cold neutral gas. We identified one region in the high-Galactic latitude Arecibo sky, G216{+}26 centered on $(\ell,b){\sim}(216\deg,{+}26\deg)$, containing filaments in the polarization gradient that are aligned with narrow {\HI} structures roughly parallel to the Galactic plane. We present a comparison of multi-phase observations and magnetic field tracers of this region, demonstrating that the warm ionized and cold neutral media are connected likely via a common magnetic field. We quantify the physical properties of a polarization gradient filament associated with {\halpha\!\!} emission, measuring a line-of-sight field strength $B_\parallel\,{=}\,6{\pm}4\,\mu{\rm G}$ and a plasma beta $\beta\,{=}\,2.1^{+3.1}_{-2.1}$. We discuss the lack of widespread multi-phase magnetic field alignments and consider whether this region is associated with a short-timescale or physically rare phenomenon. This work highlights the utility of multi-tracer analyses for understanding the magnetized ISM.

\end{abstract}


\section{Introduction} \label{sec:intro}

The diffuse interstellar medium (ISM) of our Galaxy is a complex multi-phase environment threaded with magnetic fields. This Galactic magnetic field (GMF) plays a crucial role in many astrophysical process that drive Galactic evolution. Yet, despite its significance in a variety of Galactic environments, a complete understanding of the GMF has been substantially hampered by the difficulty in interpreting its three-dimensional (3D) multi-phase geometry \citep{Haverkorn2019, Jaffe2019}. This challenge is largely due to the observational limitation that various magnetic field tracers probe the GMF in different phases of the ISM, while also providing only one- or two-dimensional projections of its full vector morphology \citep[see][for overviews on magnetic field tracers]{Jaffe2019, Ferriere2020}. As a result, understanding how the GMF is structured between different ISM phases remains poorly understood.

The ISM phases can be broadly categorized into ionized, atomic, and molecular, and have a range of temperatures and densities \citep[see][]{Ferriere2001}. A primary component of the ionized ISM is the warm ionized medium (WIM), sometimes referred to as diffuse ionized gas (DIG) in other galaxies \citep{Haffner2009}. The WIM has a typical gas temperature $T\,{\sim}\,10^4\,{\rm K}$ and volume-averaged thermal electron density $n_e\,{\sim}\,0.01\,{-}\,0.1\,{\rm cm^{-3}}$ \citep{Reynolds1990a, Weisberg2008}, and pervades much of the Galactic volume. Galactic maps of {\halpha\!\!} emission reveal many filaments, loops, and bubbles superimposed on a diffuse background that is often considered to be the WIM. The ultraviolet radiation from massive OB stars within the Galactic midplane is believed to sustain the ionization of the WIM \citep{Reynolds1984, Reynolds1990b, Haffner2009, Kado-Fong2020}, although there remain questions as to how this radiation percolates through the ISM to maintain the ionization at high Galactic latitudes \citep[e.g.,][]{Reynolds1990c, Haffner2009, Wood2010, Kim2018}.

The polarized radio sky contains complex features with no corresponding total intensity structures \citep{Wieringa1993, Gray1998, Gaensler2001, Gaensler2011}. When background polarized radio synchrotron emission passes through the foreground magnetized WIM, birefringence induces a rotation of the incident polarization angle, an effect called Faraday rotation. This process is quantified by the rotation measure (RM) and is proportional to the product of the thermal electron density ($n_e$) and line-of-sight (LOS) magnetic field strength ($B_\parallel$) integrated along the LOS. Faraday rotation alone cannot disentangle the degeneracy between $n_e$ and $B_\parallel$, and provides no information about either of their LOS distributions \citep{Ferriere2016}. 

The polarization gradient ({\polgradmax\!\!}) is a spatial gradient of the complex polarization vector $Pe^{2i\chi}$, where $P$ is the polarized intensity and $\chi$ is the polarization angle, revealing a network of small-scale filaments interpreted as turbulent-driven fluctuations in $n_e$ and/or $B_\parallel$ \citep{Gaensler2011}. Much larger features in the polarization gradient have been shown to be associated with the edges of supernova remnants and {\HII} regions \citep[e.g.,][]{Iacobelli2014}. In the diffuse ISM, Faraday rotation primarily traces the WIM \citep[e.g.,][]{Haverkorn2004b, Haverkorn2004c, Hill2008,Gaensler2011, Heiles2012}. Work by \citet{Thomson2019} has evoked some debate over whether significant Faraday rotation may occur within predominantly neutral regions with increased magnetic field strengths \citep{Bracco2020}. There is also evidence that Faraday rotation may occur within molecular clouds \citep{Tahani2018}.

The neutral atomic medium exists in two thermally-stable phases: the dense cold neutral medium (CNM) and the more rarefied warm neutral medium (WNM). The CNM ($T\,{\sim}\,10\,{-}\,100\,{\rm K}, N_{HI}\,{\sim}\,7\,{-}\,70\,{\rm cm^{-3}}$, \citealp{Mckee1977, Wolfire2003}) is often structured into sheets and filaments, occupying a small fraction of the Galactic volume \citep{Heiles1967, Heiles2003, Verschuur1970, Clark2014} while the WNM ($T\sim10^4\,{\rm K}, N_{HI}\sim0.2-0.9\,{\rm cm^{-3}}$, \citealp{Mckee1977, Wolfire2003, Marchal2021}) is more extended and fills a larger Galactic volume \citep{Heiles2003}. The CNM and WNM coexist in mutual pressure equilibrium \citep{Field1969, Goldsmith1969}. There also exists a thermally unstable neutral medium (UNM) within which a significant fraction of {\HI} gas has been shown to exist \citep{Heiles2003, Kanekar2003, Roy2013, Saury2014, Murray2015, Murray2018, Marchal2021}. Of the total {\HI} mass, it is believed that $\sim$30\% is in the CNM phase and $\sim$50\% is in the WNM phase, with the remaining $\sim$20\% occupying the thermally unstable regime \citep{Murray2018}.

Velocity-resolved {\HI} data reveal long, narrow structures deemed `{\HI} fibers' at high-Galactic latitude that are very well aligned with the plane-of-sky (POS) magnetic field orientation traced using starlight \citep{Clark2014} and dust polarization data \citep{Clark2015, Martin2015, Kalberla2016, Blagrave2017}. While some authors have interpreted these {\HI} structures as velocity caustics imprinted by the turbulent velocity field \citep{Lazarian2018}, a number of observational measurements have instead shown that they are real density structures residing in the cold ISM. Such evidence includes ratios of far-infrared to {\HI} column density \citep{Clark2019a}, NaI absorption profile equivalent widths \citep{Peek2019}, and column density power spectra \citep{Kalberla2020}.

Observational studies that provide evidence for magnetic field alignments between the ionized and neutral medium are sparse. The first of these was the 3C196 LOFAR \citep{vanHaarlem2013} field containing strong spatial alignments between radio polarized intensity structures, the POS magnetic field orientation in {\Planck} dust polarization data, and linear depolarization canals \citep{Jelic2015, Zaroubi2015, Jelic2018, Turic2021}, as well as {\HI} filaments \citep{Kalberla2016a}. These morphological alignments led the authors to conclude that the magnetized ionized and neutral media are clearly connected, and that the regular magnetic field plays an important role in confining these phases of the diffuse ISM. The Horologium and Auriga fields located within the Fan Region \citep{vandeHulst1967, Wolleben2006}, a strongly polarized Galactic structure \citep{Hill2017}, were subsequently shown to exhibit an alignment between depolarization canals \citep{Haverkorn2003a, Haverkorn2003b} and {\HI} filaments \citep{Kalberla2017}. The authors again attributed this morphological alignment to the local magnetic field confining the magnetized ionized and neutral ISM together, suggesting that the ionized medium associated with the radio polarized filament is likely wrapped around the filamentary {\HI} emission.

Despite the spatial correlations found across the multi-phased and magnetized ISM, such comparisons are often challenging and difficult to interpret \citep{Haverkorn2019, Jaffe2019}. It thus remains unclear whether the local magnetic field is shared between the Faraday-rotating plasma and filamentary neutral medium in these regions, and if so, how the GMF is morphologically connected between them. Further studies are required to understand whether such regions are representative of the mean magnetic field of the Milky Way Galaxy and our ability to identify them is observationally limited, or if they are somehow unique locations in the Galaxy.

In this work, we search for evidence that the GMF is shared between the warm ionized and cold neutral ISM using radio polarization gradients and {\HI} structures in Arecibo data. The structure of this paper is as follows. An overview of the data is presented in Section \ref{sec:data}, followed by our multi-phase results and the key morphological structures described in Section \ref{sec:results}. Our comparison of polarization gradient and {\HI} structures is presented in Section \ref{sec:analysis} and key astrophysical quantities are computed in Section \ref{sec:estimates}. Our discussion is presented in Section \ref{sec:discussion}. Our summary and conclusions are given in Section \ref{sec:summary}.


\section{Data Overview}\label{sec:data}


\subsection{1.4 GHz Radio Polarization}\label{subsec:galfacts}

We use radio polarization data from Data Release (DR) 3.1.2 of the Galactic Arecibo L-Band Feed Array Continuum Transit Survey (GALFACTS), a spectro-polarimetric sky survey \citep{Taylor2010} that provides wide-field high-resolution Stokes maps of $I_{1.4}, Q_{1.4}, U_{1.4},$ and $V_{1.4}$, where the subscripts refer to the $1.4$ GHz frequency. The primary science goal of GALFACTS is to advance our understanding of the magnetoionic properties of the Milky Way Galaxy by mapping the radio polarized sky and deriving key polarization properties for both discrete point sources and the diffuse ISM. The data have an angular resolution of {3\farcm5} \citep{Taylor2010}. The full data will cover a $\sim$300 MHz (1214-1525 MHz) bandwidth but are currently limited to a $\sim$160 MHz (1367-1525 MHz) bandwidth across 376 binned spectral channels. The observations were taken throughout the period of 2008-2016 with the 305\,m single-dish William E. Gordon (Arecibo) telescope in Puerto Rico using the Arecibo L-band feed array (ALFA). GALFACTS offers complete sky coverage of $\sim$12,700 deg$^2$ within the declination range of $-0\fdg8\,{<}\,{\delta}\,{<}\,{+}37\fdg8$. The data consist of a north survey (with fields N[1-4]), south survey (S[1-4]), and zenith survey (Z[1-4]), with each resulting data cube covering {512\arcmin$\times$512\arcmin} (8\fdg53$\times$8\fdg53) on the sky with {1\arcmin$\times$1\arcmin} pixels \citep{Guram2009, Taylor2010}. See \citet{Guram2009,Guram2011,Taylor2012,Taylor2013} for details on GALFACTS data processing.

\begin{figure*}[t!]
\centering
\vspace{-5pt}
\hspace*{-9pt}\includegraphics[align=c,height=5.75cm]{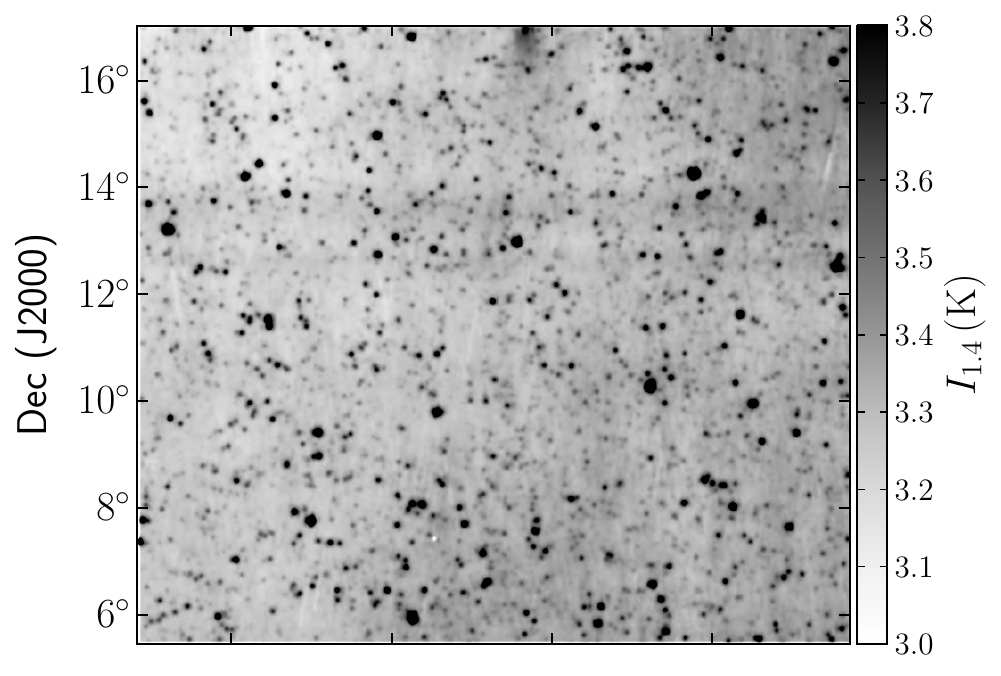}\hspace*{5pt}\includegraphics[align=c,height=5.5cm]{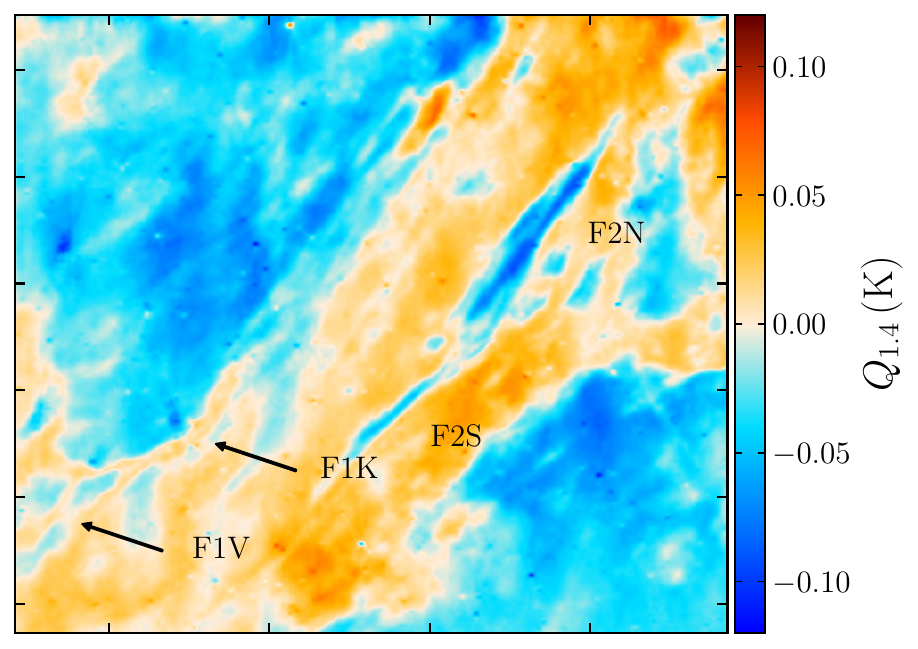}\hspace*{4pt}\vspace{-5pt}
\hspace*{-12pt}\includegraphics[align=c,height=6.5cm]{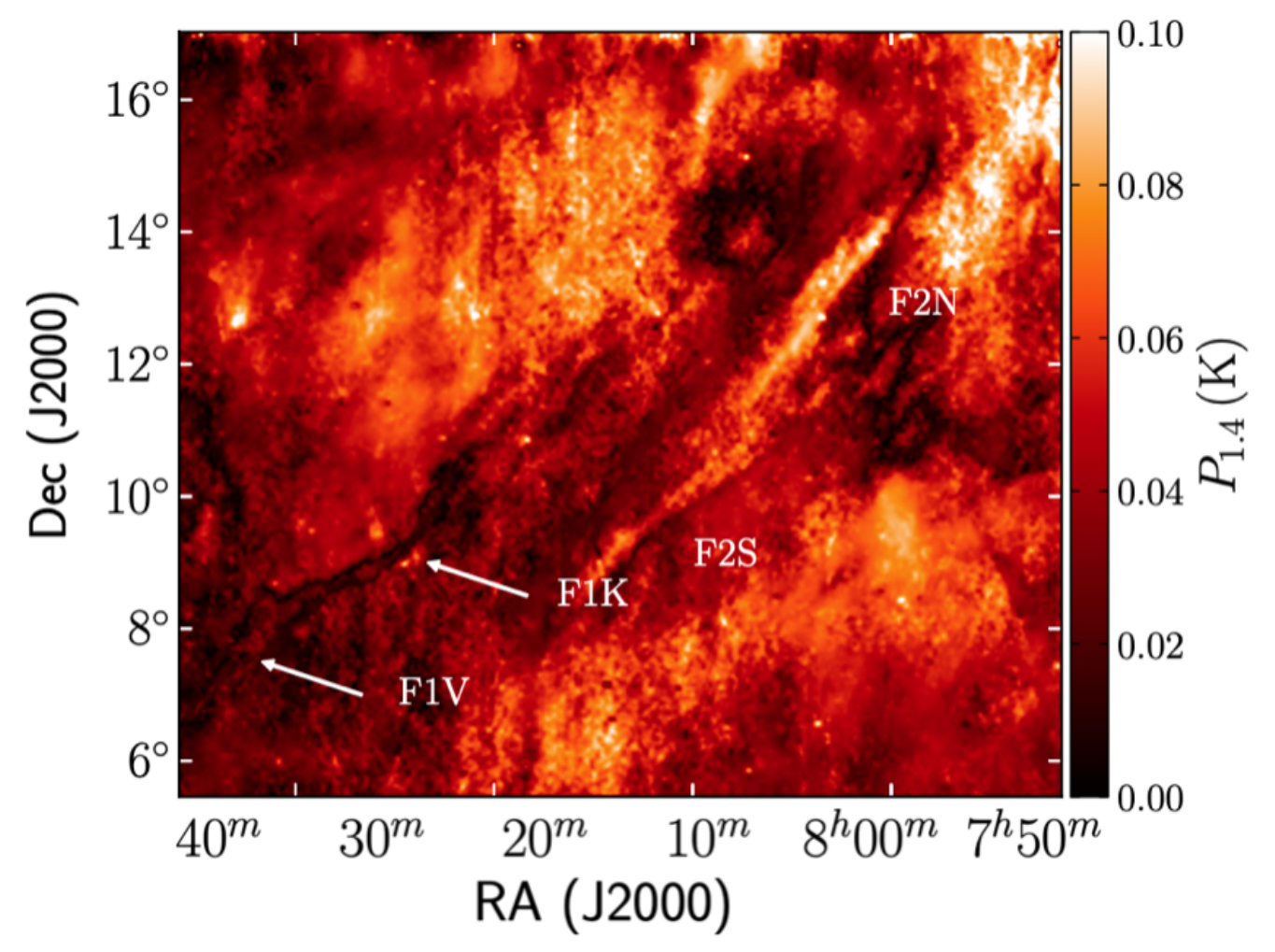}\hspace*{3pt}\includegraphics[align=c,height=6.5cm]{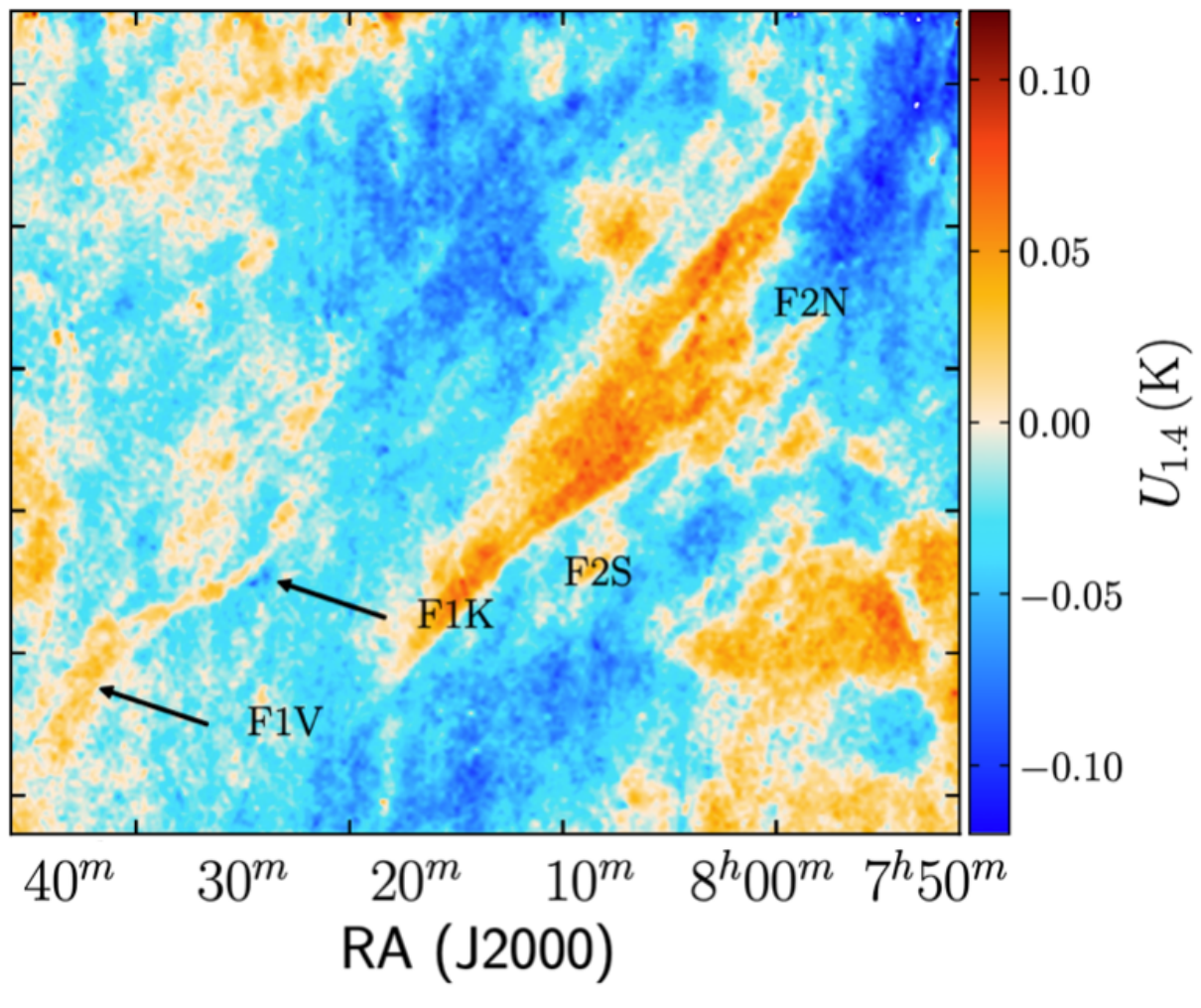}
\includegraphics[align=c,height=6.5cm]{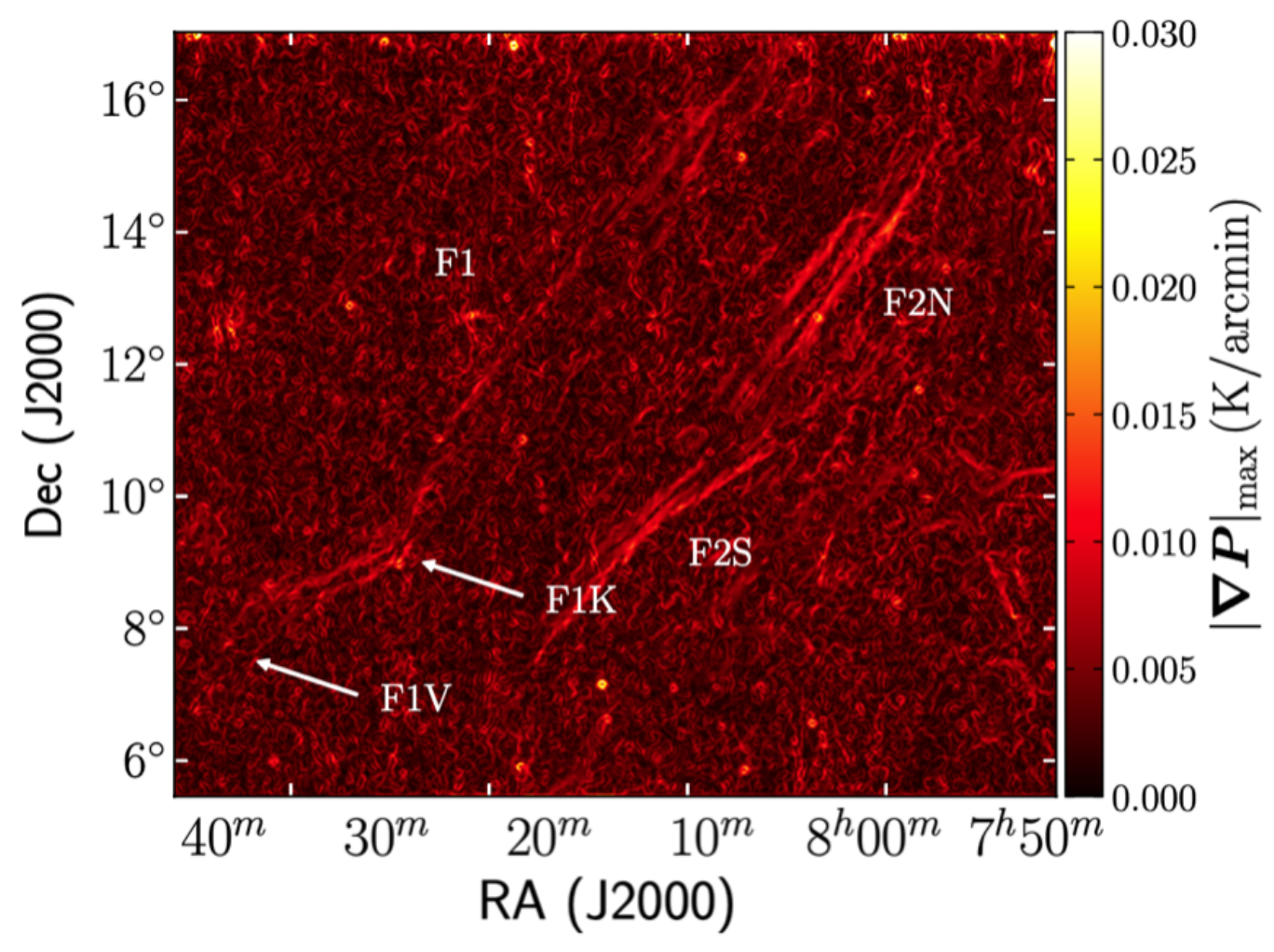}\hspace*{5pt}
\caption{1.4 GHz GALFACTS data of G216 at {5\arcmin} angular resolution. Shown are $I_{1.4}$ (top left), $P_{1.4}$ (middle left), $Q_{1.4}$ (top right), $U_{1.4}$ (middle right), and {\polgradmax\!\!} (bottom center). The polarization data have been de-striped and bright polarized sources have been removed. The locations of F1 and F2 as discussed in the text are indicated.}
\label{fig:stokes_S1}
\end{figure*}

The Arecibo data are affected by leakage-related scanning artefacts. As the telescope nods along the meridian, the Earth's rotation causes a zig-zag basket-weaving scanning pattern that results in systematic linear artefacts. While post-processing removed most of the scanning artefacts \citep{Guram2009}, they are still apparent in the data and become enhanced in the polarization gradient. There are additional artefacts that appear as clusters of short streaks along the scanning direction from intermittent broadband RFI \citep{Leahy2018}. To minimize these artefacts in the GALFACTS data, we applied a Fourier de-striping technique where artefacts are localized and removed in the Fourier domain \citep{Schlegel1998} using the \textsc{scipy} \texttt{fftpack} package \citep{Virtanen2020} to compute the Fourier transform.

We used a high-Galactic latitude noise-dominated region in the S1 footprint to locate the scanning artefacts in Fourier space. The de-striping process caused additional artefacts toward bright polarized point sources, so we masked them using a cubic-spline interpolation before proceeding with de-striping. This process was effective in minimizing the artefacts in $Q_{1.4}$ and $U_{1.4}$ but does not significantly improve $I_{1.4}$, so we only de-stripe the polarization data. The final de-striped Stokes maps were smoothed to an angular resolution of {5\arcmin} to increase the polarization gradient signal-to-noise ratio (SNR).

The maximum amplitude of the polarization gradient {\polgradmax\!\!}, sometimes referred to as the generalized polarization gradient, describes the general case in which both $P$ and $\chi$ are changing across the image plane \citep{Herron2018I}. This differs from the original form of the polarization gradient where they are assumed to be changing in the same direction \citep{Gaensler2011}. The amplitude of {\polgradmax\!\!} is given by

\begin{widetext}
    \begin{equation}\label{eq:polgradmax}
    \begin{split}
    \begin{aligned}
    |\boldsymbol{\nabla}\boldsymbol{P}_{\rm max}| &= \Bigg(\Bigg. \frac{1}{2}\left[ \left(\frac{\partial Q}{\partial x}\right)^2 + \left(\frac{\partial U}{\partial x}\right)^2 + \left(\frac{\partial Q}{\partial y}\right)^2 + \left(\frac{\partial U}{\partial y}\right)^2 \right] \\
    &+ \frac{1}{2} \sqrt{ \left[ \left(\frac{\partial Q}{\partial x}\right)^2 + \left(\frac{\partial U}{\partial x}\right)^2 + \left(\frac{\partial Q}{\partial y}\right)^2 + \left(\frac{\partial U}{\partial y}\right)^2 \right]^2 - 4\left[ \frac{\partial Q}{\partial x}\frac{\partial U}{\partial y} - \frac{\partial Q}{\partial y}\frac{\partial U}{\partial x}  \right]^2} \Bigg.\Bigg)^{1/2}
    \end{aligned}
    \end{split}
    \end{equation}
\end{widetext}

{\noindent}where $x$ and $y$ are Cartesian pixel coordinates of the image, and $Q$ and $U$ are the Stokes maps.

The polarization gradient is most sensitive to small-scale structures and can be used as an edge detector to identify sharp changes in the amplitude and/or direction of the complex polarization vector. Due to its sensitivity to small-scale structures, the polarization gradient enhances noise in the image \citep{Burkhart2012}. Convolving $Q$ and $U$ with a Gaussian kernel before computing the spatial gradient is mathematically equivalent to applying a spatial filter to the gradient operator \citep{Robitaille2015}. Smoothing the Stokes maps thus increases the SNR while simultaneously probing larger spatial scales in the polarization gradient.


\subsection{21 cm Line Emission}\label{subsec:galfa-hi}

We use velocity-resolved DR 2 Galactic Arecibo L-Band Feed Array {\HI} (GALFA-{\HI}) survey {\HI} data \citep{Peek2018}. The GALFA-{\HI} data were commensally observed with GALFACTS using Arecibo and have the same sky coverage and angular resolution. We use custom data cubes with the same sky coverage as the GALFACTS footprints over the velocity range $|v_{\rm lsr}|\leq$ 188 {\kms\!\!} with a degraded 0.8 {\kms\!\!} velocity resolution. The scanning artefacts are not prominent in the {\HI} data and do not significantly affect our analysis, so we do not de-stripe or smooth the GALFA-{\HI} data.


\subsection{H$\alpha$ Emission}\label{subsec:halpha}

We use the composite {\halpha\!\!} map \citep{Finkbeiner2003} that combines data from the Virginia Tech Spectral line Survey \citep[VTSS;][]{Dennison1998}, Southern H-Alpha Sky Survey Atlas \citep[SHASSA;][]{Gaustad2001}, and Wisconsin H-Alpha Mapper \citep[WHAM;][]{Haffner2003}. The composite data were reprocessed to remove image artefacts, calibrated to a stable zero-point on {$1\deg$} scales using WHAM, and were resampled to a common {6\arcmin} angular resolution \citep{Finkbeiner2003}.

We complement the composite {\halpha\!\!} map with the velocity-resolved WHAM data that have an angular resolution of $1\deg$, velocity resolution of $12\,\kms$, and a sensitivity to faint emission on large angular scales \citep{Haffner2003}. This combination of {\halpha\!\!} data help to provide a more complete picture of ionized gas.


\begin{figure*}[t!]
\setlength{\lineskip}{-1pt}
\centering
 \includegraphics[height=6.7cm]{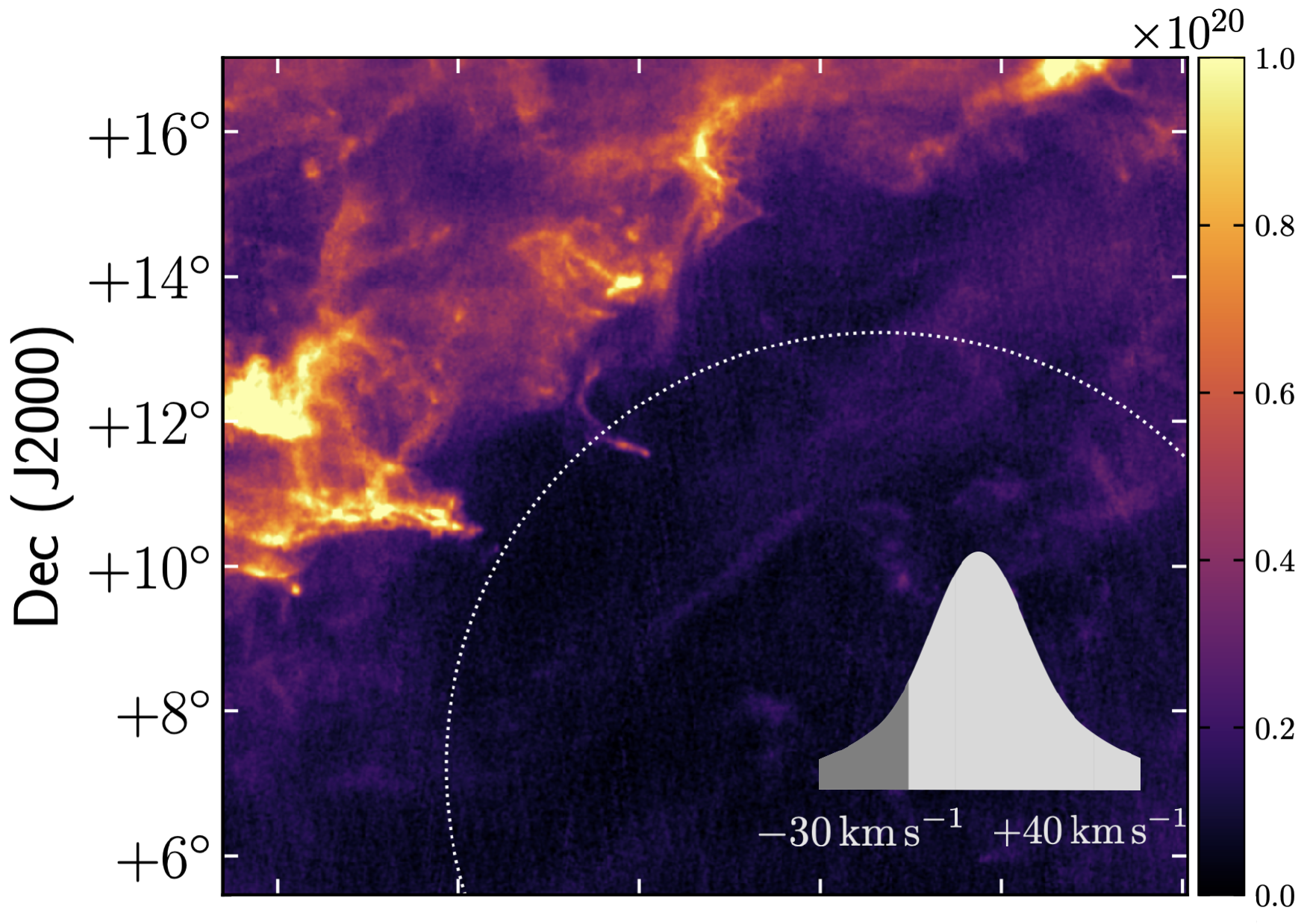}\includegraphics[height=6.7cm]{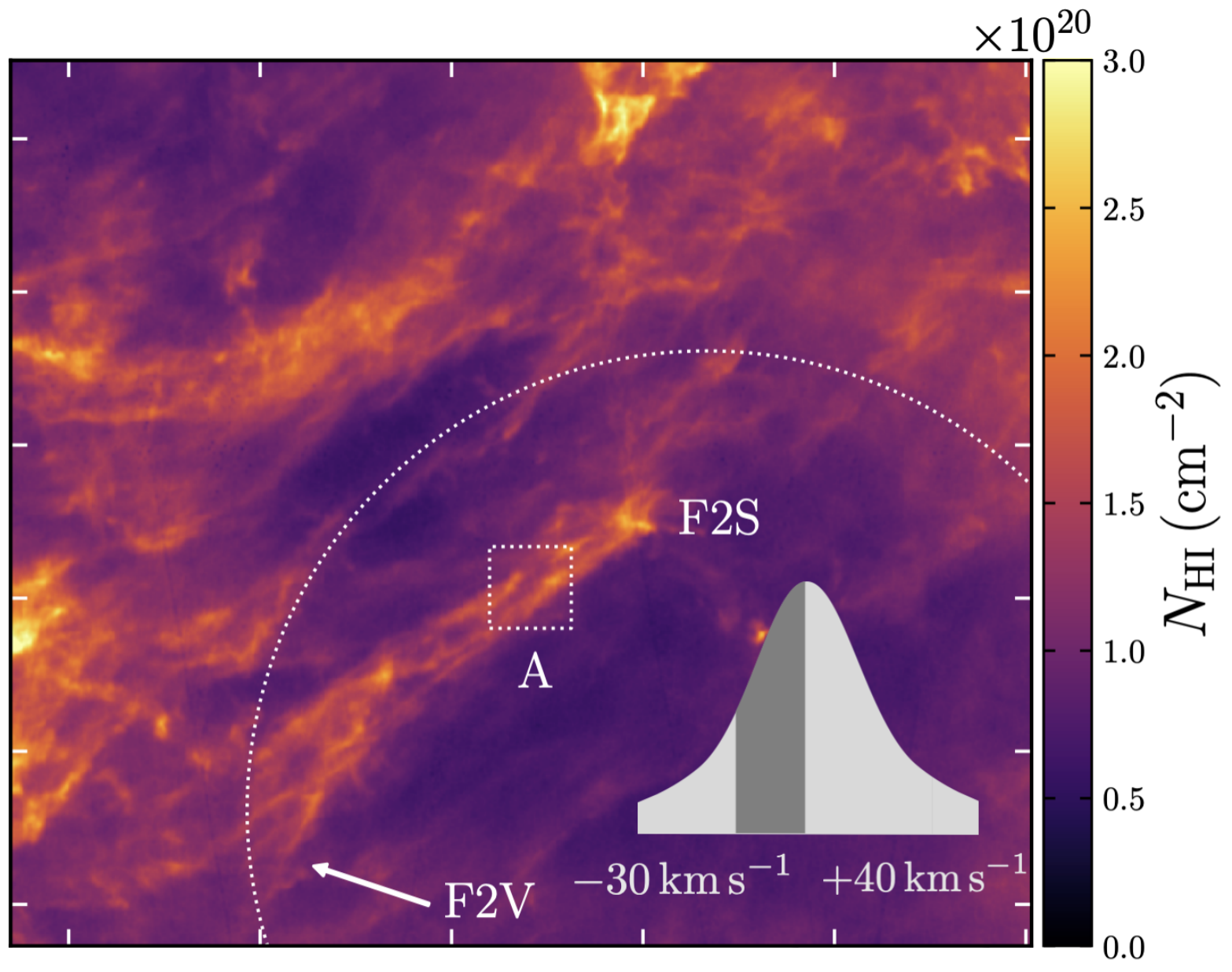}
 \includegraphics[height=7.6cm]{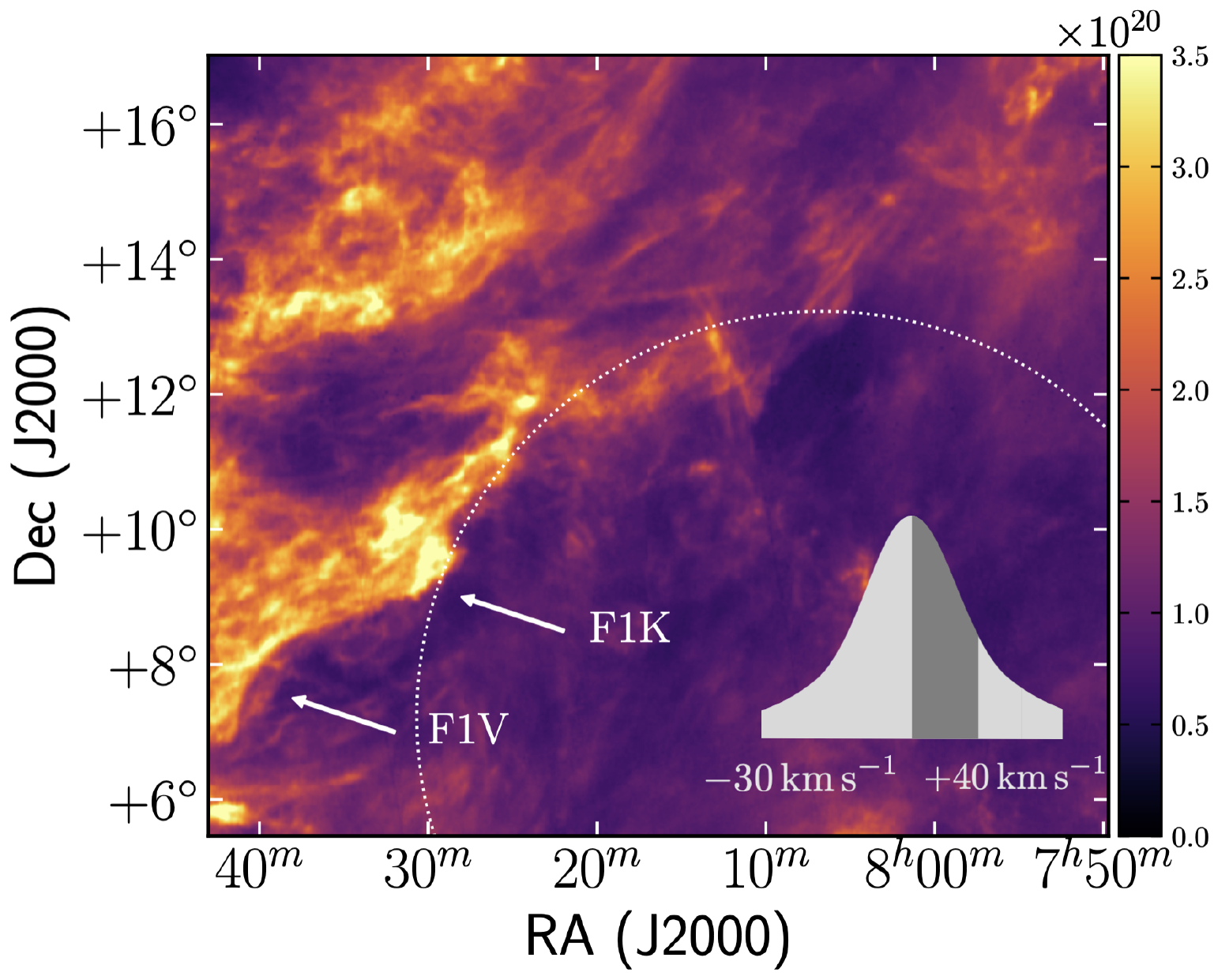}\includegraphics[height=7.6cm]{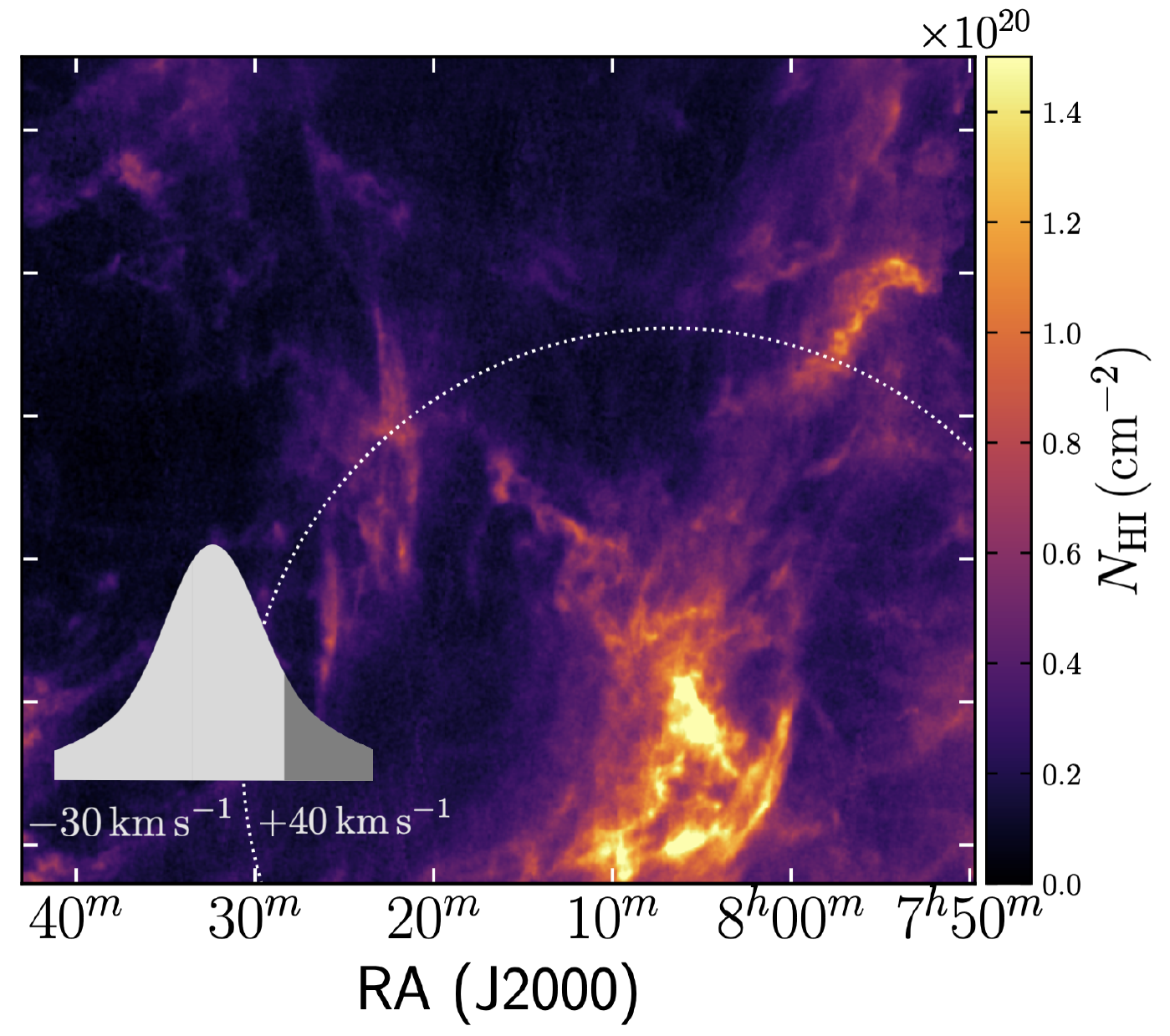}
 \caption{GALFA-{\HI} column density maps of G216 at {3\farcm5} resolution across the full velocity range. The data were integrated over ${-30}\,{\leq}\,{v_{\rm lsr}}\,{\leq}\,{-10}$\,\kms (top left), ${-10}\,{\leq}\,{v_{\rm lsr}}\,{\leq}\,{+4}$\,\kms (top right), ${+4}\,{\leq}\,{v_{\rm lsr}}\,{\leq}\,{+20}$\,\kms (bottom left), and ${+20}\,{\leq}\,{v_{\rm lsr}}\,{\leq}\,{+40}$\,\kms (bottom right) to show various features in the data. Locations of F1 and F2 as discussed in the text are indicated. Note the different intensity scales between panels. The bottom corner of each figure indicates the range in {\vlsr}. The segmented dotted circles (white) show the approximate location of the projected low column density region discussed in Section \ref{subsec:results-surroundings}. Region A marked by the dotted rectangle (white) in the top right figure indicates the subset of the field for which we apply the {\ROHSA} phase separation in Section \ref{subsec:hifibers}.}
\label{fig:HI_S1}
\end{figure*}

\subsection{Dust Emission}\label{subsec:planck}

We use the {5\arcmin} R2.00 {\Planck} 353 GHz Stokes $I_{353}$ all-sky dust map, where the subscript refers to the 353 GHz frequency. The data were post-processed with the Generalized Needlet Internal Linear Combination (\texttt{GNILC}) algorithm to remove the cosmic infrared background (CIB) anisotropies from the thermal dust emission \citep{Planck2013XI}. The High Frequency Instrument is calibrated to the cosmic microwave background (CMB) dipole and is measured as a temperature differential in units of ${\rm K_{CMB}^{-1}}$. We converted this to units of ${\rm MJy\,sr^{-1}}$ via the conversion factor of $287\,{\rm MJy\,sr^{-1}\,K_{CMB}^{-1}}$ \citep{Planck2018III}. We subtracted the CIB monopole from the $I_{353}$ map, which has an intensity of 0.13 MJy\,sr$^{-1}$ at 353 GHz \citep{Planck2018III}, and added a Galactic {\HI} offset correction to the $I_{353}$ map of 0.01035 MJy\,sr$^{-1}$ \citep{Planck2018XII}.

We complement the $I_{353}$ map with the {80\arcmin} R3.00 {\Planck} $Q_{353}$ and $U_{353}$ dust polarization data \citep{Planck2018III} that was also post-processed with \texttt{GNILC}. We subtracted the CMB monopole from $Q_{353}$ and $U_{353}$, which has an intensity of $0.64\,{\rm MJy\,sr^{-1}}$ at this frequency in polarization \citep{Planck2016VIII}. The $Q_{353}$ and $U_{353}$ maps follow the COSMO polarization angle convention \citep{Planck2018III}, so we converted the data to IAU convention\footnote{\href{https://www.iau.org/news/announcements/detail/ann16004/}{https://www.iau.org/news/announcements/detail/ann16004/}} by multiplying $U_{353}$ by $-1$.


\section{Multi-Frequency Morphology Results}\label{sec:results}

We visually compared structures in the GALFACTS polarization gradient and GALFA-{\HI} emission of the Arecibo sky, focusing on high-Galactic latitude structures that are associated with diffuse/translucent {\HI} emission and that are not spatially correlated with excess synchrotron emission. Our search identified G216+26 (hereafter G216) centered on $(\ell,b)\,{\sim}\,(216\deg,+26\deg)$ as the only region found to contain coherent filaments in the polarization gradient that are spatially correlated with linear {\HI} structures, alluding to a possible connection between their magnetic fields. The remainder of this paper investigates whether the warm ionized and cold neutral ISM are associated in this region and if they share a common magnetic field.

\subsection{1.4 GHz Synchrotron Emission}\label{subsec:results-GALFACTS}

Figure \ref{fig:stokes_S1} shows the 1.4 GHz GALFACTS spectropolarimetry results smoothed to {5\arcmin} angular resolution. This includes $I_{1.4}$ (top left), $P_{1.4}$ (middle left), $Q_{1.4}$ (top right), $U_{1.4}$ (middle right), and {\polgradmax\!\!} (bottom center). Recall that several bright polarized point sources were masked and interpolated to prevent additional artefacts in the polarization gradient. Smoothing the de-striped $Q_{1.4}$ and $U_{1.4}$ maps from an angular resolution of {3\farcm5} to {5\arcmin} before computing the gradient enhances the SNR in the polarization gradient by a factor of ${\gtrsim}\,2$ while retaining most high spatial frequencies.

This region contains two distinct filamentary structures in the polarization gradient that are separated by ${\sim}\,{2-3}\deg$, extend ${\sim}\,{10\deg}$ in length, and are roughly parallel to the Galactic plane. There is very little indication of the polarization gradient filaments using the {3\farcm5} resolution data; smoothing $Q_{1.4}$ and $U_{1.4}$ to enhance the SNR allow them to become visible. There is also a `background' of small-scale structures in the polarization gradient that likely reflects noise, unmasked point sources, and discontinuities in $n_e$ and/or $B_\parallel$.

\begin{figure*}[t!]
\setlength{\lineskip}{-1pt}
\centering
 \includegraphics[height=5.4cm]{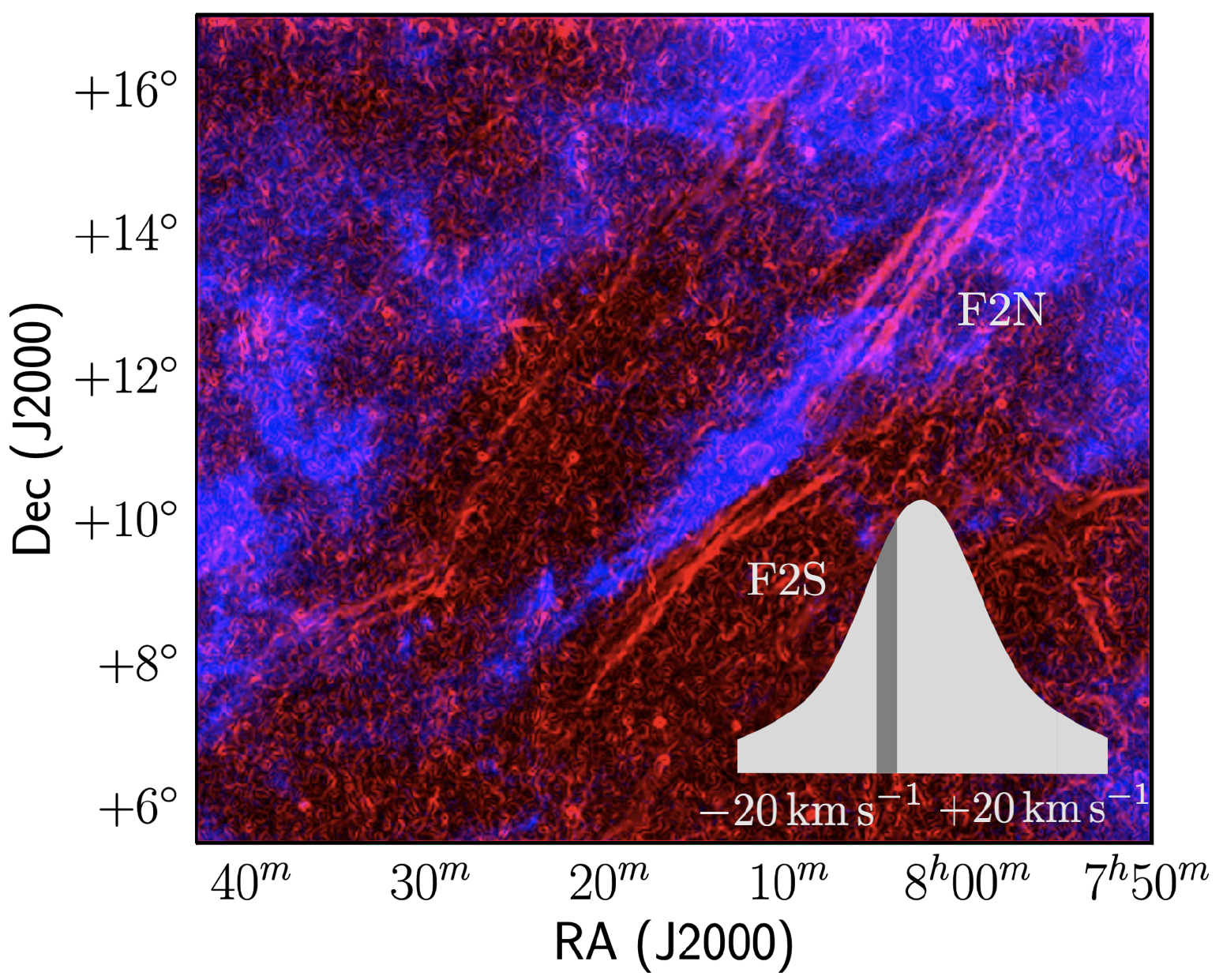}\includegraphics[height=5.4cm]{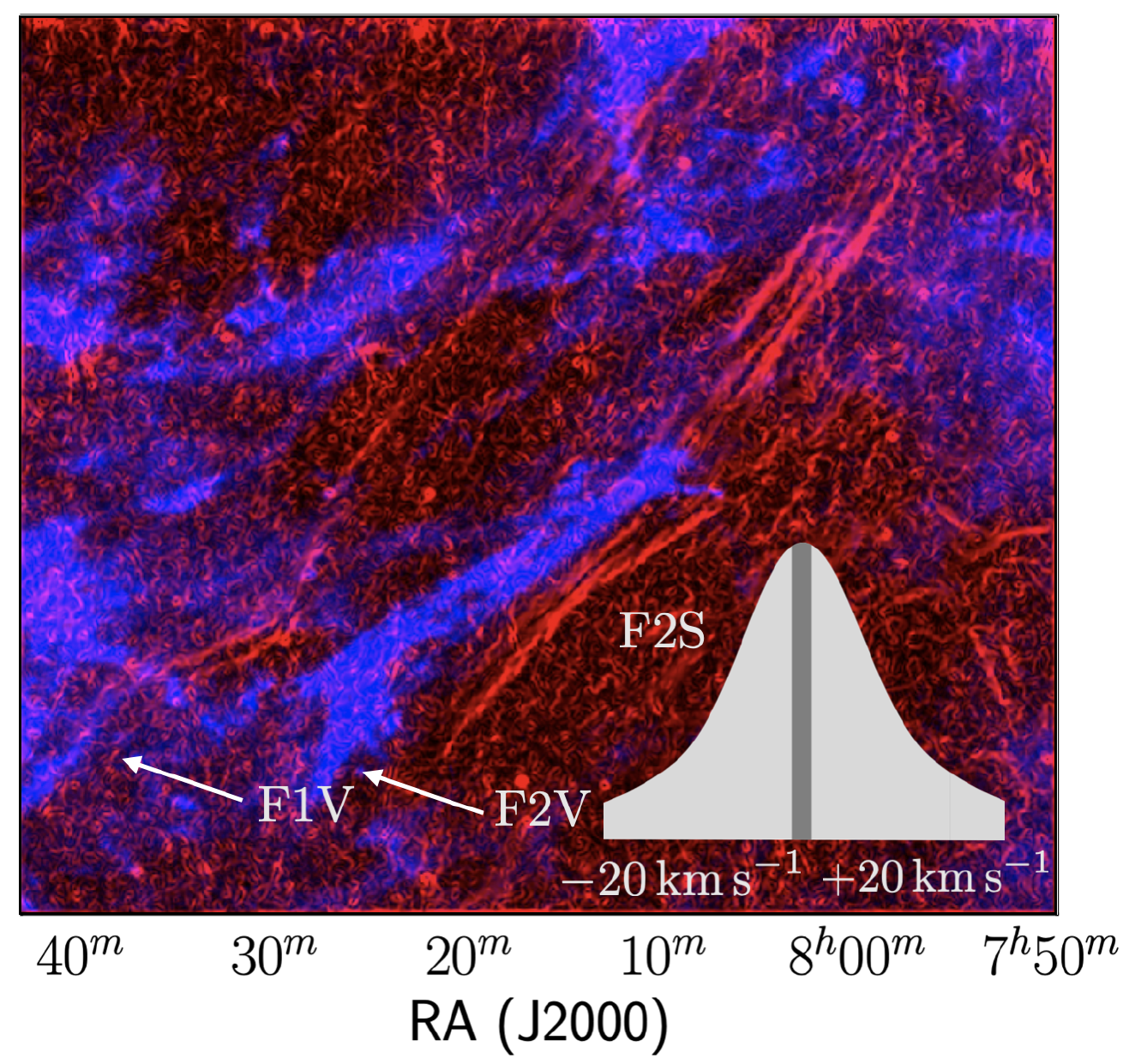}\includegraphics[height=5.4cm]{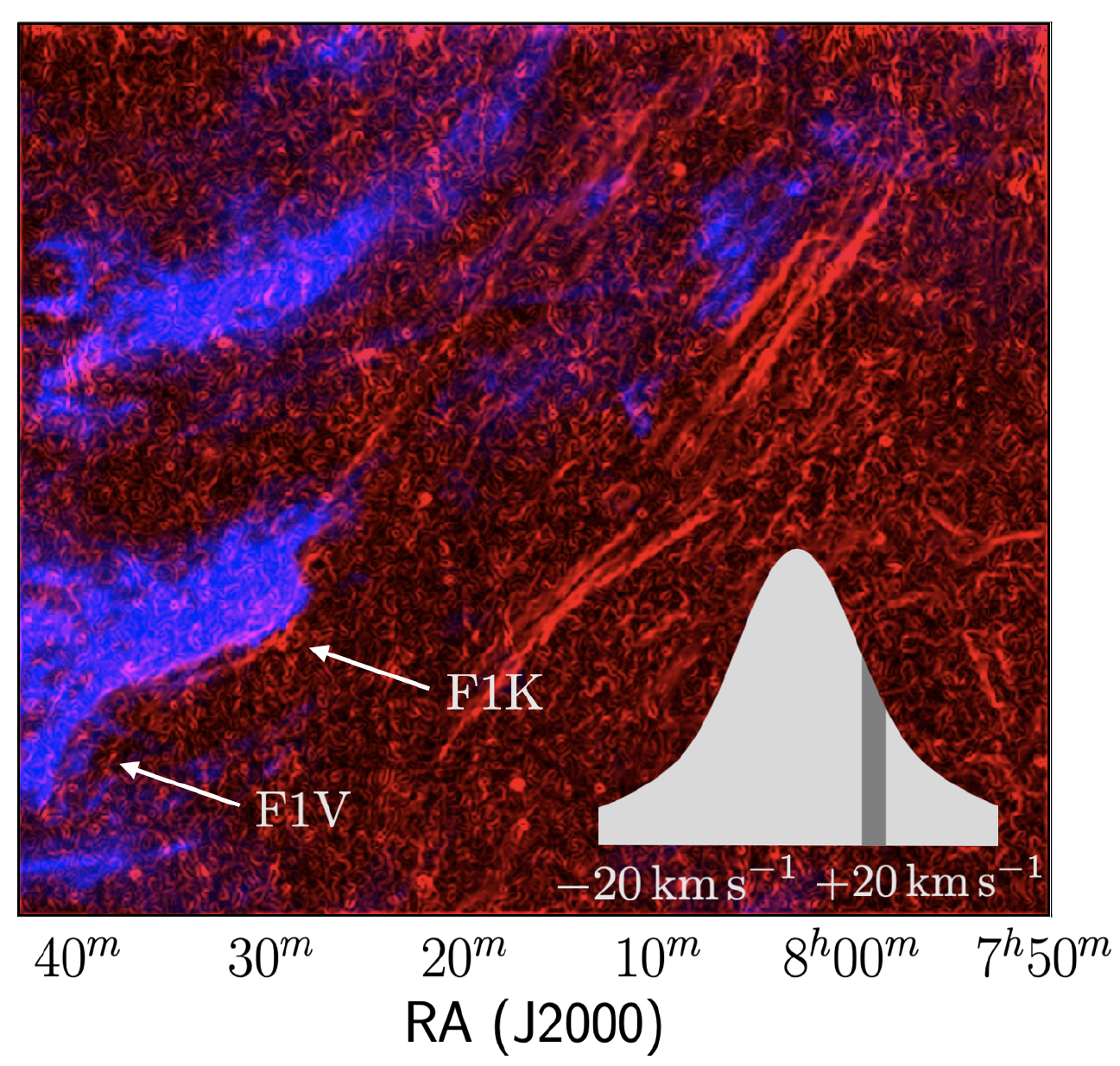}
 \caption{De-striped GALFACTS {\polgradmax\!\!} of G216 at {5\arcmin} resolution (red) with GALFA-{\HI} velocity channel maps (blue) at a $\vlsr$ of ${-}\,5.5\,\kms$ (left), ${+}0.4\,\kms$ (middle), and ${+}7.0\,\kms$ (right) overlaid. {\polgradmax\!\!} is on a linear scale spanning $0{-}0.01\,{\rm K\,arcmin^{-1}}$ and the {\HI} channel maps are on a linear scale over $2{-}4\,{\rm K}$ (left), $4{-}9\,{\rm K}$ (middle), and $5{-}18\,{\rm K}$ (right). Locations of F1 and F2 associated with {\HI} emission as discussed in the text are indicated.}
\label{fig:polgrad_HI}
\end{figure*}

The polarization gradient filament farthest from the Galactic plane (F1) is a relatively simple structure that contains a knee-shaped bend (F1K) toward the southern end. The F1K portion of F1 exhibits a clear double-jump morphology (i.e., two parallel filamentary structures). Double-jump morphologies in the polarization gradient have been attributed to a delta or top-hat profile in $n_e$ or $B_\parallel$ from strong shocks \citep{Burkhart2012}. South of F1K, the filament extends into a V-shape structure (F1V) where the filament breaks off into two filamentary structures. While F1V is poorly seen in the {5\arcmin} resolution polarization gradient, it can be seen in the Stokes maps and, with significantly greater levels of smoothing, is observed at lower resolution maps of the polarization gradient. Further smoothing does not significantly change the polarization gradient map.

The polarization gradient filament closer to the Galactic plane (F2) is more complex than F1, containing three filaments toward the north (F2N) and two filaments toward the south (F2S). While F2S contains a double-jump feature, one of these filaments is straight while the other is slightly curved, unlike the F1K parallel double-jump morphology. The F2N triple-jump morphology comprises three unequal-length parallel filaments.

The $I_{1.4}$ data contain background extragalactic point sources in addition to broad vertical and horizontal features from a combination of scanning artefacts and quasi-stationary RFI \citep{Leahy2018}. As defined by our search criteria, we find no spatial correlation between $I_{1.4}$ and the structures described here.

\subsection{21 cm {\HI} Emission}\label{subsec:results-HI}

Figure \ref{fig:HI_S1} shows integrated column density maps of the velocity-resolved GALFA-{\HI} data across the full velocity range that contains emission, calculated from the brightness temperature maps under the optically thin assumption

\begin{align}\label{eq:NHI}
    {\NHI} = 1.823\times10^{18}\,{\rm cm^{-2}} \int \left(\frac{T_B}{\rm K}\right)\left(\frac{{\rm d}\vlsr}{\kms}\right),
\end{align}

{\noindent}a reasonable assumption to make at these latitudes \citep{Murray2018}. The {\HI} data have been integrated over ${-30}\,{\leq}\,{v_{\rm lsr}}{\leq}{-10}\,\kms$ (top left), ${-10}\,{\leq}\,{v_{\rm lsr}}\,{\leq}\,{+4}\,\kms$ (top right), ${+4}\,{\leq}\,{v_{\rm lsr}}\,{\leq}\,{+20}\,\kms$ (bottom left), and ${+20}\,{\leq}\,{v_{\rm lsr}}\,{\leq}\,{+40}\,\kms$ (bottom right) to show various features in the data.

At more negative velocities (Figure \ref{fig:HI_S1}, top left) the {\HI} emission is localized to the northeast corner ($N_{HI}\,{\sim}\,3.5\,{\times}\,10^{19}\,{\rm cm^{-2}}$) and appears clumpy with finger-like projections. At negative velocities closer to zero (Figure \ref{fig:HI_S1}, top right), the {\HI} emission is more extended and contains a filament that appears similar to F2 ($N_{HI}\,{\sim}\,2\,{\times}\,10^{20}\,{\rm cm^{-2}}$) and breaks off into another V-shape structure (F2V, $N_{HI} \sim 1.7\,{\times}\,10^{20}\,{\rm cm^{-2}}$). At positive velocities closer to zero (Figure \ref{fig:HI_S1}, bottom left), the {\HI} emission exhibits the F1K morphology ($N_{HI}\,{\sim}\,3.8\,{-}\,5.4\,{\times}\,10^{20}\,{\rm cm^{-2}}$) and contains emission along the edge of F1V ($N_{HI}\,{\sim}\,2.3 \times 10^{20}\,{\rm cm^{-2}}$). At higher positive velocities (Figure \ref{fig:HI_S1}, bottom right), the {\HI} gas is primarily located in the southwest corner ($N_{HI}\,{\sim}\,4.5\,{\times}\,10^{19} - 1\,{\times}\,10^{20}\,{\rm cm^{-2}}$) and does not resemble any structures associated with the local gas. The {\HI} emission at $|{\vlsr}|\,{\gtrsim}\,20\,\kms$ (Figure \ref{fig:HI_S1}, top left and bottom right) contains non-local gas with velocities consistent with intermediate velocity clouds (IVCs).

\subsection{Comparison of Radio Polarization Gradient and {\HI} Structures}\label{subsec:polgrad-HI}

We compared structures in the polarization gradient to {\HI} emission. Figure \ref{fig:polgrad_HI} shows 1.4 GHz GALFACTS polarization gradient (red) with GALFA-{\HI} velocity channel maps (blue) overlaid at a {\vlsr} of ${-}\,5.5\,\kms$ (left), ${+}0.4\,\kms$ (middle), and ${+}7.0\,\kms$ (right). These multi-phase tracers clearly share a common orientation and similar morphologies in this region, although they are not always spatially coincident. The observed alignment is quantified later in Section \ref{subsec:polgrad_vs_hi}.

The {\HI} emission shows similar morphological structures and orientations to the polarization gradient across a range of velocities. At ${\vlsr}\,{\sim}\,{-5.5}\,{\kms}$ (Figure \ref{fig:polgrad_HI}, left), there is a prominent filamentary structure that is aligned with F2S. In light of this alignment, we consider this {\HI} filament to be morphologically associated with F2. At ${\vlsr}\,{\sim}\,{{+}0.4}\,{\kms}$ (Figure \ref{fig:polgrad_HI}, middle), only the F2S portion is seen in {\HI} and extends farther south, highlighting F2V. At ${\vlsr}\,{\sim}\,{{+}7}\,{\kms}$ (Figure \ref{fig:polgrad_HI}, right), the {\HI} emission is more extended and `fills in' F1K.

\subsection{{\halpha\!\!} Emission}\label{subsec:results-halpha}

\begin{figure*}[t!]
\centering
\hspace*{-5pt}\includegraphics[align=c,width=6.7cm]{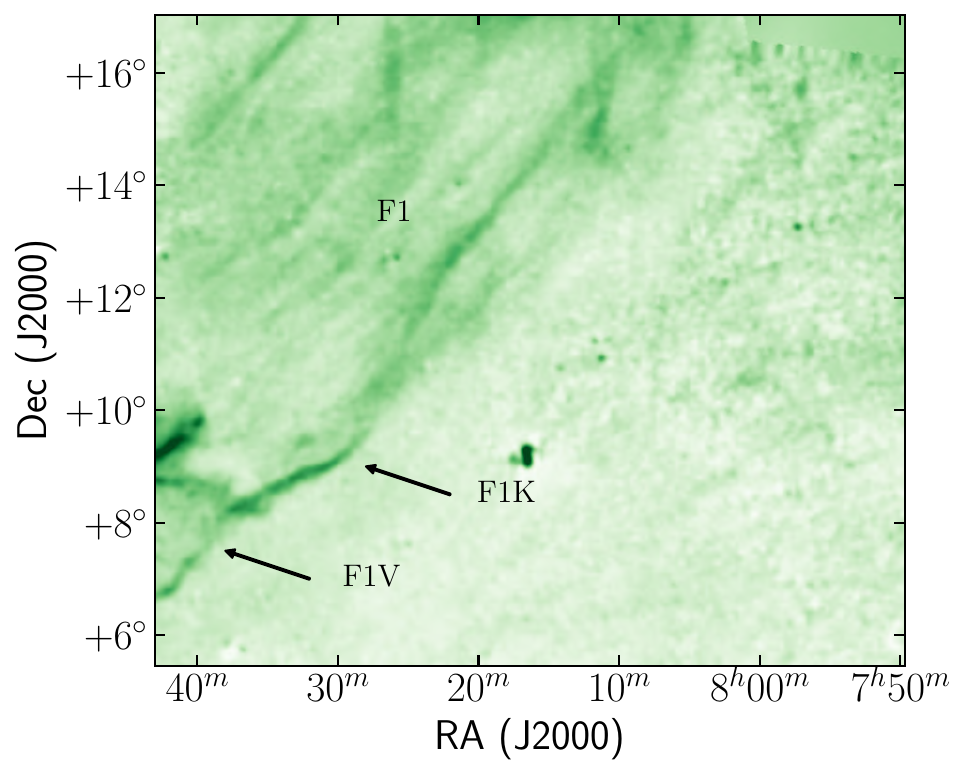}\hspace*{-5pt}\includegraphics[align=c,width=6.3cm]{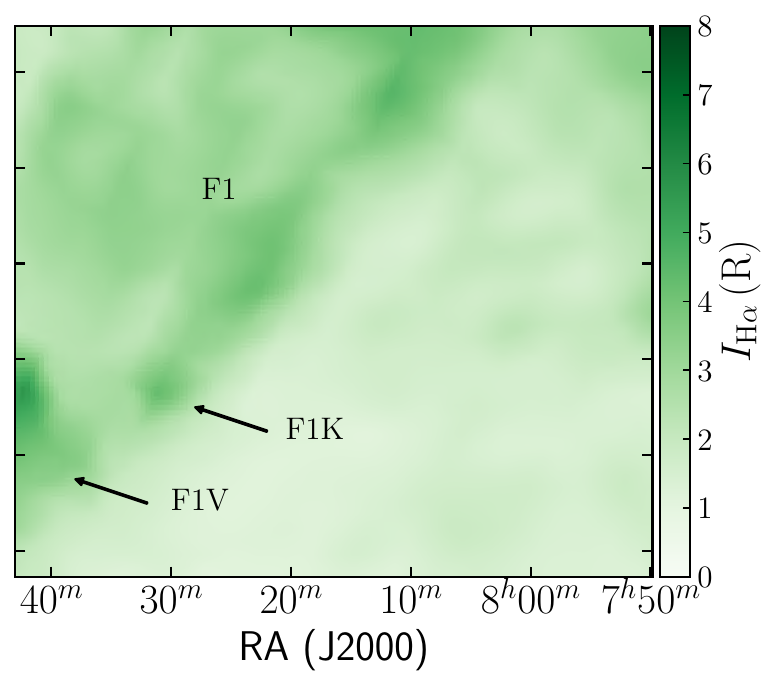}\includegraphics[align=c,width=5.5cm]{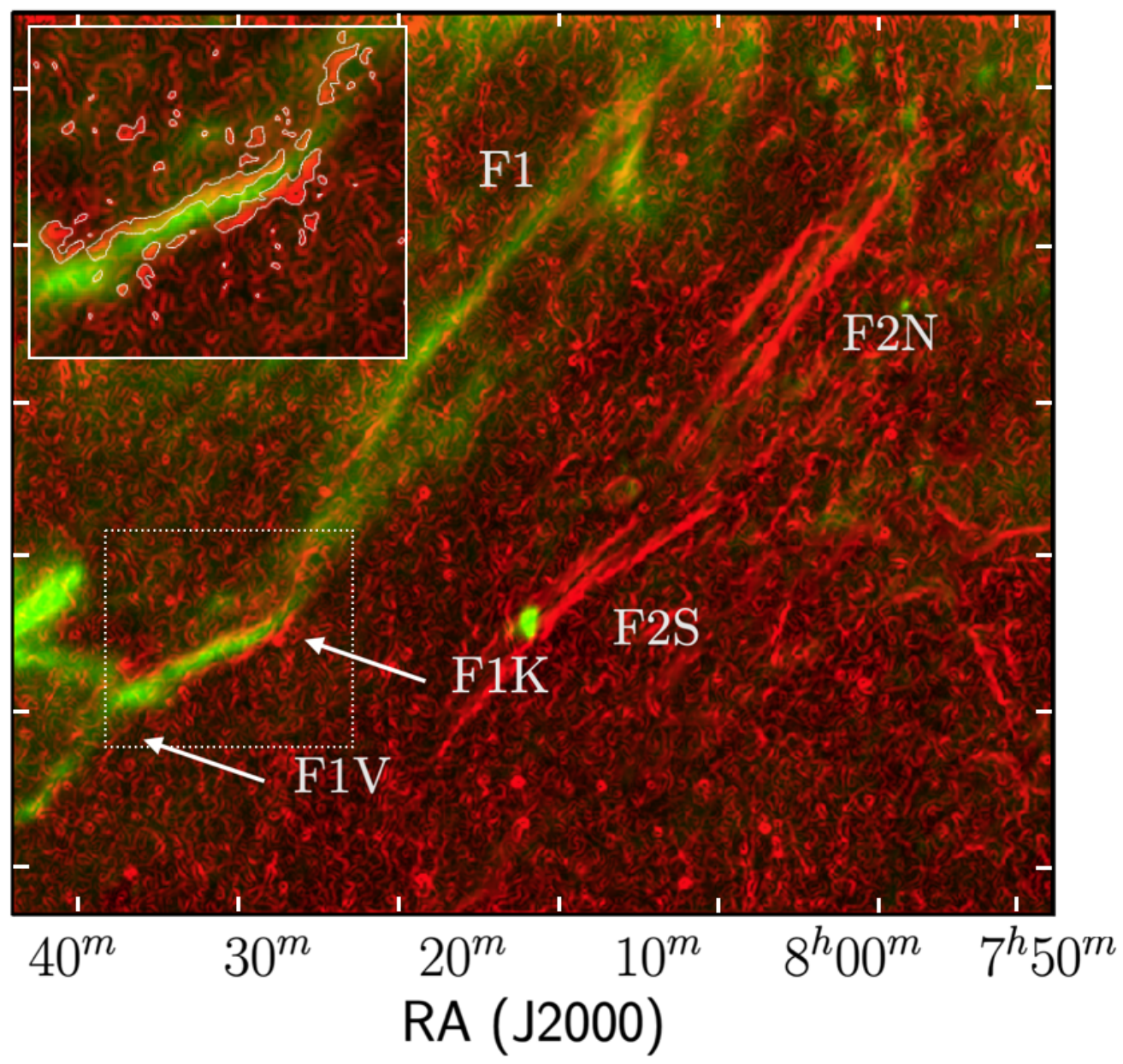}
\caption{(\textbf{Left:}) VTSS {\halpha\!\!} emission of G216. (\textbf{Middle:}) WHAM {\halpha\!\!} emission at ${\vlsr}\,{=}\,{{+}7.3\,\kms}$, the approximate peak {\halpha\!\!} velocity of F1. (\textbf{Right:}) {\polgradmax\!\!} (red) with VTSS {\halpha\!\!} emission (green) overlaid. The figure inset shows the F1K region with a 0.006 ${\rm K\,arcmin^{-1}}$ level contour of {\polgradmax\!\!} (white) overlaid. The {\polgradmax\!\!} intensity is on a linear scale from $0{-}0.01\,{\rm K\,arcmin^{-1}}$ and the {\halpha\!\!} intensity is on a squared intensity from $0{-}6\,{\rm R}$. The locations of F1 and F2 as discussed in the text are indicated.}
\label{fig:halpha}
\end{figure*}

G216 is covered by VTSS in the composite {\halpha\!\!} map \citep{Finkbeiner2003} and contains a prominent {\halpha\!\!} filament ($I_{\rm H\alpha}\,{\sim}\,3\,{-}\,7\,{\rm R}$, where $1\,{\rm R}\equiv10^6/4\pi\,{\rm photons\,cm^{-2}\,s^{-1}\,sr^{-1}}$) shown in Figure \ref{fig:halpha} (left). The VTSS {\halpha\!\!} filament clearly contains the F1K ($I_{\rm H\alpha}\,{\sim}\,3\,{-}\,5\,{\rm R}$) and F1V ($I_{\rm H\alpha}\,{\sim}\,3\,{-}\,7\,{\rm R}$) morphologies. This filament runs along the edge of extended emission ($I_{\rm H\alpha}\,{\sim}\,3\,{\rm R}$) that is primarily located in the northeast corner with an intensity that is roughly twice as high as that of the southwest corner where there is little emission. There is also a suggestion of linear features in the extended {\halpha\!\!} emission in the north east corner with roughly the same orientation as F1.

Figure \ref{fig:halpha} (middle) shows the WHAM {\halpha\!\!} emission at a velocity of ${\vlsr}\,{=}\,{{+}7.3\,\kms}$, corresponding to the approximate peak velocity of the emission toward F1. The small-scale {\halpha\!\!} filament seen in the VTSS data corresponds to a larger, more extended filament in the WHAM data ($I_{\rm H\alpha}\,{\sim}\,4\,{\rm R}$). Similar to the VTSS results, the WHAM filament lies along the edge of lower intensity diffuse emission ($I_{\rm H\alpha}\,{\sim}\,3\,{\rm R}$) in the north east corner that is roughly twice as high as that of the south west.

The VTSS filament is spatially coincident with F1 and lies between the double-jump morphology of F1K. Figure \ref{fig:halpha} (right) shows the correspondence between the polarization gradient (red) and VTSS {\halpha\!\!} emission (green).

\subsection{Thermal Dust Emission}\label{subsec:results-dust}

\begin{figure*}[t!]
\centering
\includegraphics[align=c,height=7cm]{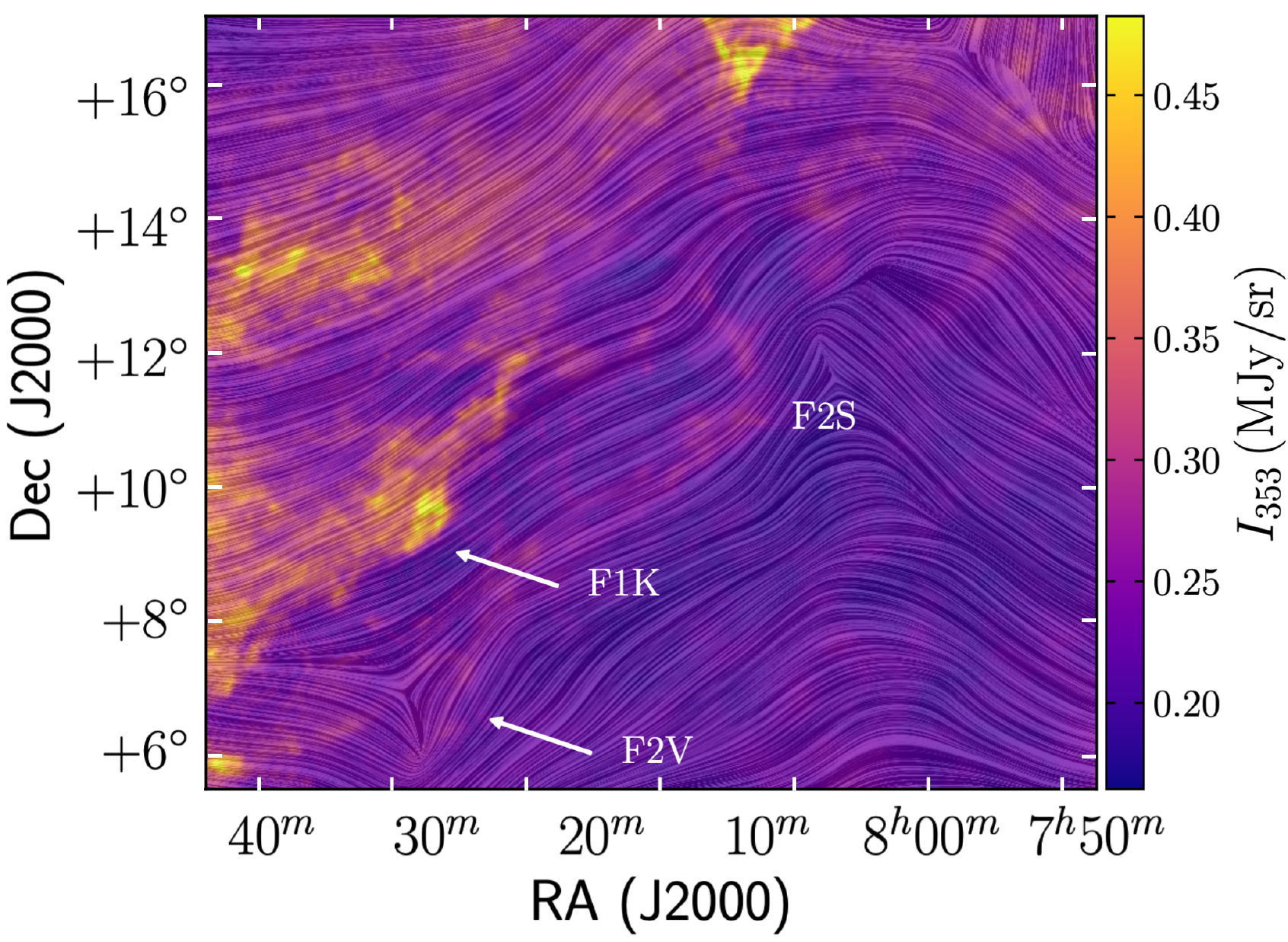}\hspace*{10pt}\includegraphics[align=c,height=7cm]{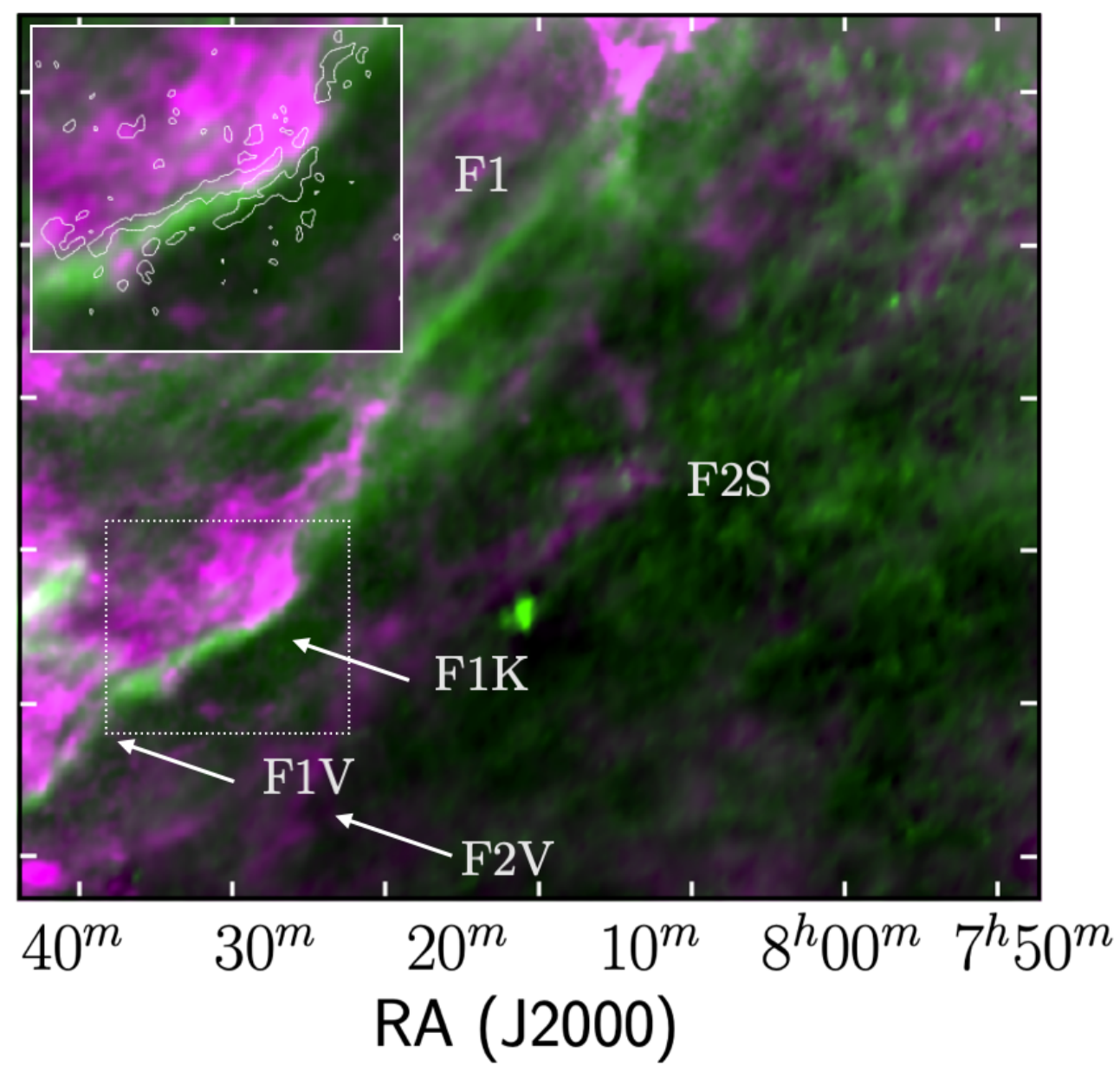}
\caption{{\Planck} $I_{353}$ thermal dust emission (magenta) of G216. (\textbf{Left:}) The POS magnetic field orientation derived using $Q_{353}$ and $U_{353}$ are overlaid as textured lines. Locations of F1 and F2 associated with dust emission as discussed in the text are indicated. (\textbf{Right:}) VTSS {\halpha\!\!} emission (green) is shown to highlight the spatial separation of the ionized and neutral media along F1. The figure inset shows the F1K region with a 0.006 ${\rm K\,arcmin^{-1}}$ level contour of {\polgradmax\!\!} (white) overlaid. The {\halpha\!\!} intensity is on a linear scale spanning $0.5{-}7\,{\rm R}$ and the $I_{353}$ intensity is on a linear scale over $0.2{-}0.4\,{\rm MJy\,sr^{-1}}$.}
\label{fig:planck-halpha}
\end{figure*}

Figure \ref{fig:planck-halpha} (left) shows the {\Planck} $I_{353}$ dust thermal emission. The dust emission is strongly correlated with the local {\HI} gas found at velocities ${-10}\,{\lesssim}\,{\vlsr}\,{\lesssim}\,{+20\,\kms}$, showing the same F1K ($I_{353}\sim0.4\,{\rm MJy\,sr^{-1}}$), F2S ($I_{353}\sim0.2\,{\rm MJy\,sr^{-1}}$) and F2V ($I_{353}\,{\sim}\,0.3\,{\rm MJy\,sr^{-1}}$) morphologies. Along F1K and F1V, the VTSS filament is slightly displaced from the edge of the dust emission toward the direction of the Galactic plane. Figure \ref{fig:planck-halpha} (right) shows the VTSS {\halpha\!\!} emission (green) with the {\Planck} $I_{353}$ dust emission (magenta) overlaid to highlight this spatial separation between the ionized and neutral medium along F1.

\subsection{The Surrounding Area}\label{subsec:results-surroundings}

\begin{figure}[t!]
\centering
\vspace*{5pt}\includegraphics[width=8.6cm]{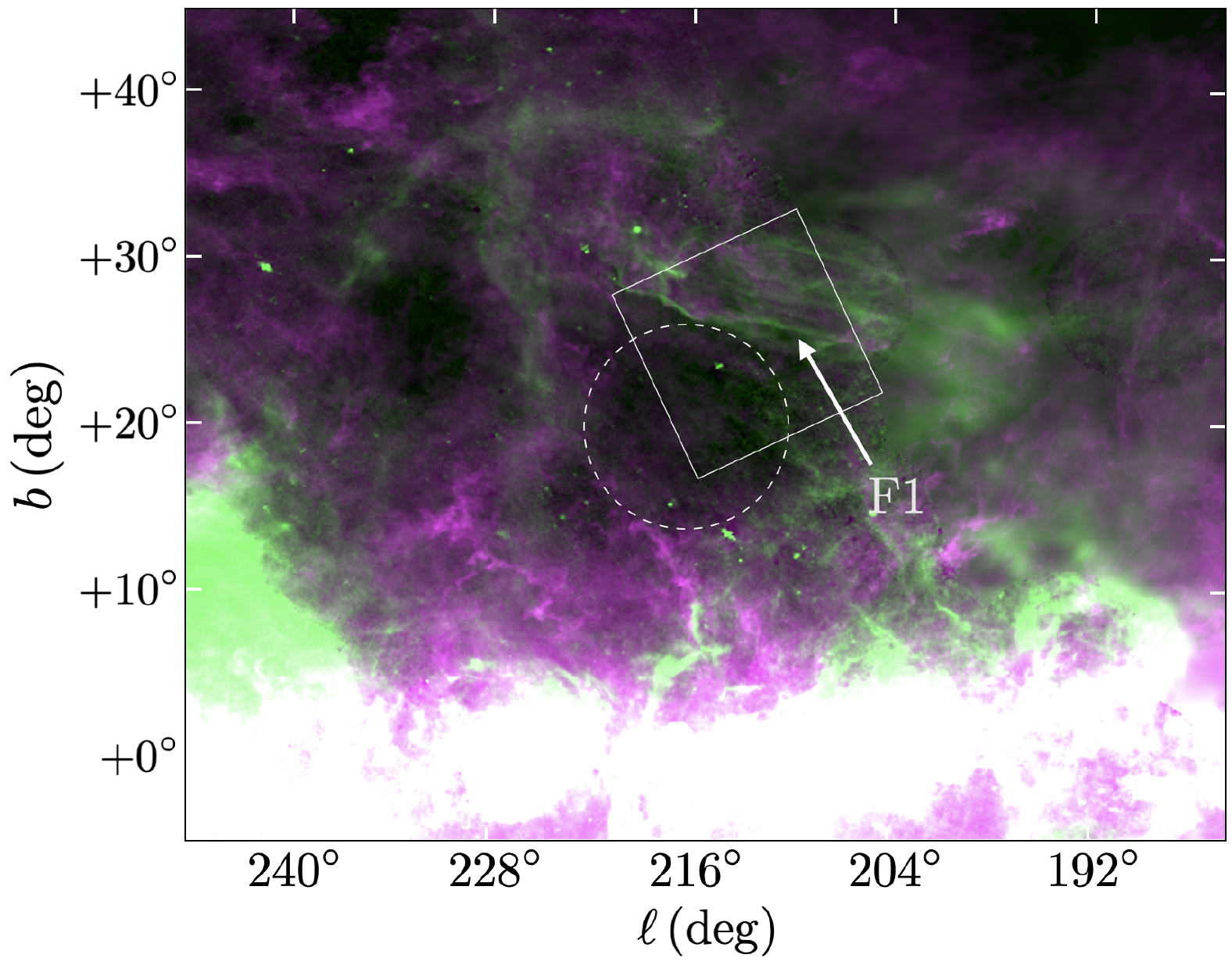}
\caption{VTSS {\halpha\!\!} (green) and {\Planck} $I_{353}$ dust (magenta) emission for a larger ${\sim}\,50\deg\,{\times}\,50\deg$ field of the surrounding area in Galactic coordinates. The footprint of the region presented in this paper and the low column density region is indicated by the rectangle (white) and dotted circle (white), respectively. The location of F1 is also indicated. The {\halpha\!\!} intensity is on a linear scale over $0.1{-}10\,{\rm R}$ and the $I_{353}$ dust intensity is on a square root scale spanning $0.2{-}1.5\,{\rm MJy\,sr^{-1}}$.}
\label{fig:field}
\end{figure}

The G216 filaments lie along the northern edge (in Galactic coordinates) of a low column density region seen in the projected VTSS {\halpha\!\!} ($I_{\rm H\alpha}\,{\sim}\,1.2\,{\rm R}$), {\HI} (${\NHI}\,{\sim}\,3\,{\times}\,10^{20}\,{\rm cm^{-2}}$), and {\Planck} dust ($I_{353}\,{\sim}\,0.2\,{\rm MJy\,sr^{-1}}$) emission just above the Galactic plane. Figure \ref{fig:field} shows a ${\sim}\,50\deg\,{\times}\,50\deg$ region of the surrounding area in VTSS {\halpha\!\!} emission (green) and {\Planck} $I_{353}$ dust emission (magenta) in Galactic coordinates. The G216 footprint investigated in this paper is indicated with a rectangle (white) and the approximate location of the projected low column density region is shown with a ${\sim}10\deg$ diameter dotted circle (white) centered on $(\ell,b)\,{\sim}\,(215\deg,+20\deg)$. The location of F1 is indicated and is clearly oriented roughly parallel to the Galactic plane. This cavity is roughly circular in dust emission and triangular in {\halpha\!\!} emission. The {\halpha\!\!} emission that is morphologically associated with our field extends as far west as $\ell\,{\sim}\,195\deg$ but this is beyond the VTSS footprint in a region where the GALFACTS are yet to be processed. We do not extend our footprint farther east as the current footprint contains the full extent of the polarization gradient filaments in this region.

There are two nearby {\halpha\!\!} filaments that were located using WHAM data (\citealt{Haffner1998}; see their Figure 1 that includes our footprint). There is a vertical (in Galactic coordinates) {\halpha\!\!} filament located at $\ell\,{\sim}\,225\deg$ centered on ${\vlsr}\,{=}\,{+20}\,\kms$ with a curved filament found at its northern end where the peak emission extends from $\ell\,{\sim}\,220\deg$ centered on ${\vlsr}\,{=}\,{-20}\,\kms$ to $\ell\,{\sim}\,235\deg$ centered on ${\vlsr}\,{=}\,{+20}\,\kms$. The magnitude of these velocities are higher than the peak velocity of ${\vlsr}\,{=}\,{7.3\,\kms}$ that we find for the F1-associated {\halpha\!\!} emission. We do not find polarization gradient structures associated with these {\halpha\!\!} filaments using the {5\arcmin} GALFACTS data. The F1K structure of our field as well as the nearby {\halpha\!\!} filaments \citep{Haffner1998} can also be seen in Figure 2 of \citet{Finkbeiner2003}.

\section{Analysis}\label{sec:analysis}


\subsection{Synchrotron Polarization}

We computed the POS synchrotron pseudo magnetic field orientation via $\theta_{1.4}\,{=}\,(1/2)\arctan(U_{1.4}/Q_{1.4})$. We first multiplied $Q_{1.4}$ and $U_{1.4}$ by $-1$ corresponding to a {90\deg} rotation in the polarization angle since the magnetic field orientation is orthogonal to that of the electric field. De-rotating the synchrotron polarization angles can in principle remove Faraday rotation effects and provide the intrinsic magnetic field direction. However, the structures in this paper are not spatially correlated with excess synchrotron emission, so de-rotation would yield the magnetic field orientation of the background synchrotron-emitting source rather than the foreground Faraday-rotating structures of interest. As a measure of the combined magnetic field orientation and Faraday rotation, $\theta_{1.4}$ remains a valuable quantity.

Figure \ref{fig:GALFACTS_PLIC} shows the {5\arcmin} polarization gradient map with $\theta_{1.4}$ overlaid as pseudo vectors (white, to distinguish this from the true magnetic field direction) with lengths that are proportional to $P_{1.4}$. The orientation of $\theta_{1.4}$ is noticeably coherent in many areas of the field, particularly in the northeast and southwest corners as well as along the length of F2. There are drastic changes in $\theta_{1.4}$ along F1K and F1V, as well as along the edges of F2 where the pseudo magnetic field orientation suddenly becomes oriented parallel to the filaments.


\subsection{Dust Polarization}

We computed the POS magnetic field orientation of the dust emission using the 80{\arcmin} {\Planck} $Q_{353}$ and $U_{353}$ maps via $\theta_{353}=(1/2)\arctan(U_{353}/Q_{353})$. The Stokes maps were both first multiplied by $-1$, corresponding to a {90\deg\!\!} rotation in the polarization angle. Figure \ref{fig:planck-halpha} (left) shows $\theta_{353}$ overlaid on $I_{353}$ as textured lines using \textsc{LicPy} \citep{Cabral1993}.\footnote{\href{https://rufat.be/licpy/}{https://rufat.be/licpy/}} The magnetic field direction is generally parallel to F1 and F2, except where it follows the morphology of F2V and changes direction around a right ascension of ${\sim}\,8^h$. 

\subsection{Magnetically-Aligned, Cold {\HI} Structures}\label{subsec:hifibers}
    
The identification of magnetically-aligned {\HI} structures was made using the Rolling Hough Transform (RHT),\footnote{\href{https://github.com/seclark/RHT}{https://github.com/seclark/RHT}} a machine vision algorithm that identifies and parameterizes coherent, linear structures in an image \citep{Clark2014}. The RHT provides an array of linear intensities binned by an angle $\theta$ measured from the vertical for each pixel, $R(\theta,x,y)$, quantifying the probability that each pixel is associated with a coherent linear structure. The RHT backprojection visualizes the resulting linear features by integrating $R(\theta,x,y)$ over $\theta$. The RHT uses three input parameters: a smoothing kernel diameter $D_K$, a window diameter $D_W$, and a linear intensity threshold $Z$. Details on the RHT procedure are further described by \citet{Clark2014}.

We applied the RHT to the GALFA-{\HI} data over $-20\,{\lesssim}\,v_{\rm lsr}\,{\lesssim}\,{+20}\,\kms$, integrating over four spectral channels corresponding to a ${\sim}\,3\,\kms$ velocity binning. Based on the results of \citet{Clark2014}, we use the RHT parameters $D_W\,{=}\,100\arcmin$, $D_K\,{=}\,10\arcmin$, and $Z\,{=}\,70\%$. The {\HI} RHT backprojections are shown in Figure \ref{fig:hi_rht}.

To investigate the thermal properties of narrow {\HI} structures, we used the Regularized Optimization for Hyper-Spectral Analysis ({\ROHSA}) Gaussian decomposition code \citep{Marchal2019}.\footnote{\href{https://github.com/antoinemarchal/ROHSA}{https://github.com/antoinemarchal/ROHSA}} The complexity of the {\HI} line varies considerably over this region, as does the number of Gaussians required to describe them. This is particularly true along LOSs that contain IVCs. As a result, the fixed number of Gaussians in {\ROHSA} presents a considerable challenge for G216: increasing the number of Gaussians to accurately fit the IVCs causes other regions become over-fit, affecting the overall phase separation. Overcoming this requires subdividing the region into several smaller sub-regions and creating a mosaic of the resulting phase-separated components. This is beyond the scope of this work, so we focus on a single sub-region toward F2 (region A) indicated with a dotted rectangle (white) in the top right panel of Figure \ref{fig:HI_S1}.

\begin{figure}[t!]
\centering
\includegraphics[width=8.7cm]{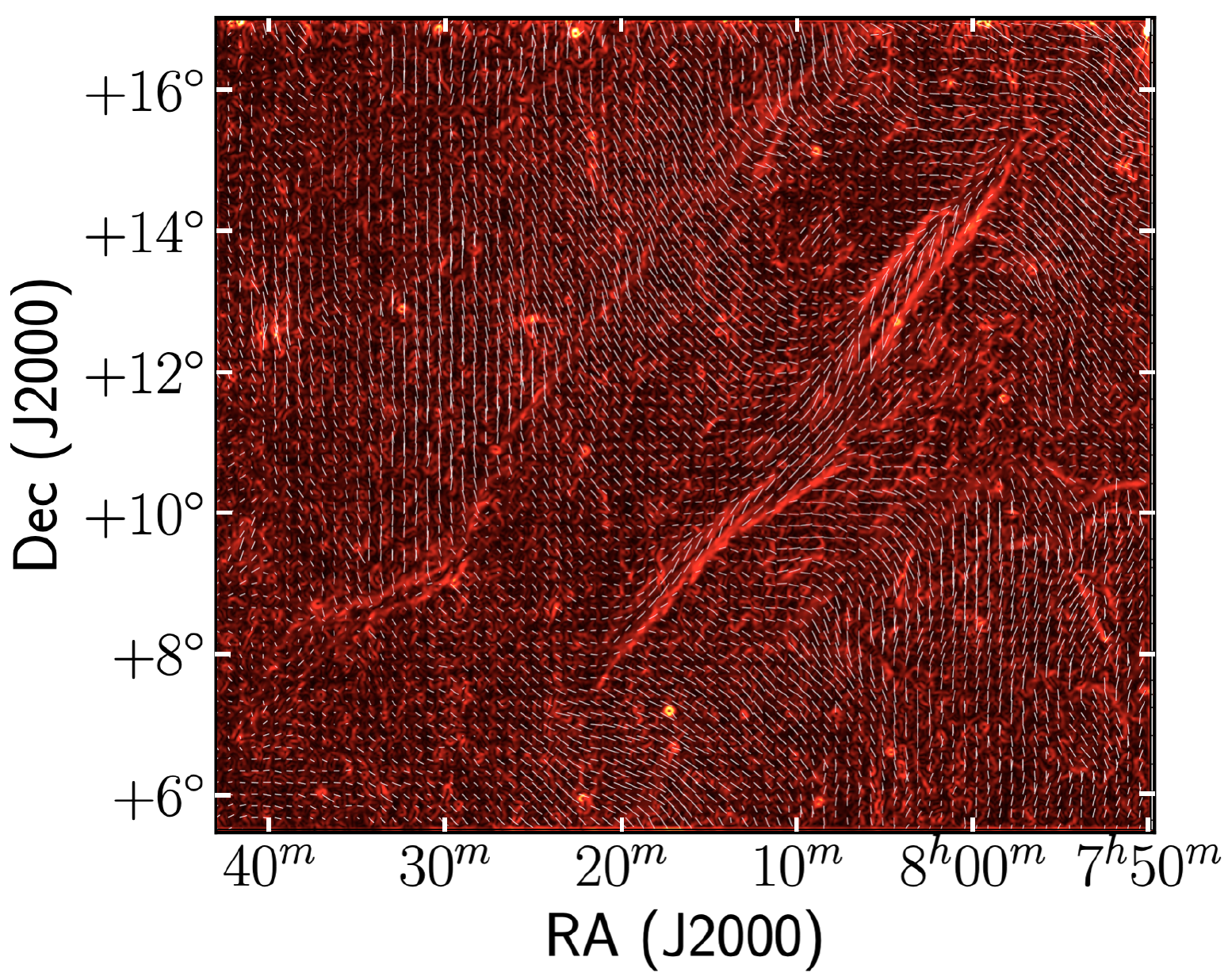}
\caption{De-striped GALFACTS {\polgradmax\!\!} of G216 at {5\arcmin} resolution, with the polarization angles derived from $Q_{1.4}$ and $U_{1.4}$ rotated by $90\deg$ shown as line segments with lengths that are weighted by $P_{1.4}$. The {\polgradmax\!\!} intensity scale is the same as that shown in figure \ref{fig:stokes_S1} (bottom right).}
\label{fig:GALFACTS_PLIC}
\end{figure}

{\ROHSA} is based on a regularized nonlinear least-square criterion that accounts for the spatial coherence of {\HI} emission by applying a Laplacian filtering to the parameter maps that penalize small spatial frequencies, and the multi-phase nature of the atomic medium by minimizing the variance in velocity dispersion. We set the maximum number of Gaussians fit with {\ROHSA} along each LOS to five, and set the four hyper-parameters that control the strength of the regularization equal to 10 (see \citealt{Marchal2019} for details). The magnitude of the hyper-parameters was empirically chosen to converge toward a noise-dominated residual that is roughly constant across the field (with an average reduced chi-square of ${\sim}\,{1.4}$) while minimizing the number of Gaussians used. A decomposition using six Gaussians does not qualitatively change our results for the local gas.

\begin{figure*}[t!]
  \centering
  \includegraphics[width=9.6cm]{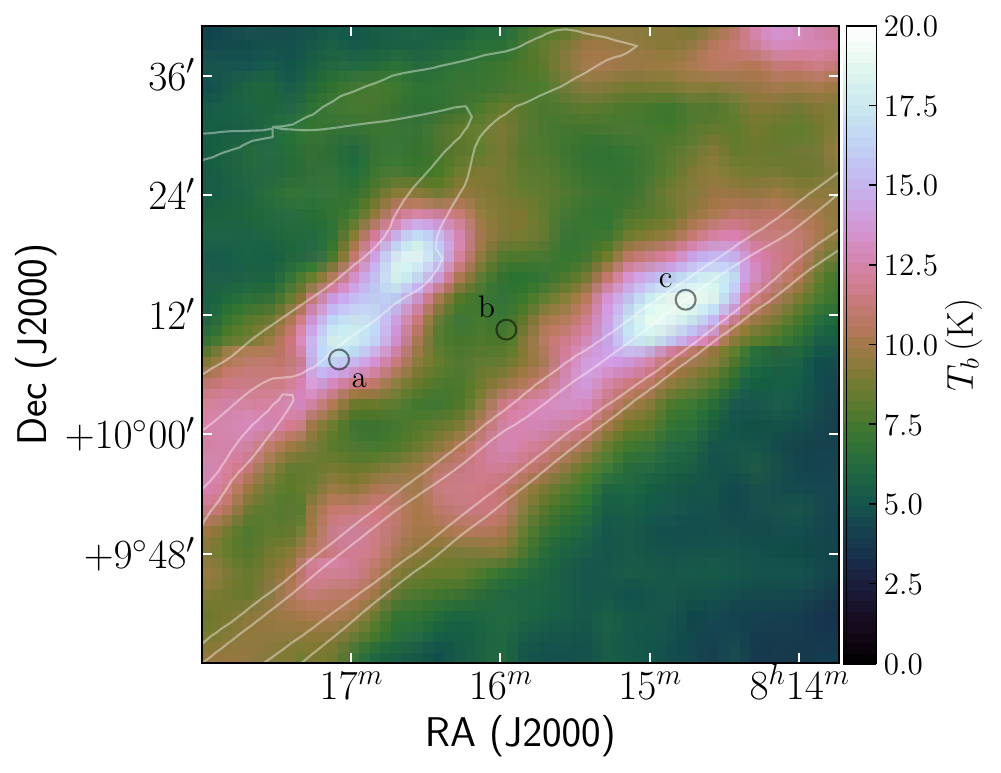}\includegraphics[width=7.8cm]{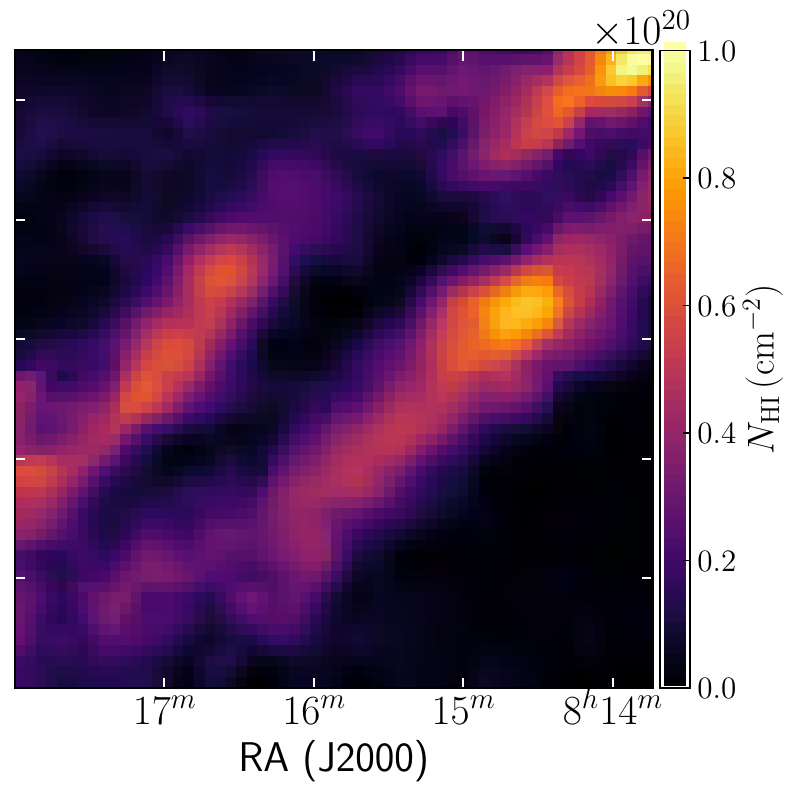}
  \caption{Small subset of the GALFA-{\HI} data toward F2, designated region A, as indicated with the dotted rectangle (white) in the top right figure of Figure \ref{fig:HI_S1}. (\textbf{Left:}) Velocity channel map at the peak of the mean spectrum corresponding to ${\vlsr}\,{=}\,{-1.8\,\kms}$. The RHT backprojection is overlaid as contours (white) using levels of 0.1 and 0.5. The circles (black) mark the lines of sight toward which we show the phase-separated spectra in Figure \ref{fig:mosaic_spectra_selected}. (\textbf{Right:}) Column density map of the local CNM emission ($G_1$) derived with {\ROHSA}.}
  \label{fig:ROHSA}
\end{figure*}

\begin{figure*}[t!]
    \centering
    \includegraphics[width=15cm]{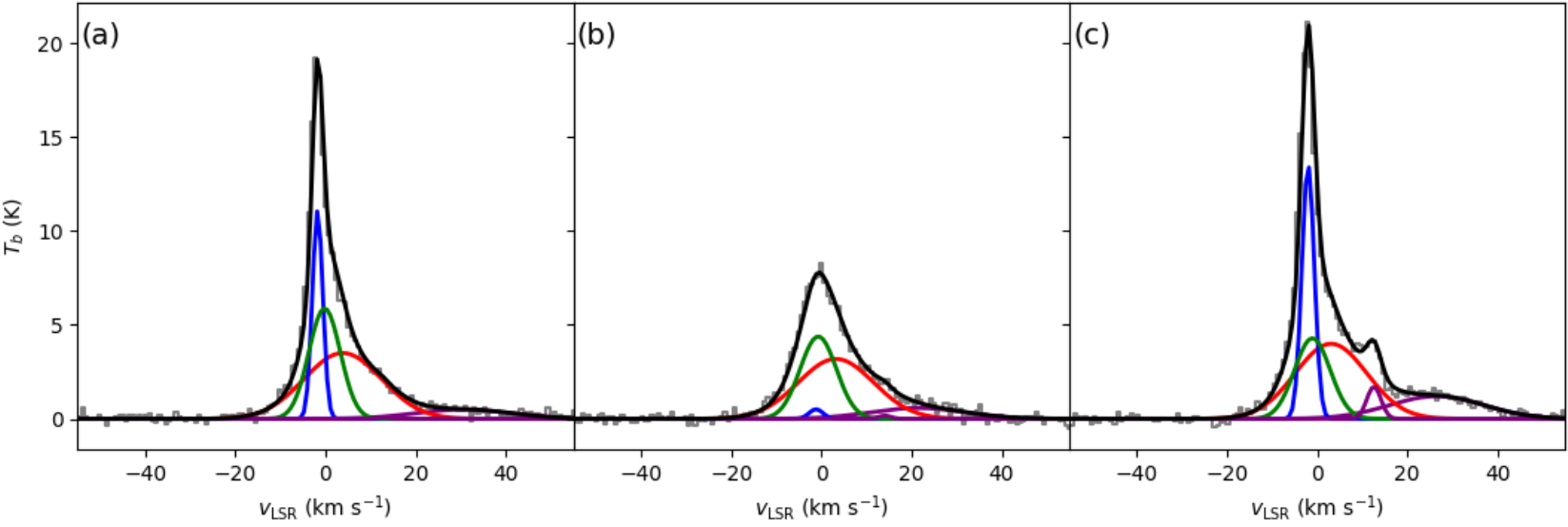}
    \caption{{\ROHSA} Gaussian decomposition for the three LOSs toward region A indicated in Figure \ref{fig:ROHSA} (left). The GALFA-{\HI} data is shown in gray while the total emission described by {\ROHSA} is shown in black. The CNM, UNM, and WNM components of the local gas are shown in blue, green, and red, respectively. The two components used to describe the IVC are shown in purple.} 
    \label{fig:mosaic_spectra_selected}
\end{figure*}

The local {\HI} emission with $|{\vlsr}|\,{\lesssim}\,{10}\,\kms$ was well represented using three Gaussian components, $G_1$, $G_2$, and $G_3$, with mean velocity dispersions weighted by column density, respectively, of ${\langle\sigma_1\rangle}\,{=}\,{1.4\,\kms}$, ${\langle\sigma_2\rangle}\,{=}\,{4.0\,\kms}$ and ${\langle\sigma_3\rangle}\,{=}\,{8.4\,\kms}$, reminiscent of the three-phase neutral ISM composed of CNM, UNM and WNM. These Doppler linewidths are a mixture of thermal and turbulent contributions and lead to upper-limit kinetic temperatures
of $T_{K,1}\,{\sim}\,250$\,K, $T_{K,2}\,{\sim}\,1990$\,K and $T_{K,3}\,{\sim}\,8600$\,K, respectively. Gaussians $G_4$ and $G_5$ were used to encode intermediate velocity gas with a velocity centroid ${\vlsr}\,{\gtrsim}\,10\,\kms$.

\begin{figure*}[t!]
\centering
 \includegraphics[width=18cm]{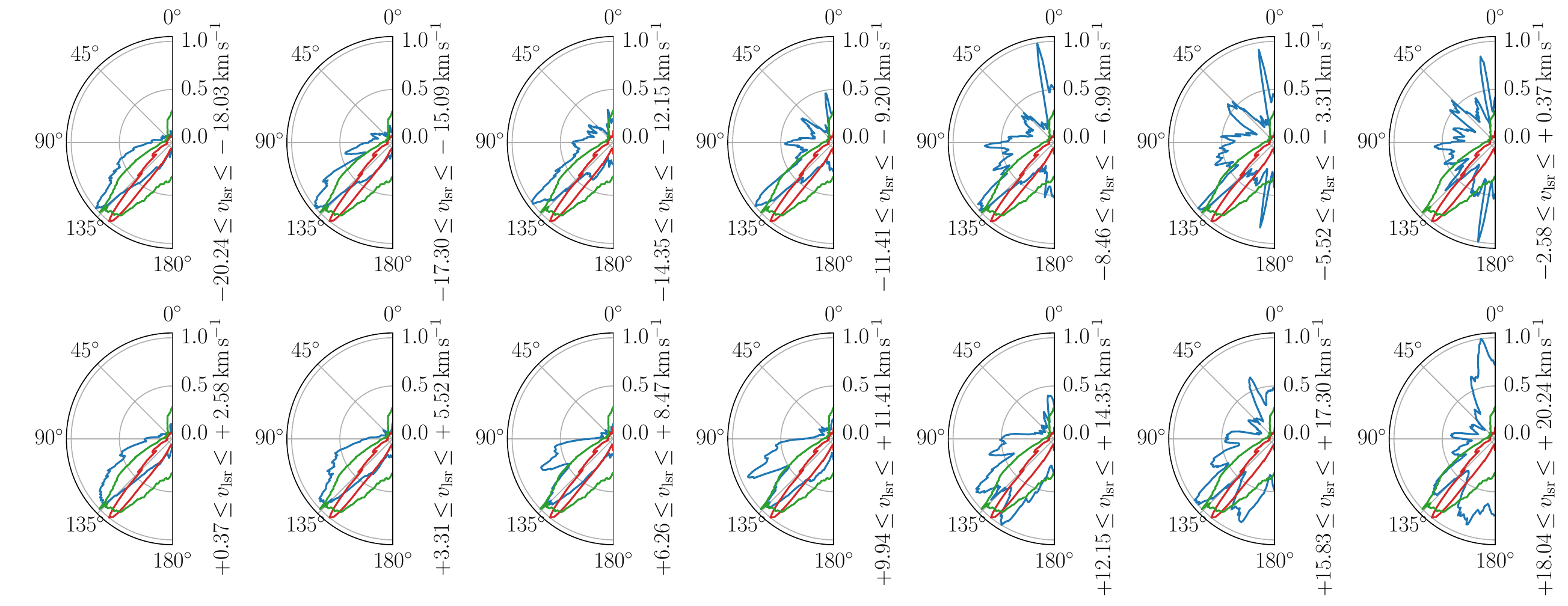}
\caption{RHT orientations of the de-striped {5\arcmin} {\polgradmax\!\!} (red), {\halpha\!\!} emission (green), and velocity-channel {\HI} fibers (blue). The distributions of {\polgradmax\!\!} and {\halpha\!\!} are the same in each panel while those of {\HI} are a function of binned {\HI} velocity. All histograms have been normalized to the same intensity scale and rotated such that an orientation of $0\deg$ is vertical in the image plane.}
\label{fig:HI_vs_rht_histograms}
\end{figure*}

Figure \ref{fig:ROHSA} (left) shows the {\HI} brightness temperature map at ${\vlsr}\,{=}\,{-1.8\,\kms}$ corresponding to the peak of the average spectrum in region A. The contours (white) show the RHT backprojection applied to the {\HI} channel map with a velocity channel width of $0.8\,\kms$ using the same parameters as those used on the binned velocity channels. The column density map of the local CNM component, of particular interest for our RHT analysis, is shown in Figure \ref{fig:ROHSA} (right) under the optically thin assumption of Equation \ref{eq:NHI}. The small-scale filamentary structures identified with the RHT in the peak channel map are strongly correlated with the CNM component. 

We inspected the spectral decomposition toward three lines of sight. Figure \ref{fig:mosaic_spectra_selected} shows the spectral decomposition for LOSs (a), (b), and (c) indicated with circles (black) in Figure \ref{fig:ROHSA}, where (a) and (c) are toward filamentary {\HI} structures and (b) is toward more extended emission between them. The LOSs toward narrow {\HI} structures clearly have significant CNM components, while the that of more extended {\HI} emission is UNM- and WNM-dominated. Toward this portion of F2, linear {\HI} structures are dominated by CNM over a velocity range of ${-}10\,{\lesssim}\,{\vlsr}\,{\lesssim}\,0\,\kms$, suggesting that narrow {\HI} structures outside of this range are likely WNM-dominated. A more complete {\ROHSA} analysis is required to determine if this applies to the entire G216 field.


\subsection{Comparison of Polarization Gradient Filament and Narrow {\HI} Structure Orientation}\label{subsec:polgrad_vs_hi}

Structures in radio polarization gradients primarily trace properties of the magnetized warm ionized gas \citep[e.g.,][]{Haverkorn2004b,Haverkorn2004c,Hill2008,Gaensler2011}, while narrow {\HI} structures tend to be associated with cold neutral gas \citep{Clark2019a,Peek2019,Kalberla2020} that is coupled to the ambient magnetic field \citep{Clark2014,Clark2015, Martin2015, Kalberla2016, Blagrave2017}. To investigate whether these very different environments of the ISM share a common magnetic field, we compare the orientation of polarization gradient filaments to that of narrow {\HI} structures.

Our visual inspection showed that structures in the GALFACTS polarization gradient and GALFA-{\HI} emission share similar morphologies and a common orientation (Section \ref{subsec:polgrad-HI}). To perform a more quantitative comparison, we compared the distributions of RHT orientations between {\polgradmax\!\!}, {\HI}, and {\halpha\!\!}. We applied the RHT to maps of the de-striped {5\arcmin} polarization gradient and {\halpha\!\!} emission using the same input parameters as applied to the binned {\HI} channel maps (Section \ref{subsec:hifibers}). The RHT backprojections of {\polgradmax\!\!} and {\halpha\!\!} emission are shown in Figures \ref{fig:polgrad_RHT} and \ref{fig:VTSS_RHT}, respectively.

Figure \ref{fig:HI_vs_rht_histograms} shows the RHT orientation distributions of {\polgradmax\!\!} (red), {\halpha\!\!} (green), and {\HI} (blue) across an {\HI} velocity range $-20\,{\lesssim}\,v_{\rm lsr}\,{\lesssim}\,{+20}\,\kms$. An angle of $0\deg$ is vertical in the image plane and increases anti-clockwise \citep{Clark2014}. The polarization gradient filaments are very well aligned with structures in {\halpha\!\!} emission, although the latter has a slightly wider distribution. In general, the peak {\HI} orientation tends to be ${\sim}\,10\deg$ clockwise from the polarization gradient filaments. At velocities $-20\lesssim v_{\rm lsr}\,{\lesssim}\,{+5}\,\kms$, the {\HI} emission is mostly single-peaked and strongly aligned with the polarization gradient filaments. The {\HI} emission becomes somewhat double-peaked at $+6\lesssim v_{\rm lsr}\,{\lesssim}\,{+11}\,\kms$ while the more dominant {\HI} orientation remains strongly aligned with the polarization gradient filaments. At $v_{\rm lsr}\,{\gtrsim}\,{+15}\,\kms$, the {\HI} orientation is less organized. These RHT results demonstrate that the filamentary structures in the polarization gradient are, in general, well aligned with the small-scale {\halpha\!\!} and {\HI} structures across a wide {\HI} velocity range.


\subsection{GALFACTS Depolarization}

\begin{figure}[t!]
  \centering
  \includegraphics[width=8.9cm]{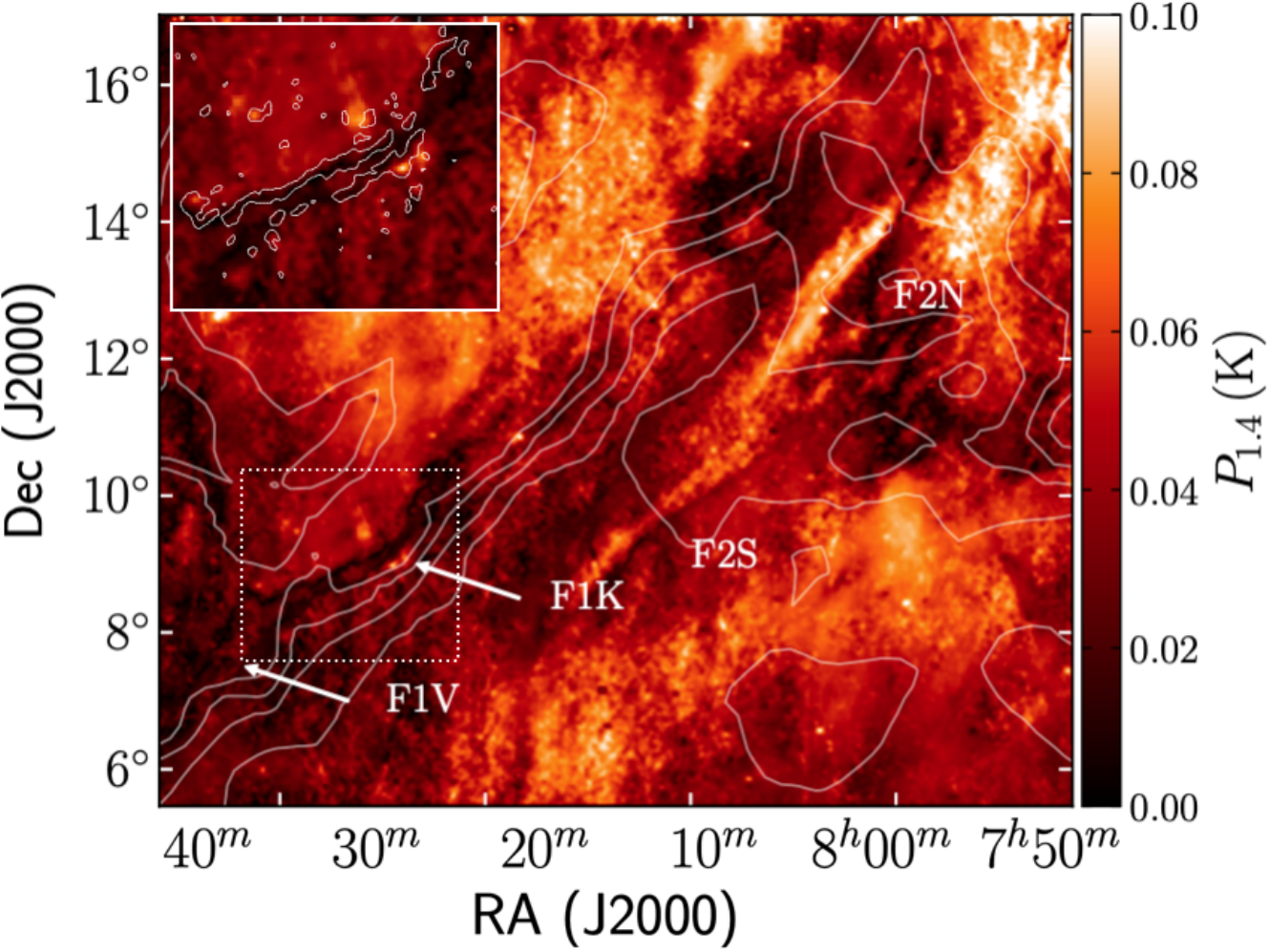}
  \caption{De-striped GALFACTS polarized intensity of G216 at $5\arcmin$ resolution. The WHAM {\halpha\!\!} contours overlaid (white) are for ${\vlsr}\,{=}\,{+7.3\,\kms}$, roughly corresponding to the peak velocity of F1K, and include levels of $1.6, 2.0, 2.4,$ and $2.8\,{\rm R}$. The figure inset shows the F1K region with a 0.006 ${\rm K\,arcmin^{-1}}$ level contour of {\polgradmax\!\!} (white) overlaid. The locations of F1 and F2 as discussed in the text are indicated.}
  \label{fig:P_vs_WHAM}
\end{figure}

We find several notable depolarized regions in the GALFACTS data. There is a ${\sim}\,2\deg$ wide depolarized band between F1 and F2 that is spatially correlated with extended WHAM {\halpha\!\!} emission. Figure \ref{fig:P_vs_WHAM} shows the GALFACTS polarized intensity with WHAM {\halpha\!\!} intensity contours at ${\vlsr}\,{=}\,{+7.3\,\kms}$ overlaid (white). This velocity channel roughly corresponds to the peak {\halpha\!\!} velocity of F1. There are also two thin filamentary features along F1K with sizes comparable to the angular resolution where the GALFACTS polarized intensity drops to zero that are consistent with `depolarization canals': narrow, elongated structures in depolarization caused by strong gradients in the RM \citep{Haverkorn2004a}. The double-jump polarization gradient feature along F1K highlights the edges of the associated RM structure and reflects the strong spatial gradient in degree of Faraday rotation.


\subsection{3D RM Synthesis}\label{subsubsec:RMSynth}

To investigate the Faraday-rotating magnetoionic medium, we applied RM synthesis \citep{Brentjens2005} to the GALFACTS polarization data. For a single rotating medium that is not emitting any polarized emission, the RM of a polarized source can be measured by the line of best fit to the observed $\chi$ versus the square of the observing wavelength $\lambda^2$,

\begin{equation}\label{eq:RM}
    \left(\frac{\chi}{\rm rad}\right) = \left(\frac{\chi_0}{\rm rad}\right) + \left(\frac{\rm RM}{\rm rad\,m^{-2}}\right) \left(\frac{\lambda}{\rm m}\right)^2,
\end{equation}

{\noindent}where $\chi_0$ is the incident polarization angle. However, the relationship shown in Equation \ref{eq:RM} does not generally hold for Galactic synchrotron emission where there are often multiple rotating foregrounds as well as a mixture between synchrotron-emitting and Faraday-rotating gas \citep{Ferriere2016}. The development of RM synthesis \citep{Burn1966, Brentjens2005} allows diffuse polarized emission to be described as a function of Faraday rotation characterized by the Faraday depth $\phi$. The Faraday depth is related to $n_e$ and $B_\parallel$ via

\begin{equation}\label{eq:FR}
    \left(\frac{\phi(L)}{\rm rad\,m^{-2}}\right) = 0.81\int_L^0 \left(\frac{n_e}{\rm cm^{-3}}\right) \left(\frac{B_\parallel}{\mu{\rm G}}\right) \left(\frac{{\rm d}\ell}{\rm pc}\right),
\end{equation}

{\noindent}where $L$ is the distance to the polarized emission and ${\rm d}\ell$ is the infinitesimal path length along the LOS. The integral is taken from the source of the polarized emission to the observer, setting the convention that a positive (negative) $\phi$ indicates a LOS magnetic field direction that is pointing toward (away from) the observer (see \citealt{Ferriere2021} for details on Faraday rotation conventions).

Many key quantities in RM synthesis are dependent on the frequency coverage of the polarization data. The Faraday depth resolution is the full-width at half-maximum (FWHM) of the primary lobe in the RM structure function (RMSF) determined by the $\lambda^2$ coverage via $\delta\phi\,{\approx}\,3.8/\Delta(\lambda^2)$, where $\Delta(\lambda^2)=\lambda_{\rm max}^2-\lambda_{\rm min}^2$. We have used $3.8$ in replacement of $2\sqrt{3}$ from Equation 61 of \citet{Brentjens2005} as a more accurate measure of the FWHM of a sinc function \citep[e.g.,][]{Schnitzeler2009}. The maximum detectable Faraday depth corresponds to a ${\sim}\,50\%$ drop in sensitivity across a single frequency channel given by $\phi_{\rm max}\,{\approx}\,1.9/\delta(\lambda^2)$ and quantifies the gradual loss in sensitivity at large Faraday depths. Lastly, the maximum Faraday depth scale is the broadest detectable Faraday depth feature given by $\phi_{\rm max{\text -} scale}\,{\approx}\,\pi/\lambda_{\rm min}^2$, where Faraday structures broader than this will be significantly depolarized.

We applied 3D RM synthesis to the GALFACTS frequency cubes using the {\rmtools} package developed by the Canadian Initiative for Radio Astronomy Data Analysis (CIRADA, \citealt{Purcell2020}).\footnote{\href{https://github.com/CIRADA-Tools/RM-Tools}{https://github.com/CIRADA-Tools/RM-Tools}} This software inputs Stokes $Q$ and $U$ frequency cubes and returns a Faraday depth cube containing a $\phi$ spectrum at every spatial pixel. Prior to performing RM synthesis, we flagged $15$ frequency channels $(4\%)$ that showed strong RFI. Flagging RFI channels does not significantly affect the Faraday depth resolution or the maximum detectable Faraday depth. The frequency channel maps were also de-striped and smoothed to {5\arcmin} angular resolution following the procedure outlined in Section \ref{subsec:results-GALFACTS}.

\begin{figure}[t!]
\centering
 \includegraphics[width=8.8cm]{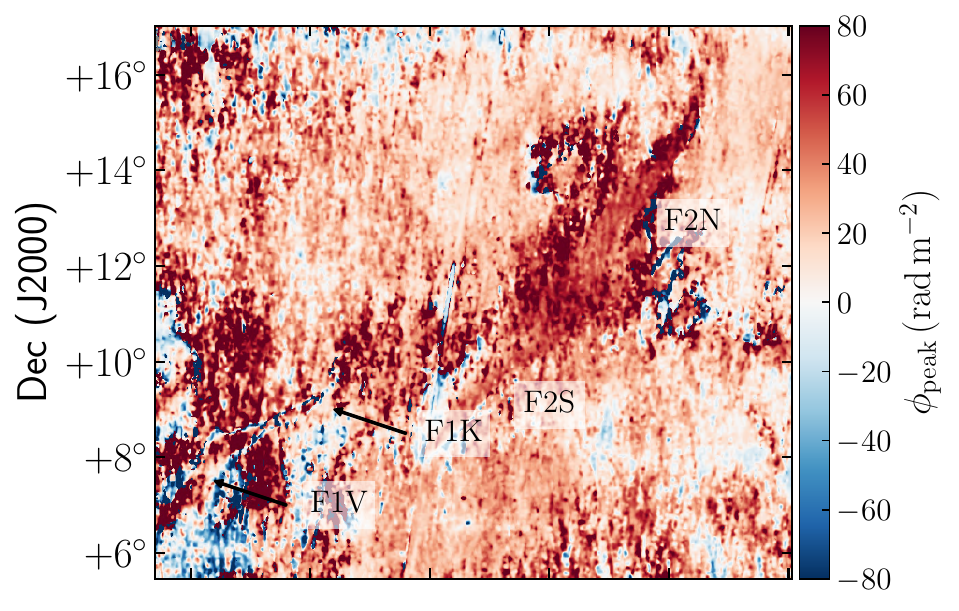}\vspace{-10pt}
 \includegraphics[width=8.8cm]{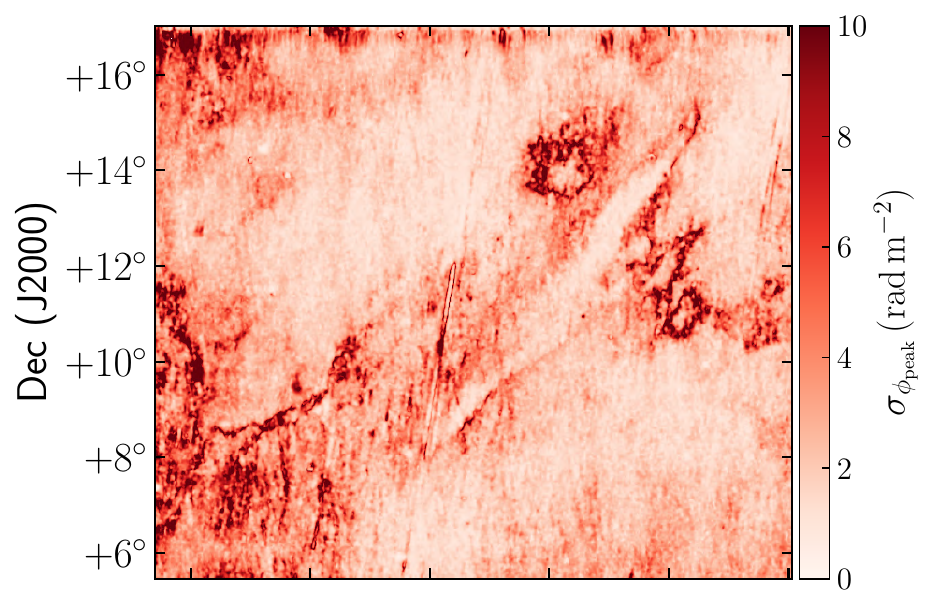}\vspace{-10pt}
 \includegraphics[width=8.8cm]{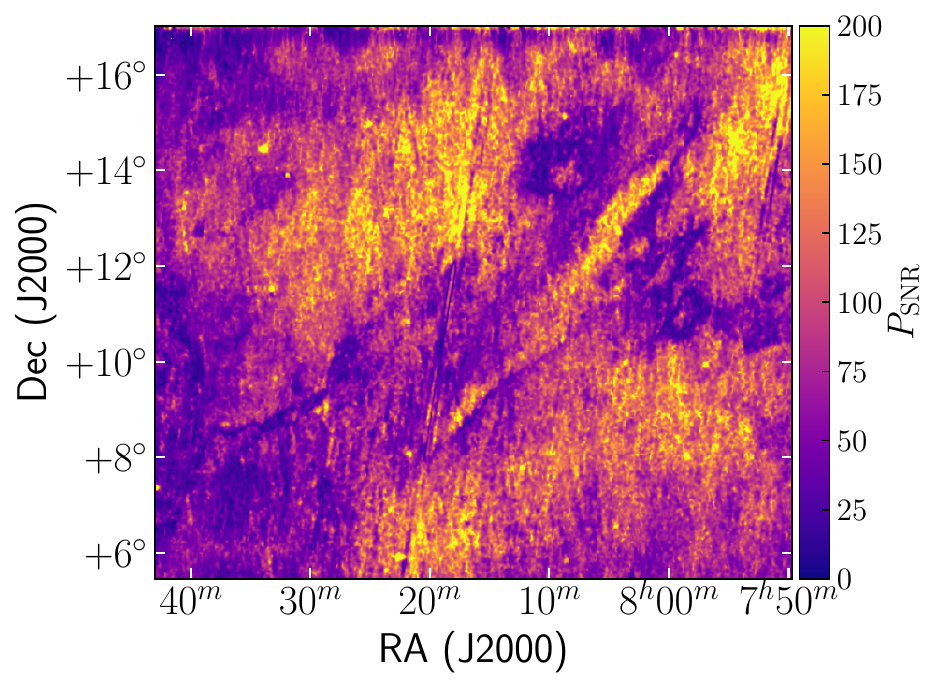}
\caption{Gaussian fitting results of the cleaned GALFACTS Faraday depth spectra. Shown are maps of the peak Faraday depth $\phi_{\rm peak}$ (top), uncertainty in polarized intensity $\sigma_P$ (middle), and SNR of the peak polarized intensity $P_{\rm SNR}$ (bottom). The locations of F1 and F2 as discussed in the text are indicated in the top panel.}
\label{fig:maxrm}
\end{figure}

We computed the $\phi$ spectra over a Faraday depth range of ${\pm}\,4000\,\radmsq$ to include a large baseline for uncertainty measurements, and used a sampling of $5$ points across the RMSF FWHM for Gaussian fitting. The GALFACTS weights are nearly equal as a function of frequency for any given spatial pixel \citep{Leahy2018}, so we applied a uniform weight to the unflagged channels. The GALFACTS frequency coverage results in a Faraday depth resolution of $\delta\phi\,{=}\,403\,\radmsq$, maximum Faraday depth $\phi_{\rm max}\,{\approx}\,7\,{\times}\,10^4\,\radmsq$, and maximum Faraday depth scale of $\phi_{\rm max{\text -} scale}\,{=}\,81\,\radmsq$. 

We used {\rmclean} \citep{Heald2009} to fit a Gaussian to the primary lobe of the RMSF to deconvolve the Faraday depth spectra. We find an RMSF FWHM of $396\,\radmsq$ with a standard deviation of $2{\times}10^{-4}\,\radmsq$ measured across the field. This is only ${\sim}\,1\%$ narrower than the expected theoretical value. We cleaned the Faraday depth spectra down to a threshold of ten times the noise level in the spectra to prevent false identification of clean components. Cleaning the Faraday depth spectra has no significant effect on our results, nor does increasing the cleaning threshold. 

We applied a Gaussian fitting procedure to each of the cleaned Faraday depth spectra. The spectral baseline was used to estimate the uncertainty in the polarized intensity over the Faraday depth range $|\phi|\,{\geq}\,2000\,\radmsq$ far from the polarized signal. We estimated the root-mean-square uncertainty for each of the $Q$ and $U$ components as $\sigma_Q$ and $\sigma_U$, respectively, along the spectral baseline and compute the corresponding uncertainty in the polarized intensity via $(1/\sigma_P)^2 = (1/\sigma_Q)^2 + (1/\sigma_U)^2$. The SNR of the brightest peak in Faraday depth, $P_{\rm SNR}$, was determined as the ratio of the peak polarized intensity derived from the Gaussian fit to the baseline uncertainty in polarized intensity. The peak Faraday depth $\phi_{\rm peak}$ corresponds to the Faraday depth at the peak polarized intensity with an estimated uncertainty of

\begin{align}
    \sigma_{\phi_{\rm peak}} = \frac{\rm RMSF\,FWHM}{2\,P_{\rm SNR}},
\end{align}

{\noindent}where we use the RMSF FWHM from \texttt{RM-clean}. Based on the results of \citet{George2012}, we applied a detection threshold of $P_{\rm SNR}\,{\geq}\,10$ to minimize false detections.

Our Gaussian fitting results of the cleaned Faraday depth spectra are summarized in Figure \ref{fig:maxrm}. This includes the peak Faraday depth $\phi_{\rm peak}$ (top), uncertainty in the peak Faraday depth $\sigma_{\phi_{\rm peak}}$ (middle), and SNR in the peak polarized intensity $P_{\rm SNR}$ (bottom). The uncertainty in polarized intensity is roughly uniform over the field with a mean value of $\langle\sigma_P\rangle\,{=}\,{(6\pm1)\,{\times}10^{-4}}\,{\rm K}$. There is increased Faraday rotation toward both F1K and F1V. The sign of $\phi_{\rm peak}$ is generally positive, indicating a LOS magnetic field direction that is mostly pointing toward the observer.


\section{Physical Properties of F1}\label{sec:estimates}

To quantify our comparison of the magnetized ionized and neutral media of this region, we determine several key astrophysical quantities. As F1K is a well-defined structure with significant detections in all of our multi-phase tracers, our quantitative analysis focuses on this portion of F1. This structure is most prominent in {\halpha\!\!} emission, so we defined an `on' region by the curved F1K portion of the filament in {\halpha\!\!} emission while the `off' region was taken to be an adjacent curved region of the same size to the immediate west of this portion of the filament. To reduce potential bias from outliers, we measure the `on' and `off' signals as the median value of the `on' and `off' regions, respectively. Unless stated otherwise, all final measurements in this section are made via subtracting the `off' from the `on' region to minimize foreground and/or background contributions.

\subsection{Distance}\label{subsubsec:distance}

\begin{figure*}[t!]
\centering
  \includegraphics[width=9.1cm]{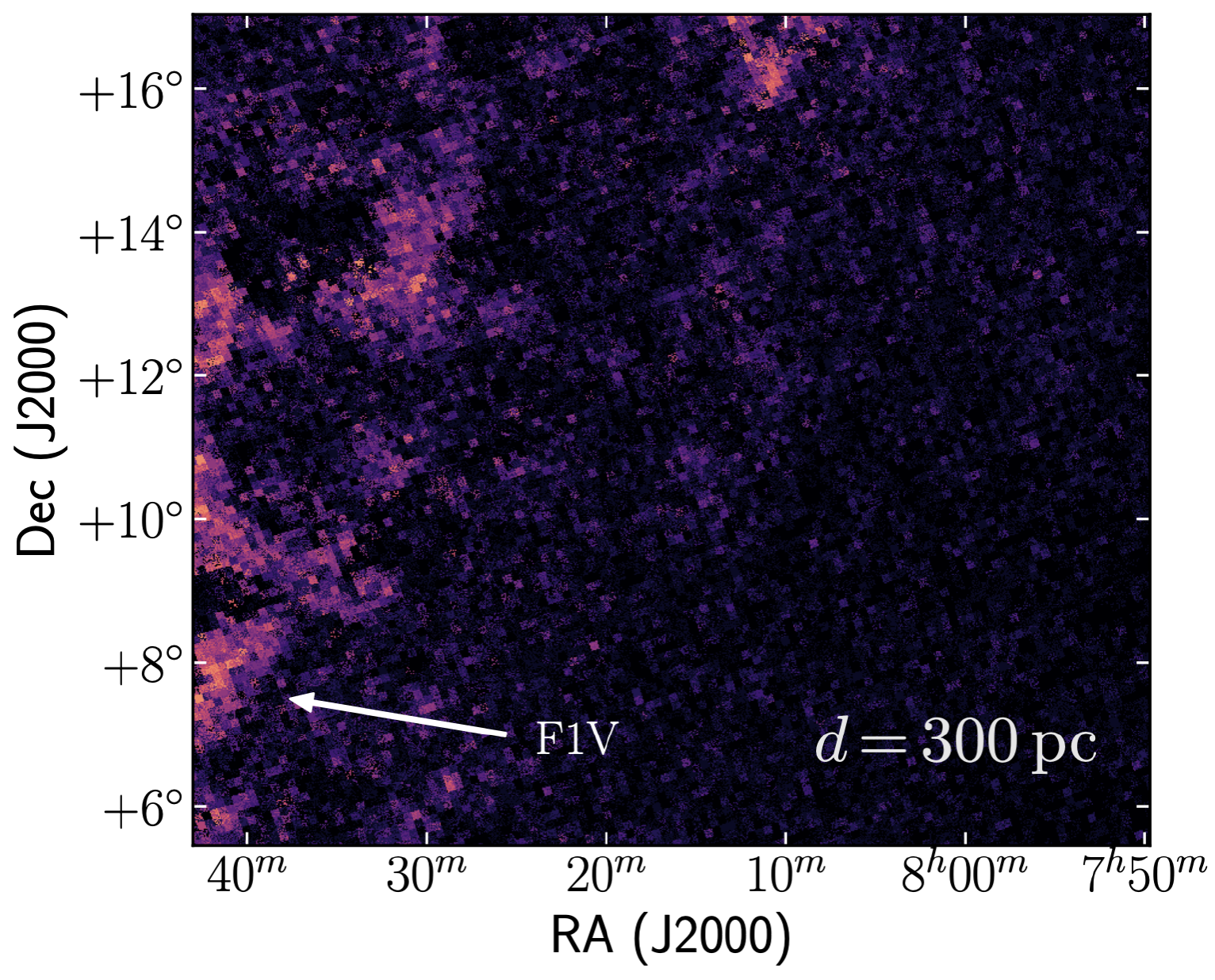}\hspace*{-2pt}\includegraphics[width=8.8cm]{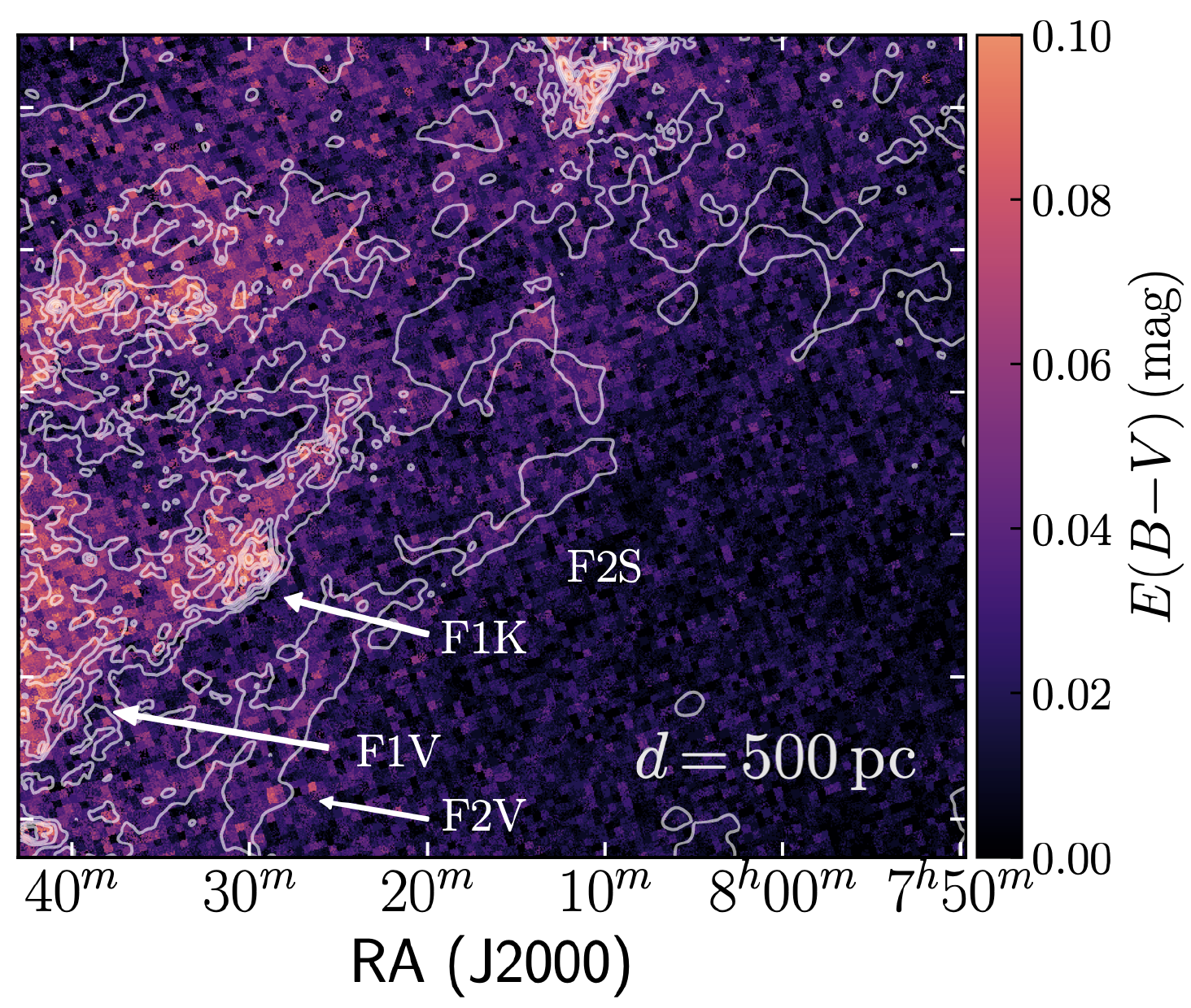}
  \includegraphics[width=8.cm]{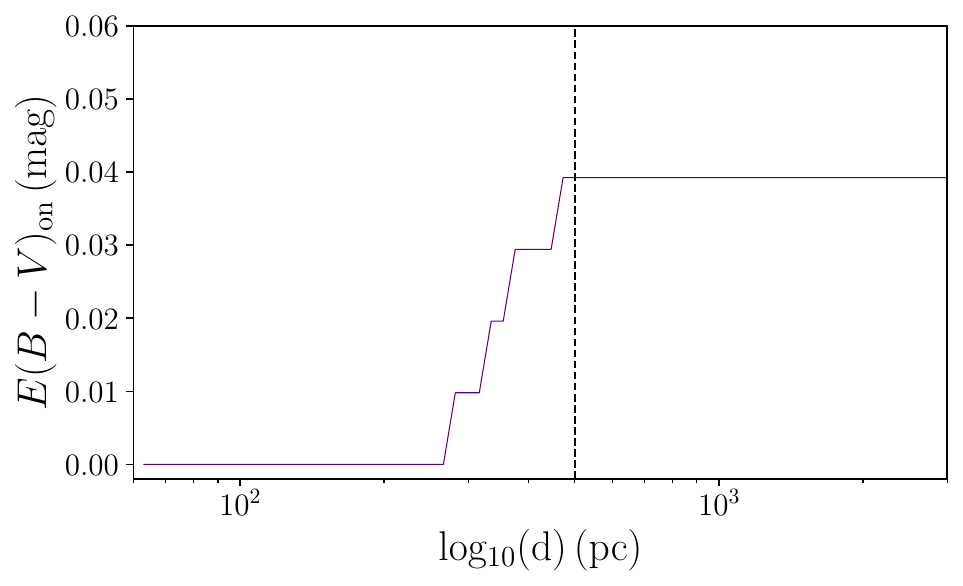}\includegraphics[width=8.cm]{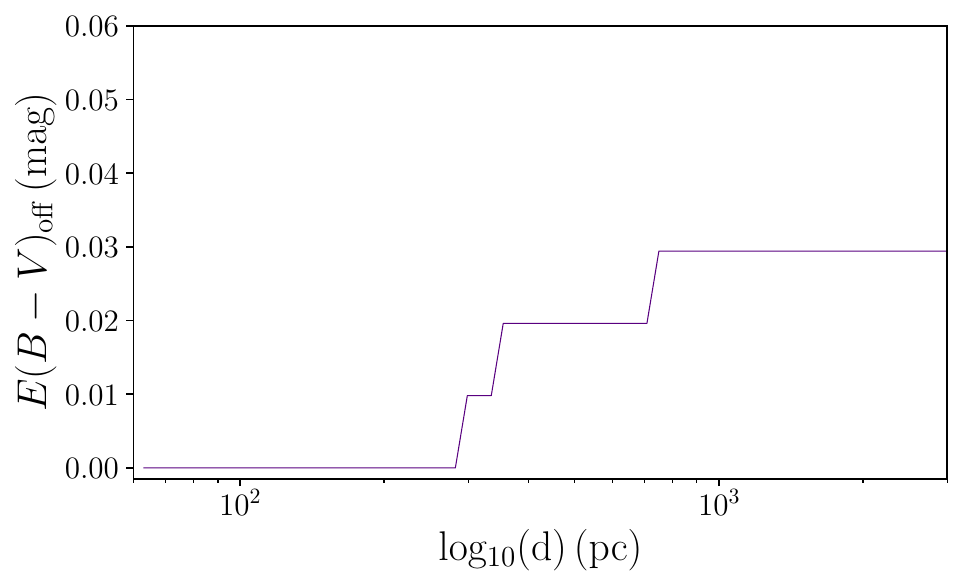}
 \caption{\citetalias{Green2019} dust reddening results of G216. (\textbf{Top:}) Integrated $E(B{-}V)$ maps at a distance of $300\,{\rm pc}$ (left) and $500\,{\rm pc}$ with the {\Planck} $I_{353}$ contours (white) overlaid representing values of 0.18, 0.24, 0.29, 0.35, 0.40, 0.45, 0.51, and 0.56 ${\rm MJy\,sr^{-1}}$ (right). The locations of F1 and F2 associated with dust emission as discussed in the text are indicated. (\textbf{Bottom:}) Median $E(B{-}V)$ reddening profiles toward F1K for the `on' region (left) with our fiducial distance of 500 pc indicated with a dashed vertical line, and the `off' region (right).}
\label{fig:bayestar}
\end{figure*}

To estimate the distance to F1, we used the 3D dust reddening map of \citet{Green2019} (hereafter ``\citetalias{Green2019}''), based on \textit{Gaia} parallaxes and stellar photometry from the Panoramic Survey Telescope and Rapid Response System 1 (Pan-STARRS 1, \citealt{Chambers2016}) and Two Micron All Sky Survey (2MASS, \citealt{Skrutskie2006}). We accessed the \citetalias{Green2019} dust map with the \texttt{dustmaps}\footnote{\href{https://dustmaps.readthedocs.io/en/latest/}{https://dustmaps.readthedocs.io/en/latest/}} Python package (v.1.0.6.) and made use of the associated online viewer.\footnote{\href{http://argonaut.skymaps.info/}{http://argonaut.skymaps.info/}} The \citetalias{Green2019} map is measured in integrated colour excess units of $E(g{-}r)$ assuming an optical extinction curve \citep{Schlafly2016}, which we convert to $E(B{-}V)$ via $E(B{-}V)\,{=}\,0.981\,E(g{-}r)$ \citep{Green2019}.

Figure \ref{fig:bayestar} (top) shows the \citetalias{Green2019} integrated dust reddening maps up to a distance of $300\,{\rm pc}$ (left) and $500\,{\rm pc}$ with the {\Planck} $I_{353}$ contours (white) overlaid (right). There is some indication of F1V in $E(B{-}V)$ at a distance of ${\sim}\,300\,{\rm pc}$, but we are only able to confidently identify F1K at a distance of ${\sim}\,{500}\,{\rm pc}$. We therefore assume a fiducial distance of $d\,{=}\,500\,{\rm pc}$ to F1. We note that this might somewhat overestimate the distance if our inability to identify F1 stems from a lack of Gaia stars with distance measurements rather than a lack of dust emission at ${\sim}\,300\,{\rm pc}$. There are insignificant changes in $E(B{-}V)$ for $d\,{\gtrsim}\,500\,{\rm pc}$ and this approaches the {\Planck} integrated dust map, suggesting that there is little dust emission beyond $500\,{\rm pc}$. Our fiducial distance is roughly consistent with the \textit{Gaia}-2MASS 3D dust extinction map \citep{Lallement2019}.

Assuming a model of Galactic rotation, kinematic distance techniques use {\vlsr} measurements of {\HI} to determine the distance of {\HI} gas. We used a Monte Carlo kinematic distance code \citep{Wenger2018} based on maser parallaxes and universal Galactic rotation model \citep{Reid2014} to estimate the kinematic distance of F1. With a peak F1K {\HI} velocity ${\vlsr}\,{\sim}\,{+}6\,\kms$, we find a kinematic distance of $500^{+400}_{-320}\,{\rm pc}$ that is consistent with our fiducial distance. At a distance of ${\sim}500\,{\rm pc}$, the vertical distance from the Galactic midplane is ${\lesssim}\,300\,{\rm pc}$, so halo lag  does not significantly affect our kinematic distance estimate and is ignored.

Figure \ref{fig:bayestar} (bottom) shows the average \citetalias{Green2019} $E(B{-}V)$ profiles for the `on' (left) and `off' (right) regions along F1K. The spatial resolution of the data defines the number of profiles within our masks and is limited by the Gaia sampling of stars \citep{Green2019}. Both sets of reddening profiles show an increasing colour excess over a similar distance range, suggesting that the dust emission is likely spatially coherent in this area.

\subsection{Scaling with Path Length and Ionized Filling Factor}

Several physical quantities of interest depend on parameters that cannot be unambiguously determined. The first of these is the LOS thickness through F1, referred to as the path length $L$. While the 3D geometry of the filament is unknown, we characterize the LOS path length as $L\,{=}\,Nw$, where $N$ is the ratio of the LOS thickness of the filament to its apparent width on the sky $w$ measured in pc. We measure an angular width of $\theta\,{=}\,0.25\deg$ of F1K in VTSS {\halpha\!\!} emission which, at our fiducial distance of $500\,{\rm pc}$, corresponds to a physical width of $w\,{=}\,2.2\,{\rm pc}$. The second parameter is the dimensionless ionized filling factor $f_{\rm ion}$, defined as the fraction of the path length that is occupied by ionized gas of uniform density. In light of this, we scale the following equations with $L_{2.2}\,{\equiv}\,L/(2.2\,{\rm pc})$ and $f_{\rm ion}$.

\subsection{Emission Measure and Thermal Electron Density}

The emission measure (EM) is estimated from the {\halpha\!\!} intensity $I_{\rm H\alpha}$ following

\begin{equation}
   \left(\frac{\rm EM}{\rm pc\,cm^{-3}}\right) = 2.75 \left(\frac{T_{\rm WIM}}{10^4\,{\rm K}}\right)^{0.9} \left(\frac{I_{\halpha\!\!}}{\rm R}\right) \left(\frac{\tau_{\rm int}}{1-e^{-\tau_{\rm int}}}\right) e^{\tau_{\rm fg}}
\end{equation}

{\noindent}\citep{Reynolds1988}, where we assume a WIM temperature of $T_{\rm WIM}\,{\sim}\,0.8\,{\times}\,10^4\,{\rm K}$ \citep{Haffner1998} and $I_{{\rm H}\alpha}$ is measured in R. There are two correction terms for the dust optical depth including contributions of the foreground dust, $\tau_{\rm fg}$, and that of the dust mixed in with the {\halpha\!\!} emission, $\tau_{\rm int}$ \citep{Finkbeiner2003}, that does not include the foreground contribution. We use the {\Planck} $E(B{-}V)$ map of the diffuse, high Galactic latitude sky \citet{Planck2013XI}. We measure the foreground contribution using the `off' region, while the internal contribution is measured using the `on' minus `off' region, corresponding to $E(B{-}V)_{\rm fg}\,{=}\,0.034{\pm}0.004\,{\rm mag}$ and $E(B{-}V)_{\rm int} = 0.020{\pm}0.017\,{\rm mag}$, respectively. The dust optical depths are derived using 

\begin{align}
    \tau = \left(\frac{2.65}{1.086}\right)\,E(B{-}V)
\end{align}

{\noindent}\citep{Finkbeiner2003}, where the factor of $2.65$ describes the shape of the extinction curve and the factor of $1.086$ converts the colour excess to dust optical depth. This yields foreground and internal dust optical depths, respectively, of $\tau_{\rm fg}\,{=}\,0.084{\pm}0.009$ and $\tau_{\rm int}\,{=}\,0.05{\pm}0.04$.

We used the {\Planck} colour excess map over that of \citetalias{Green2019} due to its more complete spatial coverage, although both give similar results. These dust extinction corrections assume that there is no dust emission behind F1K, resulting in an an upper-limit of EM. At these Galactic latitudes, it is reasonable to assume that most of the dust is foreground to F1. This is consistent with our results in Section \ref{subsubsec:distance} where we found insignificant changes in $E(B{-}V)$ at distances beyond $500\,{\rm pc}$.

The thermal electron density is related to the EM via

\begin{align}\label{eq:ne}
    \left(\frac{n_e}{\rm cm^{-3}}\right) = \frac{1}{\sqrt{2.2}}\left(\frac{\rm EM}{\rm pc\,cm^{-6}}\right) (f_{\rm ion}L_{2.2})^{-0.5}
\end{align}

{\noindent}assuming that the ionized plasma is distributed in clumps of uniform density filling a total length $f_{\rm ion}L_{2.2}$ called the occupation length. We find an {\halpha\!\!} intensity $I_{\halpha}\,{=}\,2.2{\pm}0.8\,{\rm R}$, ${\rm EM}\,{=}\,6{\pm}2\,{\rm pc\,cm^{-6}}$, and $n_e\,{=}\,(1.5{\pm}0.3)\,(f_{1}L_{2.2})^{-0.5}\,{\rm cm^{-3}}$.

\subsection{Magnetic Field Strength}

The LOS magnetic field strength can be estimated by introducing $L_{2.2}$ and $f_{\rm ion}$ to Equation \ref{eq:FR},

\begin{align}\label{eq:Bparallel}
    \left(\frac{B_\parallel}{\mu{\rm G}}\right) = \frac{1}{(0.81)(2.2)} \left(\frac{\rm RM}{\rm rad\,m^{-2}}\right) \left(\frac{n_e}{\rm cm^{-3}}\right)^{-1} (f_{\rm ion}L_{2.2} )^{-1}.
\end{align}

{\noindent}Similar to Equation \ref{eq:ne}, this assumes an occupation length $f_{\rm ion}L_{2.2}$ for the ionized plasma through which the magnetic field is threaded. We make use of our map of $\phi_{\rm peak}$ from Section \ref{subsubsec:RMSynth}, measuring a Faraday depth of $\phi\,{=}\,{+}19{\pm}13\,\radmsq$. This, along with our measurements of EM from the previous section, yields a LOS magnetic field strength of $B_{\parallel}\,{=}\,({+}6{\pm}4)\,(f_{\rm ion}L_{2.2})^{-0.5}\,\mu{\rm G}$. Since the measured EM and $n_e$ are upper limits, $B_\parallel$ is a lower-limit.

\subsection{Plasma Beta}

The relative importance of magnetic energy can be quantified using the ratio of thermal gas ($P_{\rm th}$) to magnetic ($P_{\rm mag}$) pressure, commonly referred to as the plasma beta parameter,

\begin{align}
        \beta = \frac{P_{\rm th}}{P_{\rm mag}}.
\end{align}

{\noindent}To reflect the observed spatial separation between the ionized and neutral medium along F1K, we assume that the thermal gas pressure is dominated by ionized gas,

\begin{align}
        \left(\frac{P_{\rm th}/k_B}{\rm K\,cm^{-3}}\right) = 2 \left(\frac{n_e}{\rm cm^{-3}}\right) \left(\frac{T_{\rm WIM}}{\rm K}\right),
\end{align}

{\noindent}where $k_B$ is the Boltzmann constant. We again assume a WIM temperature of $T_{\rm WIM}\,{\sim}\,0.8\,{\times}\,10^4\,{\rm K}$ \citep{Haffner1998}. The contribution of gas pressure from the neutral medium is approximately an order of magnitude lower than that of the ionized medium and does not significantly affect our results. This yields a thermal gas pressure of $P_{\rm th}/k_B\,{=}\,(2.5\,{\pm}0.5){\times}10^4\,(f_{\rm ion}L_{2.2})^{-0.5}\,{\rm K\,cm^{-3}}$.

The magnetic pressure is determined via

\begin{align}
        \left(\frac{P_{\rm mag}/k_B}{\rm K\,cm^{-3}}\right) = \frac{1}{8\pi} \left(\frac{B_{\rm tot}}{\mu{\rm G}}\right)^2,
\end{align}

{\noindent}where $B_{\rm tot}$ is the total field strength in $\mu{\rm G}$. The 3D magnetic field geometry of F1, which sets the relationship between $B_\parallel$, $B_\perp$, and $B_{\rm tot}$, is unknown. However, it is unlikely that the total magnetic field of F1 is predominantly along the LOS, implying that $B_{\rm tot}\,{>}\,B_\parallel$. While we use the estimate of $B_\parallel$ for $B_{\rm tot}$, we note that this under-estimates the magnetic field strength and correspondingly over-estimates plasma beta.

We find a magnetic pressure $P_{\rm mag}/k_B = ([1.1^{+1.5}_{-1.1}]\times10^4)\,(f_{\rm ion}L_{2.2})^{-0.5}\,{\rm K\,cm^{-3}}$ and plasma beta $\beta\,{=}\,(2.1^{+3.1}_{-2.1}) \allowbreak (f_{\rm ion}L_{2.2})^{-1}$. 


\section{Discussion}\label{sec:discussion}


\subsection{Physical Interpretation}\label{subsec:physical-interpretation}

We consider two possible 3D geometries for F1: (I) a filament and (II) an edge-on sheet. For (I), we set the path length equal to the filament width. Assuming the filament makes an angle of $60\deg$ with the LOS, the median value for a random distribution, this corresponds to a ratio of path length to apparent width on the sky $N\,{=}\,{1.2}$. For (II) we assume $N\,{=}\,15$ such that the resulting path length is approximately equal to the ${\sim}30\,{\rm pc}$ typical scale length of the WIM \citep{Ferriere2020}. Since the {\HI} emission is not coincident with the ionized filament, we adopt a filling factor of $f_{\rm ion}\,{=}\,1.0$. We find that the internal pressure of F1 (discussed below) can be explained using a filament geometry. With no strong evidence to suggest that this structure is sheet-like, we adopt a filament geometry for F1. Our assumptions and observables are summarized in Table \ref{table:observables}.

We find a thermal electron density $n_e\,{=}\,1.5{\pm}0.3\,{\rm cm^{-3}}$, a LOS field strength $B_\parallel\,{=}\,6{\pm}4\,\mu{\rm G}$, and plasma beta $\beta\,{=}\,2.1^{+3.1}_{-2.1}$. The thermal electron density is roughly one order of magnitude higher than what is typically found in the diffuse WIM \citep[e.g.,][]{Ferriere2001,Gaensler2008}. The magnetic field strength of F1 may help to stabilize it against thermal pressure, and a significant $B_\perp$ component would reduce $\beta$ toward a more magnetically-dominated regime. Our model-dependent results are shown in Table \ref{table:estimates}.

We estimate the theoretical synchrotron intensity of F1 to determine if our estimate of the magnetic field strength is consistent with the lack of correlated excess synchrotron emission in the GALFACTS data (Figure \ref{fig:stokes_S1}). The GALFACTS total intensity data have not been calibrated to a stable zero-point as a result of scanning artefacts \citep{Leahy2018}. For this reason, we used reprocessed 1.4 GHz continuum data from the HI Parkes All-Sky Survey (CHIPASS, \citealt{Calabretta2014}).

The synchrotron total intensity depends on $B_\perp$ and the cosmic ray density $n_{\rm cr}$ via

\begin{align}
    I\,{\propto}\,\int B_\perp^\frac{p+1}{2} n_{\rm cr}\,{\rm d}\ell,
\end{align}

{\noindent}where $p\,{=}\,3$ is the spectral index of the cosmic ray distribution \citep{Jaffe2010}. We measure the ratio of the expected synchrotron intensity from F1 to that of the Galactic background, $I_{\rm F1}/I_{\rm MW}$, assuming $B_\perp\,{\sim}\,B_\parallel$, a constant $n_{\rm cr}$ along the LOS, a typical Galactic magnetic field strength of ${\sim}6\,\mu{\rm G}$ \citep{Rand1989} projected to the POS of $(2/\sqrt{3})6\,\mu{\rm G}$, and a Galactic synchrotron path length of ${\sim}10\,{\rm kpc}$ in this direction \citep{Sun2008}. This yields the quantity $I_{\rm F1}/I_{\rm MW}\,{\sim}\,2{\times}10^{-4}$ which can be multiplied by the local background synchrotron intensity to yield $I_{\rm F1}$. We find an average value of $I_{\rm MW}\,{\sim}3.5\,{\rm K}$ in the direction of G216, yielding a theoretical value of $I_{\rm F1}\,{\lesssim}\,1\,{\rm mK}$ that is roughly equal to the root-mean-square uncertainty in the GALFACTS Stokes $I$ map. Therefore, the magnetic field strength that we measure does not necessitate a corresponding detectable structure in GALFACTS total intensity.

{
\begin{table}
\centering
\begin{threeparttable}
\begin{tabular}{lc}
\hline
\multicolumn{2}{c}{Assumptions} \\
\hline
$d$ & $500\,{\rm pc}$ \\
$f_{\rm ion}$ & $1.0$ \\
$T_{\rm WIM}$ & $8000\,{\rm K}$ \citep{Haffner1998} \\
\hline
\multicolumn{2}{c}{Observables} \\
\hline
$\theta$             & $0.25\deg$ \\
$I_{\halpha}$        & $2.2{\pm}0.8\,{\rm R}$ \\
$E(B{-}V)_{\rm fg}$  & $0.034{\pm}0.004\,{\rm mag}$ \\
$E(B{-}V)_{\rm int}$ & $0.020{\pm}0.017\,{\rm mag}$ \\
$\phi_{\rm peak}$    & ${+}19{\pm}13\,{\rm rad\,m^{-2}}$ \\
\hline
\end{tabular}
\end{threeparttable}
\caption{Assumptions (top) and observational results (bottom) of our quantitative analysis for F1K including: $d$ the LOS distance, $f_{\rm ion}$ the ionized filling factor , $T_{\rm WIM}$ the WIM gas temperature, $\theta$ the angular width, $I_{\halpha}$ the {\halpha\!} intensity, $E(B{-}V)_{\rm fg}$ the foreground colour excess, $E(B{-}V)_{\rm int}$ the internal colour excess, EM the emission measure, and $\phi_{\rm peak}$ the peak Faraday depth.}
\label{table:observables}
\end{table}
}

Magnetic field alignments between different ISM phases are to be expected in regions where dynamical events such as supernova explosions, stellar winds, and expanding {\HII} regions have swept up the ambient medium. Such events can lead to cavities in the ISM identified as {\HI} shells where the swept-up material is contained in the cavity wall \citep[e.g.,][]{McClure-Griffiths2003}. In light of our region being found at the northern edge of a large low column density region seen in both {\halpha\!\!} and dust emission, we consider the possibility that it is related to a shell or bubble in the ISM.

{
\begin{table*}
\centering
\begin{threeparttable}
\begin{tabular}{lc|cccc}
\cline{2-6}
 & \multicolumn{1}{|c|}{\rm This\,work} & \multicolumn{4}{c}{Other work} \\
\cline{2-6}
 & \multicolumn{1}{|c|}{\rm F1\,of\,G216} & 3C196$^*$ & Sh2-27 NC$^{\dag}$ & Sh2-27 FC$^{\dag}$ & IVF$^{\ddagger}$\\
\hline
\multicolumn{1}{l|}{$L\,(\rm pc)$}                & $2.6\,(N\,{=}\,1.2)$ & $1.5{\times}10^3$  & ${\sim}30$    & ${\sim}30$ & $18{\pm}9$ \\
\multicolumn{1}{l|}{$n_e\,(\rm cm^{-3})$}         & $1.5{\pm}0.3$        & ${\sim}\,0.03$     & ${\sim}0.02$  & ${\sim}0.02$ & ${\sim}0.2$ \\
\multicolumn{1}{l|}{${B_\parallel}\,(\rm \mu G)$} & ${+}6{\pm}4$         & ${+}0.3\pm0.1$     & ${\sim}{-}15$ & ${\sim}{+}30$ & $2.8{\pm}0.8$ \\
\cline{3-5}
\multicolumn{1}{l|}{$\beta$}                      & $2.1^{+3.1}_{-2.1}$    & $31{\pm}21^{\#}$ & ${\sim}0.08^{\parallel}$ & ${\sim}0.02^{\parallel}$ & ${\sim}0.1-1$ \\
\hline
\end{tabular}
\begin{tablenotes}\footnotesize
    \item[*] Diffuse LOFAR field \citep{Jelic2015}.
    \item[$\dag$] Dust clouds foreground to the Sharpless 2-27 (Sh2-27) {\HII} region with a near (NC) and far cloud (FC) component \citep{Thomson2019}.
    \item[$\ddagger$] Intermediate Velocity Filament (IVF, \citealt{Stil2016}).
    \item[$\#$] Computed here assuming $f_{\rm ion}\,{=}\,1$, $T_{\rm WIM}\,{=}\,8{\times}10^3\,{\rm K}$, and $B_{\rm tot}\,{\sim}\,\sqrt{3}B_\parallel$.
    \item[$\parallel$] Computed here assuming $f_{\rm ion}\,{=}\,10^{-3}$, $T_{\rm CNM}\,{=}\,80\,{\rm K}$, and $B_{\rm tot}\,{\sim}\,\sqrt{3}B_\parallel$.
\end{tablenotes}
\end{threeparttable}
\caption{Physical parameters derived for F1 with a comparison of similar studies from the literature. These include: $L$ the path length, $n_e$ the thermal electron density, $B_\parallel$ the LOS magnetic field strength, and $\beta$ the plasma beta.}
\label{table:estimates}
\end{table*}
}

The IVC {\HI} gas at ${\vlsr}\,{\lesssim}\,{-20}\,\kms$ contains finger-like projections directed toward the Galactic plane (Figure \ref{fig:HI_S1}, top left) that strongly resemble Rayleigh-Taylor instabilities found along the walls of Galactic {\HI} shells \citep[e.g.,][]{Dawson2011, McClure-Griffiths2003}. We searched catalogues of known {\HI} shells near this part of the outer Galaxy \citep{Heiles1979, Hu1981, Heiles1984} and found no previously-identified shell that is morphologically associated with G216. Using data from the all-sky {\HI} $4\pi$ (HI4PI) survey \citep{HI4PI2016} based on EBHIS \citep{Winkel2016} and the Parkes Galactic All-Sky Survey (GASS, \citealt{McClure-Griffiths2009, Kalberla2010, Kalberla2015}), we inspected the {\HI} structure in velocity channel maps to search for evidence of an {\HI} shell. Based on the criteria described by \citet{McClure-Griffiths2002}, we find no strong evidence of the existence of an {\HI} shell. The morphology of 353 GHz {\Planck} dust polarization data can be used to identify a swept-up shell wall \citep[e.g.,][]{Soler2018}, however, we do not find strong evidence for a large-scale shell in the {\Planck} dust magnetic field orientation.

Despite the lack of evidence for an {\HI} shell, we consider the possibility of a recently-propagated shock through this region. Using soft X-ray data from the all-sky ROSAT survey \citep{Snowden1997} over the energy range $0.1{-}2.4\,{\rm keV}$, we searched for excess X-ray emission associated with G216. We measured an average X-ray count consistent with zero photons per pixel, finding no evidence of X-ray emission that is spatially correlated with structures in this paper.

We used the X-ray spectral fitting program {\xspec} \citep{Xspec1996}, accessed with the online tool {\webspec},\footnote{\href{https://heasarc.gsfc.nasa.gov/webspec/webspec.html}{https://heasarc.gsfc.nasa.gov/webspec/webspec.html}} to model the expected X-ray photon count of a typical shock propagating through this region. {\webspec} assumes an absorbing foreground column ${\NHI}$, and thermal electron density $n_e$ (which we assume to be equal to the {\HI} density), plasma temperature $T_p$, and radius $R$ of the emitting region. Based on our measurements in Section \ref{sec:estimates}, we converted the foreground dust optical depth $\tau_{\rm fg}$ to a foreground {\HI} column density of ${\NHI}\,{=}\,1.9{\times}10^{20}\,{\rm cm^{-2}}$. We assume a radius corresponding to an angular size of $10\deg$, roughly equal to the size of the low column density region toward G216, at our fiducial distance of $500\,{\rm pc}$. We measured an expected X-ray count that is roughly consistent with little-to-no average X-ray emission per ROSAT pixel over a range in electron density $2\,{\times}10^{-3}\,{<}\,n_e\,{<}\,0.01\,{\rm cm^{-3}}$ and plasma temperature $10^5\,{<}\,{T_p}\,{<}\,10^7\,{\rm K}$. This is likely due to the absorbing foreground {\HI}; without it, we find average X-ray photon counts that are several orders of magnitude higher. We are therefore unable to rule out the possibility of a previous shock. If a shock did occur, this could have triggered the formation of dense CNM gas adjacent to F1K where several dense clumps are found. 


\subsection{Multi-Phase Structure of F1 and F2}\label{subsec:multiphase-structure}

The F1 polarization gradient filament has clear associations with tracers of the multi-phase ISM. While F1 is spatially correlated with a small-scale VTSS filament, it is possibly embedded in the more extended WHAM filament. The ${\sim}2\deg$ wide depolarized band between F1 and F2 in the GALFACTS data is spatially correlated with extended WHAM emission along the edge of the diffuse ionized and neutral gas emission. This suggests that the small-scale {\halpha\!\!} filament may be associated with the edge of a larger, more diffuse structure. The spatial separation between the ionized and neutral media along F1K suggests an ionized layer around a neutral gas cloud. The ionized layer of F1 is spatially offset from the neutral cloud in the direction toward the Galactic plane where the photoionizing flux from e.g., massive OB stars is expected to be higher.

While F1 is spatially correlated with a bright {\halpha\!\!} filament and a sharp edge in the {\HI} column density and {\Planck} dust emission, the correspondences between F2 and other ISM tracers are more subtle. Careful inspection of the VTSS and WHAM data reveals that F2 is spatially correlated with edges of very faint {\halpha\!\!} emission. This is reminiscent of small-scale structures in the polarization gradient ascribed to turbulence-driven fluctuations that were too faint to be observed in {\halpha\!\!} emission \citep{Gaensler2011}. Numerical simulations of magnetohydrodynamic (MHD) turbulence reproduce extended filamentary structures in the polarization gradient similar to those found in real data \citep[e.g.,][]{Iacobelli2014}, suggesting that F2 and similar structures are an expected consequence of turbulence.

\begin{figure}[t!]
\centering
  \includegraphics[width=9cm]{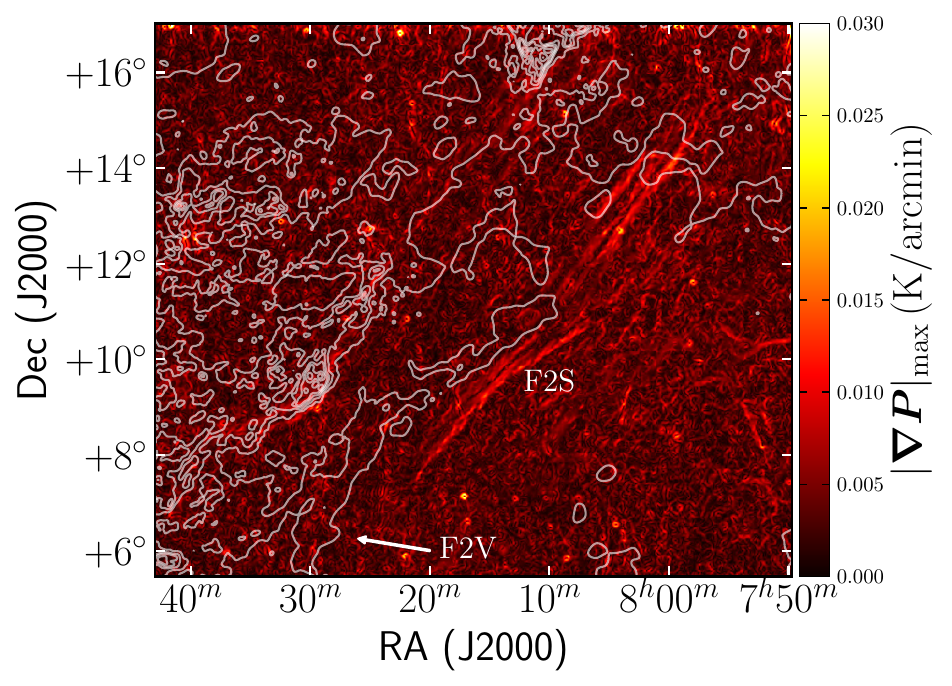}
 \caption{GALFACTS {\polgradmax} of G216 at {5\arcmin} resolution with the same {\Planck} $I_{353}$ contours as Figure \ref{fig:bayestar} overlaid (white). The locations of F2 associated with dust emission as discussed in the text are indicated.}
\label{fig:polgradvsPlanck}
\end{figure}

Despite the lack of {\HI} or dust emission spatially coincident with F2, there is a filament in the neutral ISM that extends along the eastern edge of F2S with a similar curvature. Figure \ref{fig:polgradvsPlanck} shows the GALFACTS polarization gradient (red) with {\Planck} $I_{353}$ dust emission contours overlaid (white), clearly showing the alignment. This dust filament strongly resembles an {\HI} filament in the local gas at velocities ${+4}\,{\lesssim}\,{\vlsr}\,{\lesssim}\,{+20\,\kms}$. The strong morphological similarity but lack of spatial coincidence between the neutral filament and F2S suggests an ionized-neutral boundary along F2.

\subsection{Comparison to Other Studies}

We compare our results to radio polarization studies of discrete structures in the high-Galactic latitude diffuse/translucent ISM, restricting this to a few studies.

The 3C196 LOFAR field contains polarized intensity structures that are aligned with the {\Planck} dust POS magnetic field direction \citep{Jelic2015}. Assuming that the observed RM is produced by the WIM over a path length of 1.5 kpc, the authors find ${\langle}B_{\parallel}{\rangle}\,{=}\,0.3{\pm}0.1\mu{\rm G}$. The estimated ${\langle}n_e{\rangle}$ is lower than typical values for the diffuse WIM and our results for F1 by roughly one and two orders of magnitude, respectively. Based on models of the GMF \citep{Sun2008}, the authors note that the regular magnetic field is primarily in the POS toward this direction. The same GMF models predict both large-scale LOS and POS magnetic field components toward G216, consistent with our observations. The authors suggest that a prominent, polarized filamentary structure in 3C196 is an ionized filament wrapped around neutral {\HI} gas, similar to our interpretation of the ionized layer of F1. In Section \ref{subsec:frsource}, we discuss the possibility that LOFAR polarized intensity structures trace the predominantly neutral, rather than ionized, ISM.

The foreground Faraday rotation toward the nearby {\HII} region Sharpless 2-27 (Sh2-27) was modelled using data from the 64 m Parkes Radio Telescope as part of the Global Magnetoionic Medium Survey (GMIMS, \citealt{Wolleben2009}) \citep{Thomson2019}. This revealed near (NC) and far cloud (FC) components with $B_\parallel\,{\sim}{-}15\,\mu{\rm G}$ and $B_\parallel\,{\sim}{+}30\,\mu{\rm G}$, respectively, over equal path lengths of ${\sim}30\,{\rm pc}$. These foreground clouds are interpreted as CNM-bearing dust clouds that dominate the RM along the LOS. Assuming $T_{\rm CNM}\,{=}\,80\,{\rm K}$ and an ionization fraction ${\sim}10^{-3}$ \citep{Ferriere2020} for the CNM, and $B_{\rm tot}\,{\sim}\,\sqrt{3}B_\parallel$, we find that the NC and FC components have magnetically-dominated plasma betas of $\beta\,{\sim}\,0.08$ and $\beta\,{\sim}\,0.02$, respectively. 

\citet{Stil2016} investigated the magnetic properties of an ionized intermediate-velocity {\halpha\!\!} filament (designated as IVF here) in VTSS \citep{Finkbeiner2003} and WHAM \citep{Haffner2003} data. Using RM measurements based on National Radio Astronomy Observatory VLA Sky Survey (NVSS) data \citep{Taylor2009}, excess RM is found to be spatially correlated with extended WHAM emission. This is in contrast to our results of F1 where we find an excess RM that is correlated with the small-scale VTSS filament. Using an empirical relationship between the EM and dispersion measure \citep{Berkhuijsen2006}, the authors find $B_\parallel\,{=}\,2.8{\pm}0.8\,\mu{\rm G}$ and ${\beta}\,{\sim}\,0.1{-}1$. The authors conclude that their IVF is completely ionized and magnetically dominated, similar to our results of F1.

These radio polarization studies (summarized in Table \ref{table:estimates}) highlight the importance of magnetic fields in structuring the high-Galactic latitude ISM. Magnetic fields play an important role in shaping both discrete ionized and neutral structures in the diffuse/translucent ISM. However, it remains unclear if and how the magnetic field in these phases are associated, emphasizing the need for further multi-phase magnetic field analyses.


\subsection{Multi-Phase Magnetic Field Comparison}\label{subsec:coupling}

The alignment found between GALFACTS polarization gradient filaments and narrow GALFA-{\HI} structures across a wide velocity range suggests that the ionized and neutral gas share a common magnetic field in this region. Since the structures analyzed in this paper are not spatially correlated with excess synchrotron emission, we are unable to de-rotate the synchrotron polarization angles to directly compare magnetic field orientations in the ionized and neutral medium. The observed synchrotron polarization angles, however, provide information about the combination of magnetic field orientation and Faraday rotation. If the observed polarization angle is coherent, either the magnetic field orientation and degree of path-length integrated Faraday rotation is constant as a function of position, or these two quantities change in the amount required to yield no net change in the observed polarization angle. While the latter is possible, the former is more plausible. The strong degree of coherence in the observed GALFACTS polarization angles in Figure \ref{fig:GALFACTS_PLIC} highlights large areas of this region where the path-length integrated LOS magnetic field orientation in the ionized gas is roughly constant. This is supported by the roughly constant $\phi_{\rm max}$ that we measure toward these LOSs. Comparison with the dust polarization data in Figure \ref{fig:planck-halpha} (left) suggests that the POS field component is also strongly coherent.

We showed that F1 may be a magnetically-dominated ionized filament. Significant magnetic pressure would prevent gas flow perpendicular to the filament, resulting in a magnetic field in the ionized medium that is parallel to F1 \citep[e.g.,][]{Stil2016}. If the {\halpha\!\!} filament is magnetically dominated, the alignment that we find with the {\Planck} dust POS magnetic field orientation and narrow {\HI} structures would support our interpretation of multi-phase magnetic field alignment. 

If the ionized and neutral medium do share a common magnetic field, it is reasonable to expect that they have consistent magnetic field directions and comparable field strengths. Zeeman splitting measurements can be used to measure the LOS magnetic field direction and strength in cold neutral gas. We identified a Zeeman splitting measurement using {\HI} absorption \citep{Heiles2004} toward a single LOS at the far east of our field at coordinates $(\ell,b)\,{=}\,(213\deg,{+}30.1\deg)$ with two magnetic field strength estimates: ${-1.9\,{\pm}\,2.2}\,\mu{\rm G}$ and ${-3.2\,{\pm}\,3.7}\,\mu{\rm G}$. While the magnetic field direction measured using Zeeman splitting is negative, the sign convention is opposite to that of RM \citep{Robishaw2018}, resulting in a consistent magnetic field direction between the ionized and neutral gas along this LOS.

To compare our magnetic field strength estimates in the ionized gas to that of the predominantly neutral medium, we apply the Davis-Chandrasekhar-Fermi (DCF) method \citep{Davis1951, Chandrasekhar1953} to the {80\arcmin} {\Planck} dust polarization angles, $\tilde{\chi}_{353}$, to measure the total magnetic field strength. We use the form of the DCF as modified by \citet{Heitsch2001},

\begin{align}
    \langle{B}\rangle^2 = \xi4\pi\rho \frac{\sigma_v^2}{\sigma(\tan\delta_\chi)^2},
\end{align}

{\noindent}where $\sigma_v$ is the turbulent linewidth, $\rho$ is the mass density of the neutral medium, $\xi\,{=}\,0.5$ is a correction factor that reflects the ratio of turbulent magnetic to turbulent kinetic energy, and $\delta_\chi$ is the dispersion in dust polarization angle from the mean value given by $\delta_\chi \equiv \chi_{353} - \langle\chi_{353}\rangle$.

To apply the DFC method to F1, we used the same `on' mask in Section \ref{sec:estimates} at {5\arcmin} resolution for use with the {\HI} data and degraded to {80\arcmin} resolution for the dust polarization data. Assuming a filamentary geometry of F1 with a path length of 2.2 pc (Section \ref{subsec:physical-interpretation}), we measured an {\HI} number density of $n_{\rm H\,I}\,{=}\,{\NHI}/2.2\,{\rm pc}\,{\sim}\,10\,{\rm cm^{-3}}$ over a velocity range $0{\lesssim}\,{\vlsr}\,{\lesssim}\,{+}10\,\kms$ using Equation \ref{eq:NHI}. This corresponds to a mass density of $\rho\,{=}\,1.4m_Hn_{\rm H\,I}\,{=}\,2{\times}10^{-23}\,{\rm g\,cm^{-3}}$. We assumed that $\sigma_v\,{=}\,\sigma_{\rm turb}\,{=}1.4\,\kms$, where $\sigma_{\rm turb}$ is the turbulent velocity \citep{McClure-Griffiths2006, Clark2014}, found to be equal to the CNM component of F2 found with {\ROHSA}. We find a dust polarization angle dispersion of $\delta_\chi\,{\sim}\,5\deg$, yielding a total magnetic field strength of ${\sim}21\,\mu{\rm G}$ in the neutral medium, which may only be accurate to within a factor of a few \citep{McClure-Griffiths2006, Yoon2019, Cho2016, Skalidis2021}. While this is higher than the total magnetic field strength in the ionized gas inferred from our Faraday depth measurements, $B_{\rm tot}\,{\sim}\,6{\pm}4\,\mu{\rm G}$, this is consistent with increased magnetic field strengths found in the predominantly neutral medium \citep[e.g.,][]{Wolleben2009, Crutcher2010, McClure-Griffiths2006, Clark2014, Tritsis2019, Thomson2019}. The magnetic field strength that we estimate using the DFC method along F1 is also higher than that of the far east of our field away from the filament found using Zeeman splitting measurements, further suggesting magnetic field compression along F1.

Our {\ROHSA} results (Section \ref{subsec:hifibers}) suggests that we may be seeing a magnetic field alignment between the WNM and CNM. While we identify narrow {\HI} structures over an {\HI} velocity range of ${-}20\,{\lesssim}\,{\vlsr}\,{\lesssim}\,{+}20\,\kms$ for the entire region, the WNM is found to dominate the {\HI} emission outside of the velocity range ${-}10\,{\lesssim}\,{\vlsr}\,{\lesssim}0\,\kms$ toward the small portion of F2 that we analyze. If the WNM dominates narrow {\HI} structures in a similar velocity range in the rest of G216, then the roughly coherent {\HI} orientation over the ${\sim}40\,\kms$ velocity range (Figure \ref{fig:hi_rht}) may reflect a common POS magnetic field orientation in the WNM and CNM. A more complete {\ROHSA} analysis on the entire field is required to further understand over what velocity range the CNM and WNM dominate narrow {\HI} structures. 


\subsection{Interpreting Spatial Correlations Between Radio Polarization Structures and {\HI} Emission}\label{subsec:frsource}

Spatial correlations found between radio polarization structures and {\HI} emission can be interpreted in two ways. Either the ionized and neutral media, and potentially their magnetic fields, are correlated, or the polarized emission is dominated by emission from the {\HI} structure itself. We discuss the two possibilities here.

Faraday rotation by the multi-phase ISM has significant implications for the interpretation of radio polarimetric observations, particularly at low frequencies. Faraday rotation is linearly proportional to $n_e$ and typically discussed as an effect from the WIM \citep[e.g.,][]{Haverkorn2004b,Haverkorn2004c,Hill2008,Gaensler2011,Heiles2012}. Assuming a typical local thermal electron density and $B_\parallel\,{=}\,2\,\mu{\rm G}$, \citet{VanEck2017} showed that the WIM produces an amount of Faraday rotation per unit path length of ${\sim}\,0.32\,{\rm rad\,m^{-2}\,pc^{-1}}$ while the WNM and HIM, respectively, only produce ${\sim}\,0.016\,{\rm rad\,m^{-2}\,pc^{-1}}$ and ${\sim}\,0.008\,{\rm rad\,m^{-2}\,pc^{-1}}$. A similar estimate for the CNM, assuming an electron density of $n_e\,{\sim}\,0.02\,{\rm cm^{-3}}$ \citep{Ferriere2020} and the same LOS magnetic field strength, yields ${\sim}\,0.03\,{\rm rad\,m^{-2}\,pc^{-1}}$ \citep{Thomson2019}. As a result, observable Faraday rotation from phases other than the WIM requires significant LOS magnetic field strengths and/or path lengths. The typical length scale of the CNM is ${\sim}10\,{\rm pc}$ \citep{Ferriere2020}, rendering significant path lengths unlikely. Increased magnetic field strengths have been shown to produce measurable Faraday rotation in CNM-associated dust clouds \citep{Thomson2019} and in molecular clouds \citep{Tahani2018}.

The 3C196 LOFAR field \citep{vanHaarlem2013} contains radio polarized structures that are spatially aligned with the dust POS magnetic field orientation in {\Planck} dust polarization \citep{Jelic2015, Zaroubi2015}. Recent work by \citet{Bracco2020} found spatial correlations between LOFAR polarized intensity structures and {\HI} emission using data from the Effelsberg-Bonn {\HI} Survey (EBHIS) toward the 3C196 field, hypothesizing that this observational correlation may be due to Faraday rotation from the CNM. However, \citet{VanEck2017} showed that at the very low LOFAR frequencies (${\sim}150\,{\rm MHz}$), Faraday rotation from the WIM causes significant depth depolarization, resulting in polarized intensity structures dominated by the Faraday-thin neutral ISM. Furthermore, the existence of LOFAR polarization structures aligned with cold {\HI} emission alone is not evidence for a significant RM contribution from the CNM. Strong evidence for Faraday rotation from the CNM would require a correlation between cold {\HI} gas and RM structures signifying enhanced Faraday rotation. The observed correlation found by \citet{Bracco2020} is consistent with the explanation provided by \citet{VanEck2017} and therefore does not constitute evidence that CNM provides a significant contribution to the measured RM toward 3C196. The results of \citet{VanEck2017} suggest that the spatial correlation between LOFAR polarized intensity structures and the {\Planck} dust magnetic field orientation presented by \cite{Jelic2015} reflects a common magnetic field orientation in the predominantly neutral ISM.

Spatial correlations found between LOFAR depolarization canals, cold {\HI} structures, and POS magnetic field orientation in {\Planck} dust emission have been described as another example of multi-phase magnetic field alignment \citep{Jelic2018, Turic2021}. The small spatial scale of these depolarization canals indicate that they are likely caused by beam depolarization, where fluctuations in the observed polarization angle become averaged within the telescope beam. The fluctuation in the observed polarization angle may either be caused by changes in the degree of foreground Faraday rotation or by changes in the intrinsic magnetic field orientation within the polarized emitting region. If changes in the foreground RM are responsible for these canals, they are likely to be WIM-dominated structures and their alignment with tracers of the magnetized neutral ISM may be evidence for multi-phase magnetic field alignment. However, if these canals trace changes in the intrinsic magnetic field orientation, they are likely to be neutral-dominated structures and the observed alignment of magnetic field tracers may again reflect a common field orientation in the neutral medium. 

The Faraday rotation presented in this paper is likely a consequence of ionized gas. The RM excess of F1 is spatially coincident with an {\halpha\!\!} filament and is clearly offset from the edge of {\HI} and dust emission, rendering Faraday rotation from within the neutral medium unlikely. The RM enhancement toward F2 is also found to be spatially correlated with very faint {\halpha\!\!} emission. Assuming typical thermal electron densities and $B_\parallel\,{=}\,6\,\mu{\rm G}$ (Section \ref{subsec:physical-interpretation}), the WIM and CNM are expected to produce an amount of Faraday rotation per unit path length of $0.97\,{\rm rad\,m^{-2}\,pc^{-1}}$ and $0.097\,{\rm rad\,m^{-2}\,pc^{-1}}$, respectively. With an enhanced RM of ${\sim}\,40\,\radmsq$ toward F2, this requires a WIM and CNM path length of ${\sim}\,40\,{\rm pc}$ and ${\sim}\,400\,{\rm pc}$, respectively. This is typical of the length scale of the WIM but roughly one order of magnitude greater than that of the CNM \citep{Ferriere2020}. The 1.4 GHz GALFACTS polarization data has a maximum Faraday depth scale that is ${\sim}30$ times greater than that of LOFAR \citep{VanEck2017} and is therefore less severely depolarized by the WIM; while the observing frequency can strongly affect the interpretation of polarized intensity structures, Faraday rotation generally remains dominated by ionized gas. The comparison of GALFACTS polarization structures and {\HI} emission presented in this paper is therefore a comparison of two tracers that genuinely trace different phases of the ISM.


\subsection{Why are Associations Between Multi-Phase Magnetic Field Tracers not more Widespread?}

Multi-phase structures and their magnetic fields are not widely associated. \citet{VanEck2019} compared Faraday depth cubes from the LOFAR Two-meter Sky Survey (LOTSS, \citealt{Shimwell2017}), covering 568 deg$^2$ of the high-Galactic latitude sky, to other multi-frequency tracers, finding only one LOFAR filament that is spatially correlated with {\HI} emission (see their Figure 11). Their {\HI} filament lies along the boundary of a linear gradient in Faraday rotation, suggested to be due to an envelope of ionized gas surrounding the predominantly neutral {\HI} filament. Of the four targeted studies that compared LOFAR polarized intensity structures to {\HI} emission \citep{Jelic2015, Bracco2020}, only two showed strong morphological correlations. If LOFAR polarized intensity structures are dominated by Faraday-thin structures in cold gas \citep{VanEck2017}, the lack of global spatial correlations with {\HI} emission only further demonstrates the complexity of radio polarization studies. For instance, the polarization horizon of LOFAR is significantly closer than that of GALFACTS and may not sample the full polarized-emitting Galactic volume. More recently, \citet{Ogbodo2020} compared magnetic field strengths derived using OH masers as part of the Mapping the Galactic Magnetic field through OH masers (MAGMO) project \citep{Green2012} to pulsar Faraday rotation measurements collated from \citet{Nota2010}. These measurements probe the neutral and ionized medium, respectively, and were chosen toward {\HII} regions that should dominate the environment in both tracers. While the statistics are limited and do not necessarily reflect the large-scale GMF properties, they found no strong correlation between these magnetic field measurements. 

The alignment found between GALFACTS polarization gradient filaments and GALFA-{\HI} structures presented in this paper is not a widespread occurrence in the high-Galactic latitude Arecibo sky. The polarization gradient filaments analyzed here are highly coherent and significantly isolated from other more complex structures in the polarization gradient compared to those typically found in the GALFACTS data. It may be that polarization gradient and {\HI} emission structures are widely correlated, but observational effects make observing them difficult. Alternatively, associations between warm ionized and cold neutral gas physically may only exist in rare circumstances.

An important observational effect is the polarization horizon, the distance beyond which polarized emission is completely depolarized \citep{Uyaniker2003}. The distance to the polarization horizon is given by

\begin{align}\label{eq:polhorizon}
    D_{\rm ph} = \frac{\pi}{0.81\lambda^2n_eB_\parallel},
\end{align}

{\noindent}resulting in polarized intensity structures that generally sample varying Galactic volumes as a function of observing frequency that also depends on angular resolution and direction. Assuming mean values of $B_\parallel\,{\sim}\,2\,\mu{\rm G}$ and $n_e\,{\sim}\,0.02\,{\rm cm^{-3}}$, the distance to the GALFACTS polarization horizon is $D_{\rm ph}\,{\sim}\,2\,{\rm kpc}$, a factor of ${\sim}5$ smaller than the total path length in this direction \citep{Sun2008}. The distance to the polarization horizon as given in Equation \ref{eq:polhorizon} is for the idealized case of a uniform medium, and its true distance is likely much less than this due to turbulence. Even if there is widespread multi-phase magnetic field alignment, the GALFACTS polarization horizon may render them unobservable and could help to explain the lack of observed alignments. The polarization horizon is not a well-defined boundary in distance and information beyond the polarization horizon may still affect observables \citep{Hill2018}.

An alignment between structures in the polarization gradient and narrow {\HI} structures necessitates both a LOS magnetic field component in the ionized medium and a POS magnetic field component in the neutral medium, respectively, and may therefore be unlikely to occur in regions that contain predominantly ionized or neutral gas. Our visual comparison of polarization gradient filaments in the Arecibo data and other multi-frequency tracers revealed that there is not a one-to-one correspondence between polarization gradient and {\halpha\!} filaments. Some examples of this in G216 include the F1V structure that is prominently observed in VTSS {\halpha\!\!} emission but not in the {5\arcmin} polarization gradient, and the filamentary F2 structure that is prominently seen in the {5\arcmin} polarization gradient but does not have a corresponding VTSS {\halpha\!\!} filament (recall Figure \ref{fig:halpha}). The nearby {\halpha\!\!} filaments \citep{Haffner1998} also lack counterparts in the {5\arcmin} polarization gradient. The polarization gradient highlights structures at particular angular scales \citep{Herron2018I}, so more extended {\halpha\!\!} structures \citep[e.g.,][]{Haffner1998} may only be highlighted by the polarization gradient using a larger kernel. In contrast, some small-scale {\halpha\!\!} structures do not have corresponding features in the polarization gradient. These ionized structures may not have significant LOS magnetic field strengths to induce an enhanced RM. 


\section{Summary and Conclusions}\label{sec:summary}

We present a multi-phase analysis of the magnetized ISM in search of evidence that the high-Galactic latitude magnetic field is shared between the ionized and neutral ISM. We visually compared structures in the 1.4 GHz GALFACTS polarization gradient to narrow, velocity-resolved GALFA-{\HI} structures, focusing on those that are associated with diffuse/translucent {\HI} emission and not spatially correlated with excess synchrotron emission. We identified a single region, G216 centered on $(\ell,b)\,{\sim}\,(216\deg,{+}26\deg)$, that contains coherent polarization gradient filaments clearly aligned with narrow {\HI} structures. We compared multi-phase observations and magnetic field tracers to investigate whether the ionized and neutral media of this region (1) are associated with one another and (2) share a common magnetic field.

This region is characterized by two filamentary structures in the polarization gradient that are roughly parallel to the Galactic plane. The polarization gradient filament farthest from the plane (F1) is spatially correlated with a bright {\halpha\!\!} filament along the edge of a large, dusty {\HI} cloud. The other filament (F2) is spatially correlated with the edges of very faint {\halpha\!\!} emission and lies along the edge of a dusty {\HI} filament.

We showed that the polarization gradient filaments are aligned with narrow {\HI} structures over a ${\sim}40\,\kms$ {\HI} velocity range using the RHT to identify coherent, linear structures. Using 3D RM synthesis, we showed that F1 may be a magnetically-dominated ($B_\parallel\,{=}\,6{\pm}4\,\mu{\rm G}$, $\beta\,{=}\,2.1^{+3.1}_{-2.1}$) filament with a magnetic field that is parallel to the filament. Since the 3D magnetic field geometry of F1 is unknown, this is an upper-estimate of $\beta$ based only on $B_\parallel$ only. The alignment with narrow {\HI} structures and dust polarization angles supports our interpretation of multi-phase magnetic field alignment. The LOS and POS magnetic field orientation in the ionized and neutral gas, respectively, are both strongly coherent across the region. Our work is consistent with filamentary structures in the ISM being preferentially parallel to the Galactic plane along the mean magnetic field \citep[e.g.,][]{Shajn1958, Soler2020}.

We used the Davis-Chandrasekhar-Fermi method to estimate the total magnetic field strength of ${\sim}21\,\mu{\rm G}$ in the neutral medium using {\Planck} dust polarization data. This is higher than the total magnetic field strength in the ionized gas inferred from our Faraday depth measurements ($B_{\rm tot}\,{=}\,11{\pm}8\,\mu{\rm G}$), suggesting an enhanced field strength in the neutral medium toward F1. We find two Zeeman splitting measurements along a single LOS toward the far east of our field \citep{Heiles2004} with magnetic field strengths of ${-1.9\,{\pm}\,2.2}\,\mu{\rm G}$ and ${-3.2\,{\pm}\,3.7}\,\mu{\rm G}$. These Zeeman splitting measurements have the same LOS magnetic field direction as our Faraday depth measurement, however, more Zeeman splitting measurements are needed to make a statistically significant statement about a common LOS field direction between the ionized and neutral medium.

We discuss the lack of widespread alignments found between multi-phase magnetic field tracers in the high-Galactic latitude Arecibo sky. We find no strong evidence for an {\HI} shell, but are unable to rule out the possibility of a shock. We consider the possibility that G216 is associated with a short-timescale or physically rare phenomenon, and suggest the possibility that a compressive event may have triggered the formation of a transient layer of dense ionized gas around a neutral gas cloud while aligning their respective magnetic fields. Further in-depth analyses on the alignment between GALFACTS polarization gradient filaments and {\HI} structures, or tracers of the ionized and neutral structures in general, are required to better understand the association between these phases of the magnetic ISM.


\section{Acknowledgments}\label{sec:acknowledgments}

We are grateful to Peter Martin for helpful discussions, Josh Speagle for advice on 3D dust maps, and the anonymous referee for their kind and constructive comments that greatly improved our paper.
J.L.C. acknowledges support from the Ontario Graduate Student Scholarship.
J.L.C. and B.M.G. acknowledge the support of the Natural Sciences and Engineering Research Council of Canada (NSERC) through grant RGPIN-2015-05948, and of the Canada Research Chairs program.
S.E.C. acknowledges support by the National Science Foundation under Grant No. 2106607. 
J.M.S. acknowledges the support of the Natural Sciences and Engineering Research Council of Canada (NSERC), 2019-04848.
The Dunlap Institute is funded through an endowment established by the David Dunlap family and the University of Toronto. The University of Toronto operates on the traditional land of the Huron-Wendat, the Seneca, and most recently, the Mississaugas of the Credit River; we are grateful to have the opportunity to work on this land.

This publication utilizes data from Galactic ALFA {\HI} (GALFA HI) survey data set obtained with the Arecibo L-band Feed Array (ALFA) on the Arecibo 305m telescope. The Arecibo Observatory is operated by SRI International under a cooperative agreement with the National Science Foundation (AST-1100968), and in alliance with Ana G. M\'endez-Universidad Metropolitana, and the Universities Space Research Association. The GALFA {\HI} surveys have been funded by the NSF through grants to Columbia University, the University of Wisconsin, and the University of California.

\software{Astropy \citep{Astropy2013,Astropy2018}, CARTA \citep{CARTA2019}, dustmaps \citep{Green2018}, healpy \citep{Zonca2019}, matplotlib \citep{Hunter2007}, numpy \citep{Harris2020}, RHT \citep{Clark2014}, RM Tools \citep{Purcell2020}, scipy \citep{Virtanen2020}.}


\clearpage
\appendix
\renewcommand\thefigure{\thesection.\arabic{figure}} 
\setcounter{figure}{0}


\section{RHT Backprojections}
\label{appendix:rht}

\begin{figure*}[h!]
    \centering
        
        {\includegraphics[width=8cm]{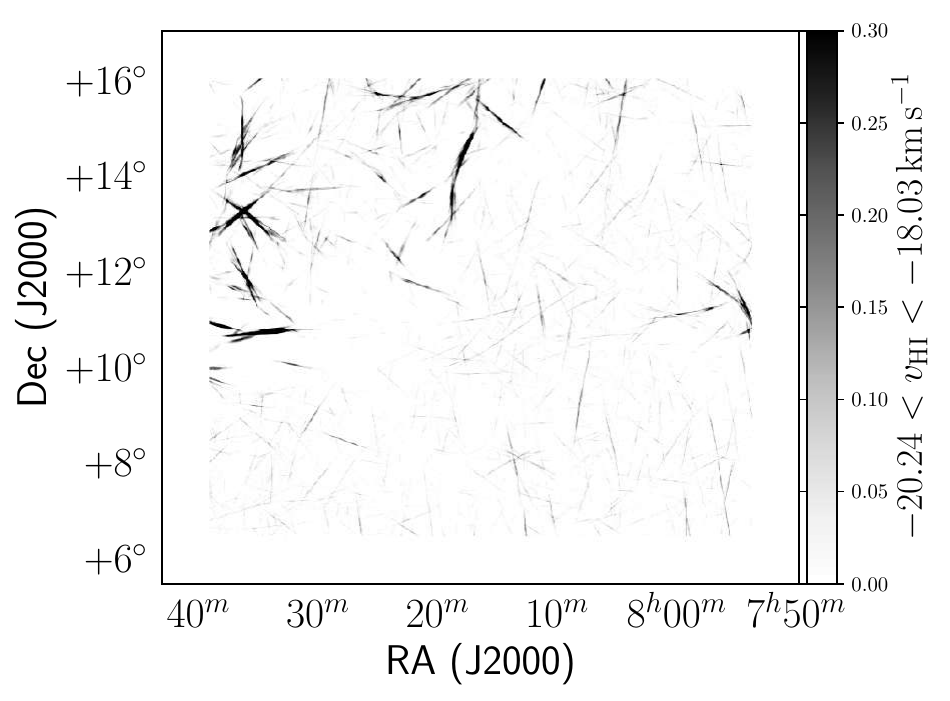}\includegraphics[width=8cm]{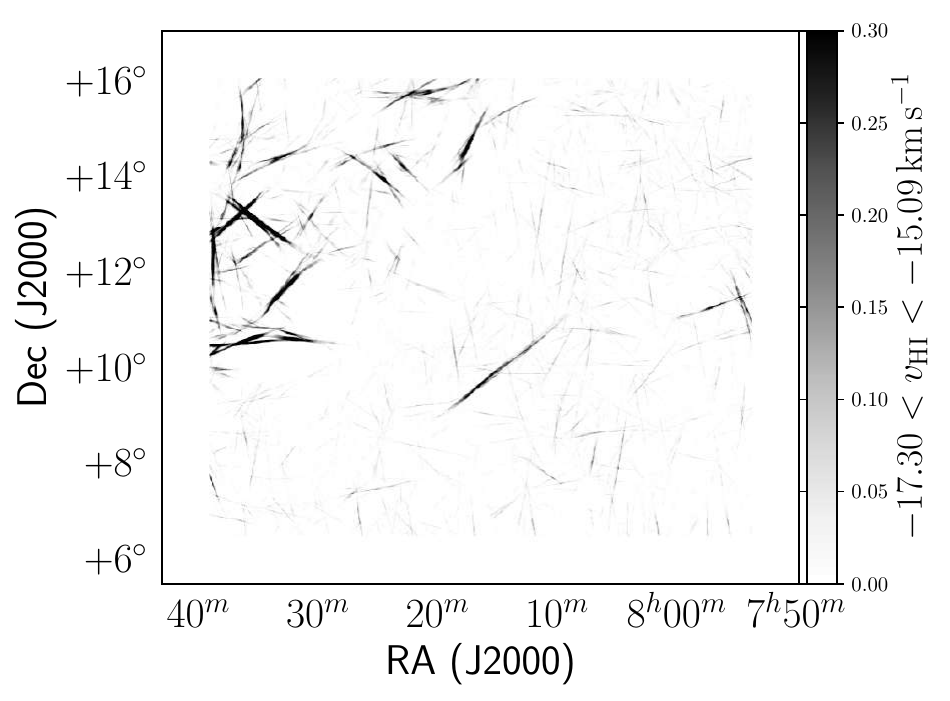}}\vspace{-15pt}
        {\includegraphics[width=8cm]{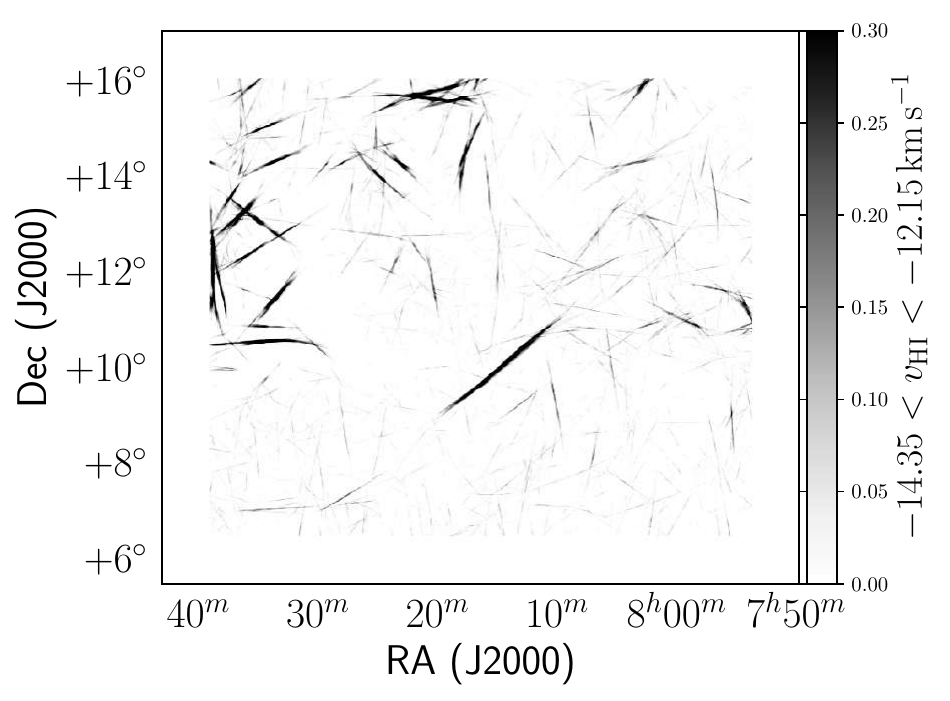}\includegraphics[width=8cm]{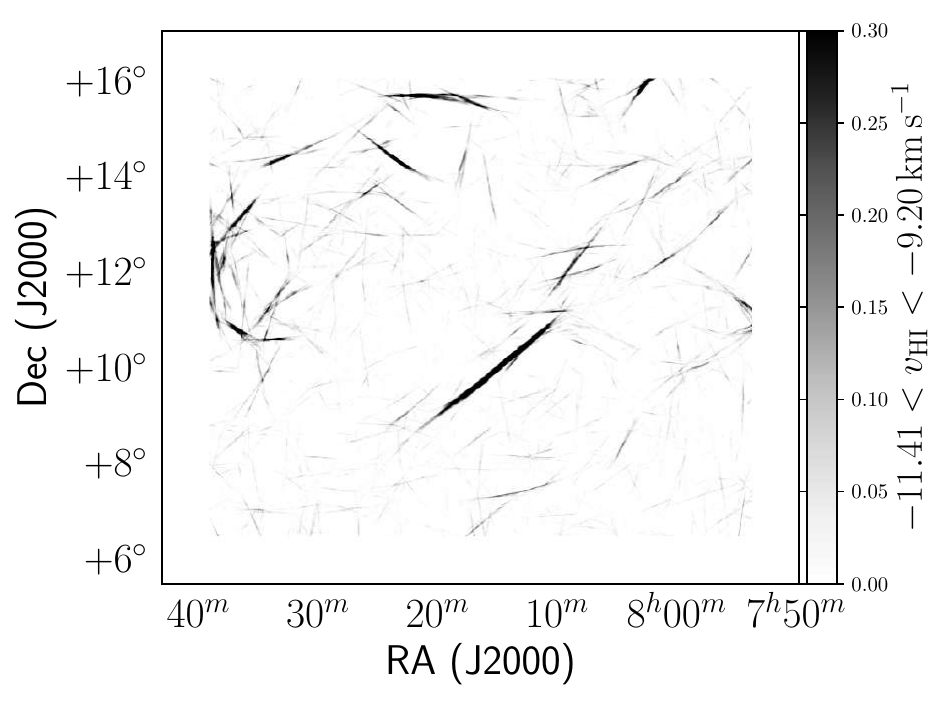}}\vspace{-15pt}
        {\includegraphics[width=8cm]{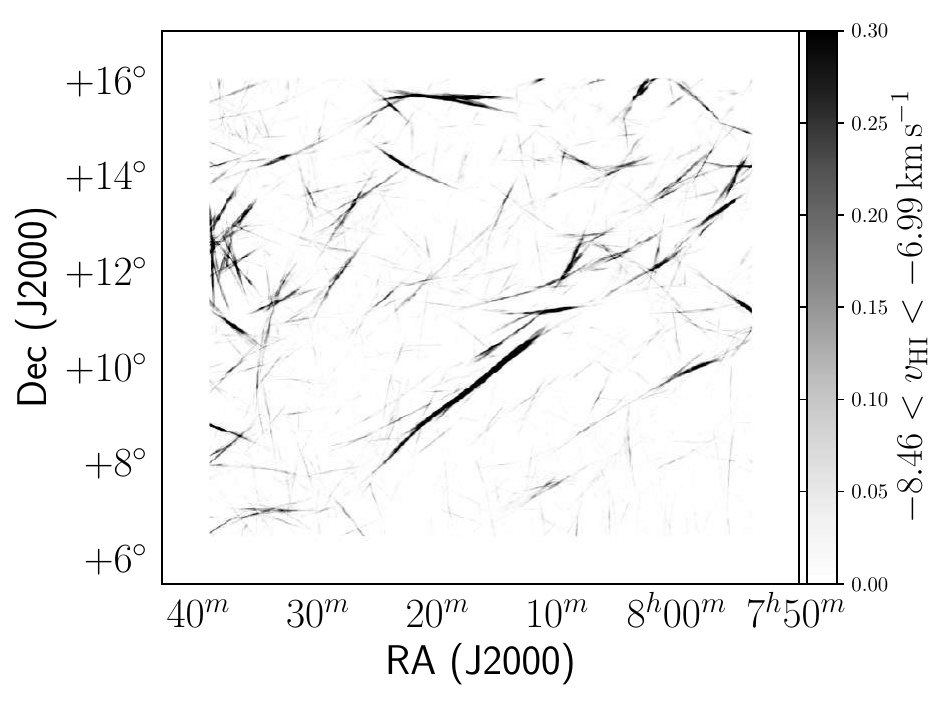}\includegraphics[width=8cm]{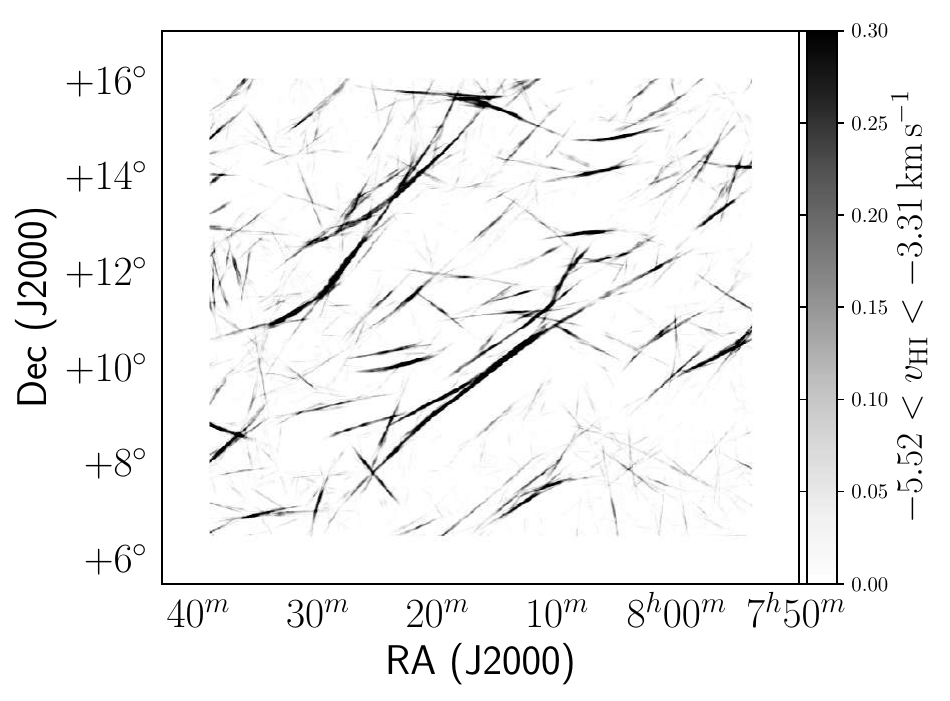}}

\caption{RHT backprojections for the GALFA-{\HI} velocity slices within the range $-20.24<v_{\rm lsr}<-3.31\,\kms$.}
\label{fig:hi_rht}
\end{figure*}

The RHT backprojections of the GALFA-{\HI} data are shown in Figure \ref{fig:hi_rht}, and reveal numerous {\HI} structures across the entire field.

\addtocounter{figure}{-1}
\begin{figure*}[t!]
    \centering
    
    {\includegraphics[width=8cm]{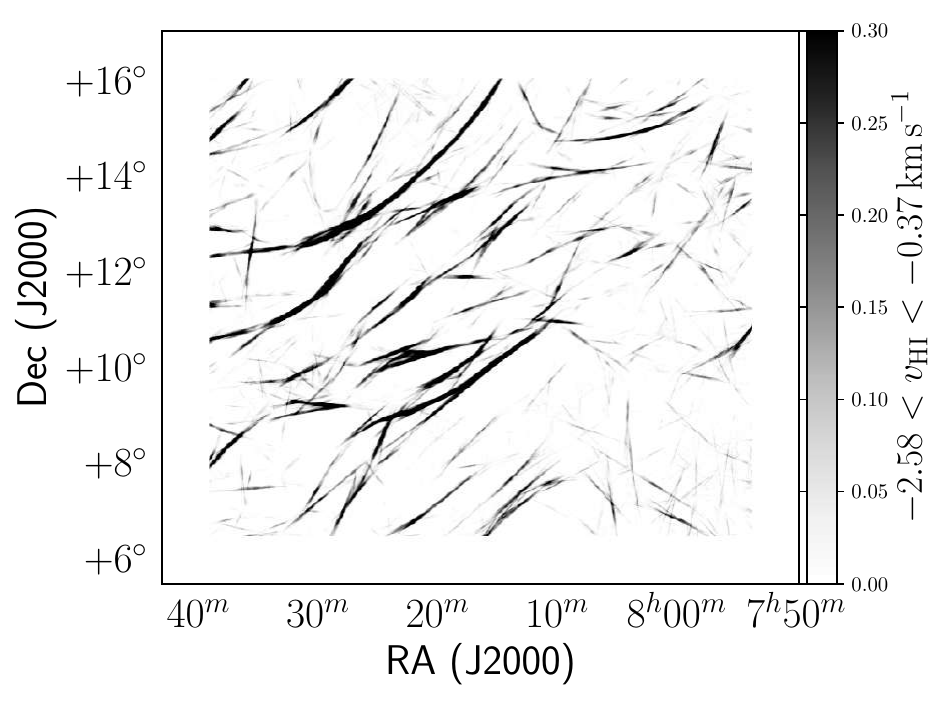}\includegraphics[width=8cm]{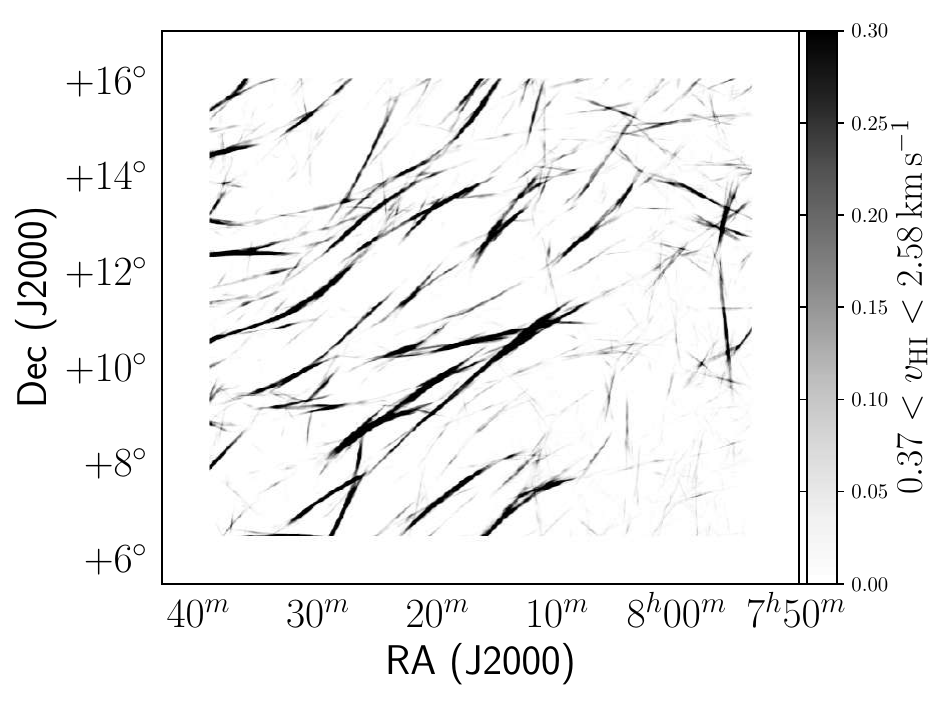}}\vspace{-15pt}
    {\includegraphics[width=8cm]{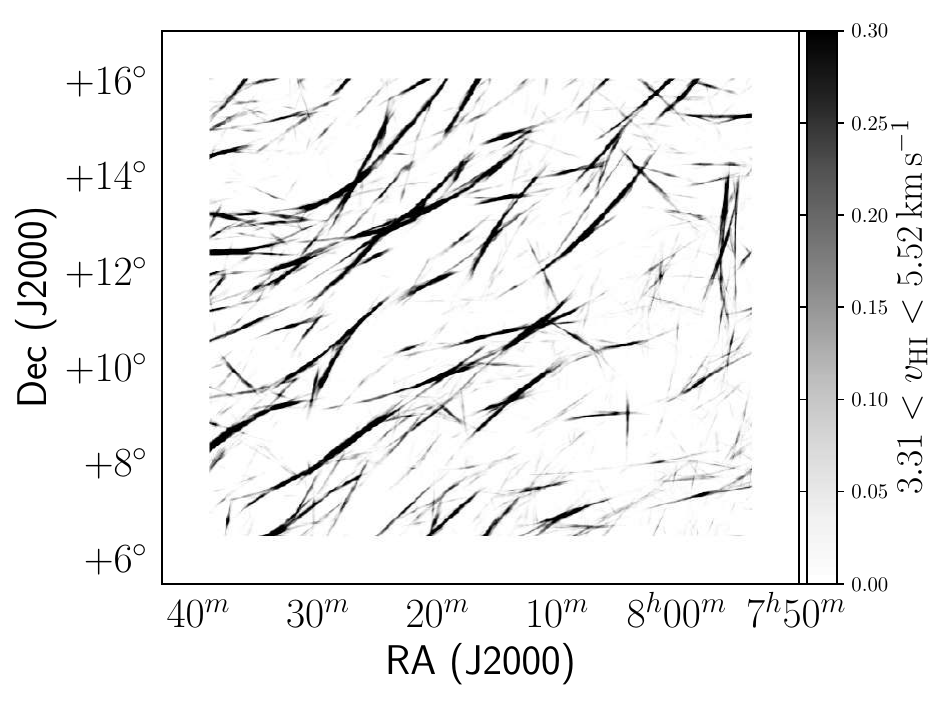}\includegraphics[width=8cm]{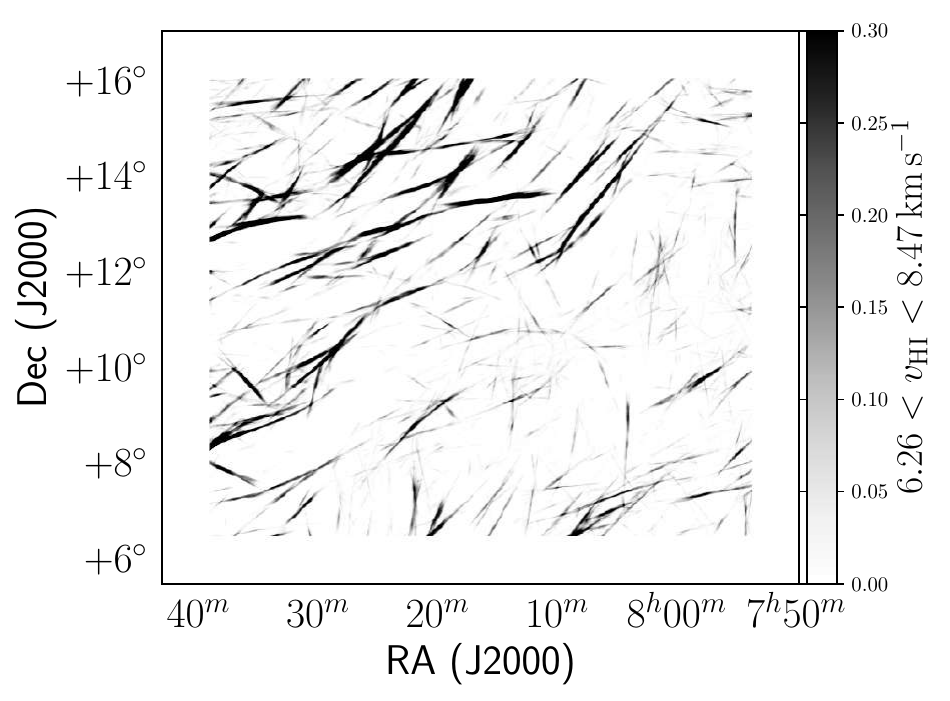}}\vspace{-15pt}
    {\includegraphics[width=8cm]{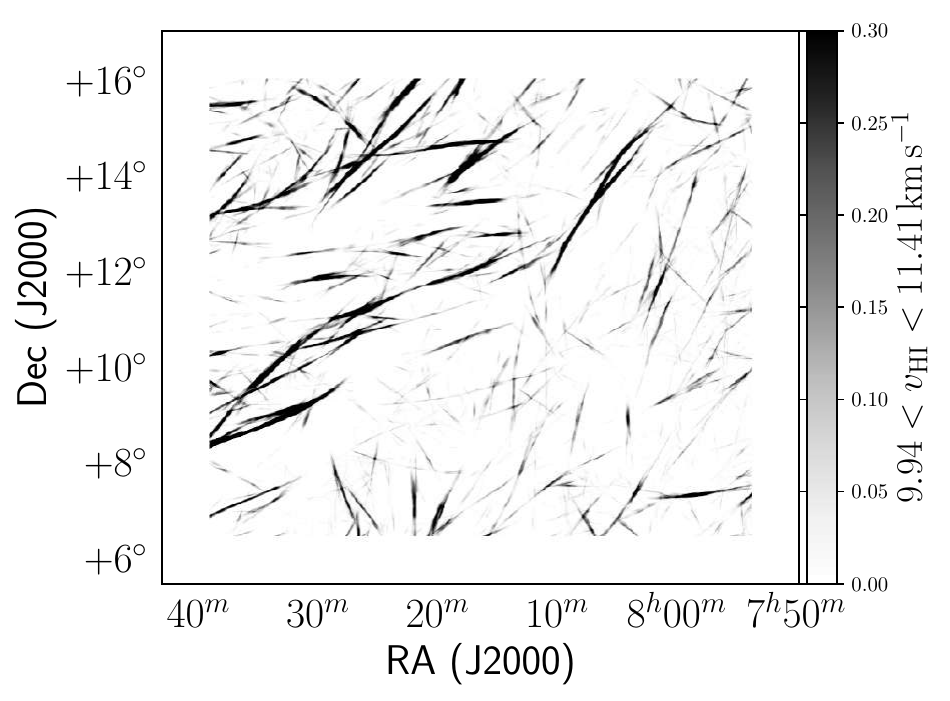}\includegraphics[width=8cm]{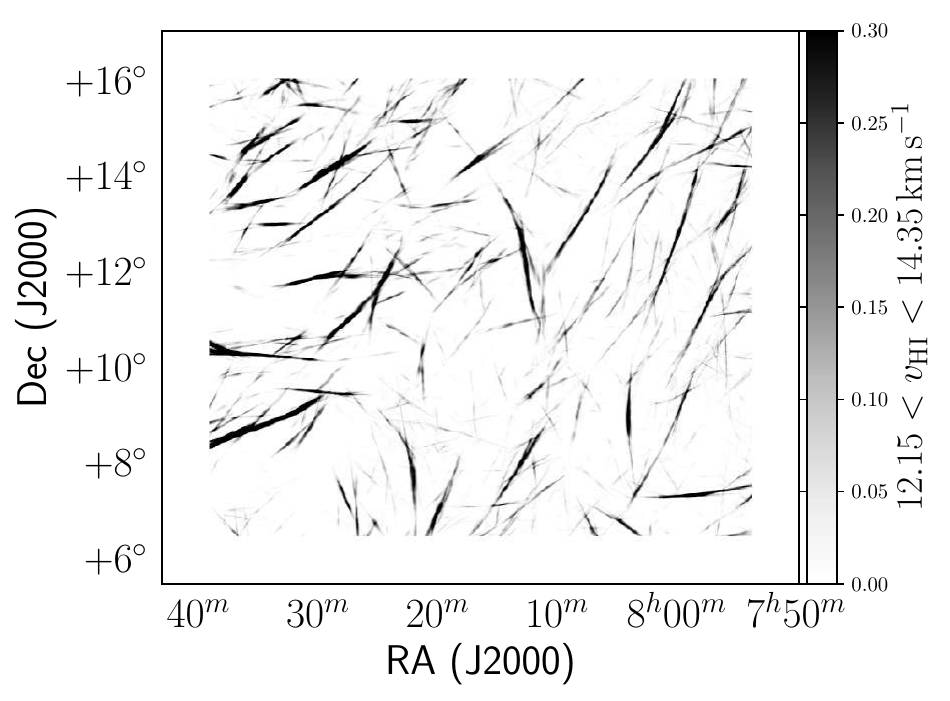}}\vspace{-15pt}
    {\includegraphics[width=8cm]{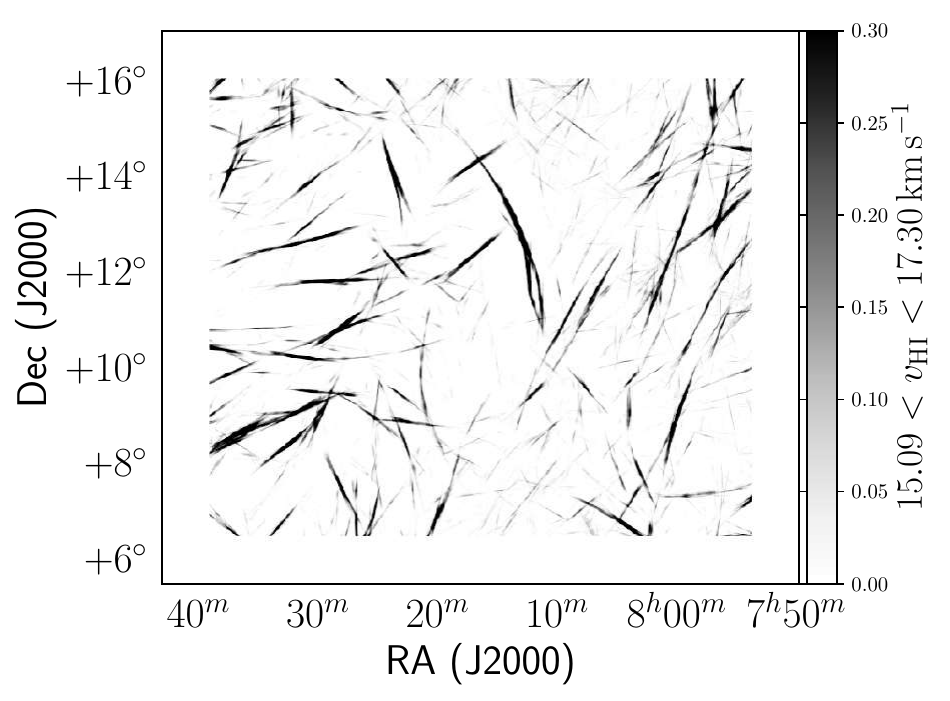}\includegraphics[width=8cm]{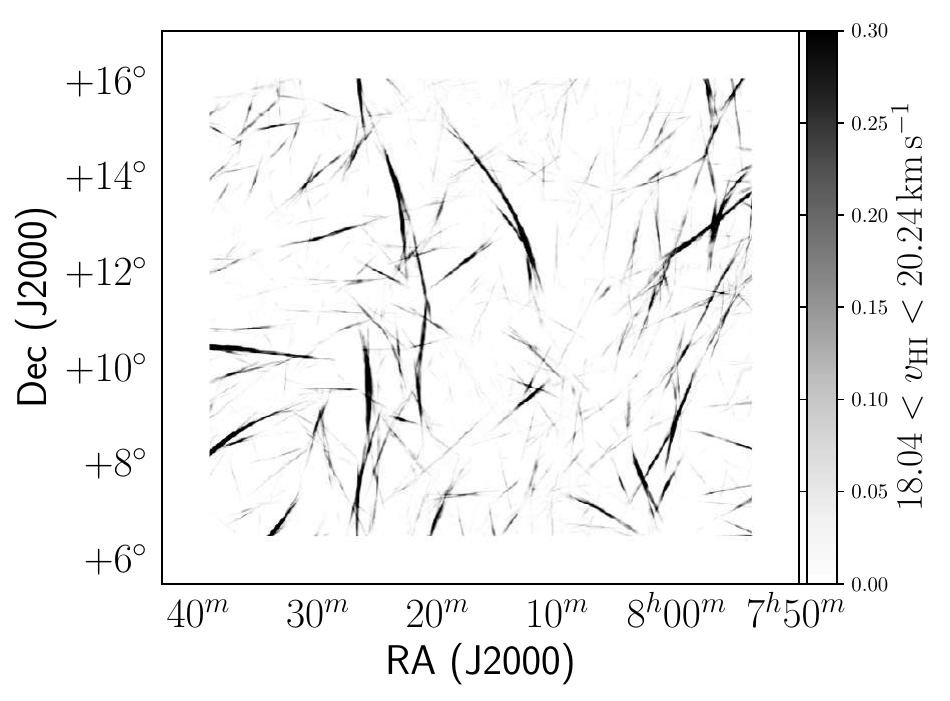}}
    
\caption{RHT backprojections for the GALFA-{\HI} velocity slices within the range $-2.58<v_{\rm lsr}<+20.24\,\kms$.}
\end{figure*}

\clearpage

\begin{figure}[h!]
\centering
 \includegraphics[width=8.5cm]{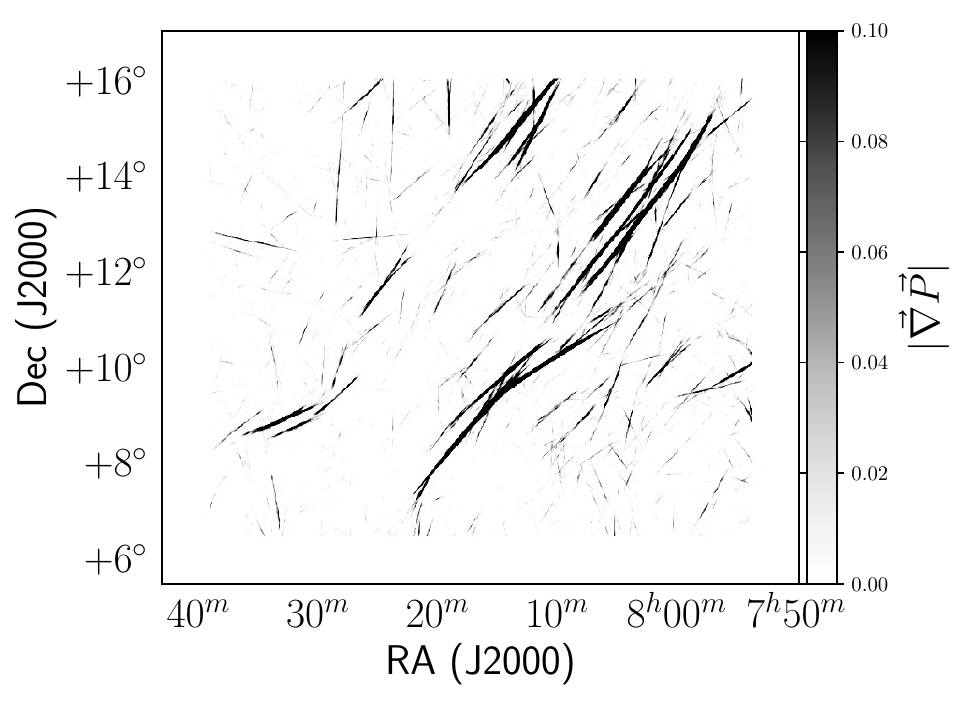}
 \caption{RHT backprojection of the de-striped {5\arcmin} GALFACTS {\polgradmax\!\!} map.}
\label{fig:polgrad_RHT}
\end{figure}

Figure \ref{fig:polgrad_RHT} shows the RHT backprojection of the {5\arcmin} GALFACTS polarization gradient. The RHT does exceptionally well at reconstructing the filamentary features in the polarization gradient, reproducing the single- and double-jump morphology of F1 along with the double- and triple-jump morphology of F2. Inspection of the RHT backprojection reveals that there are still signatures of scanning artefacts, however, they are ${\sim}\,10$ times fainter than the polarization gradient filaments.

\begin{figure}[h!]
\centering
 \includegraphics[width=8.5cm]{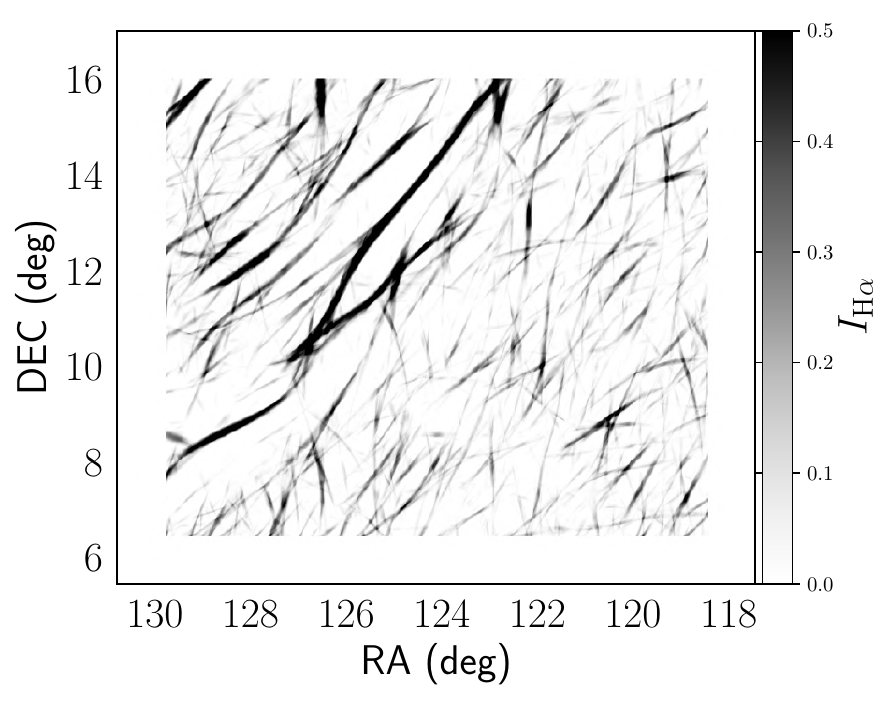}
 \caption{RHT backprojection of the VTSS {\halpha\!\!} map.}
\label{fig:VTSS_RHT}
\end{figure}

Figure \ref{fig:VTSS_RHT} shows the RHT backprojection of the VTSS {\halpha\!\!} emission.

\bibliographystyle{apj_long_etal}
\bibliography{bibliography}

\begin{thebibliography}{150}
\expandafter\ifx\csname natexlab\endcsname\relax\def\natexlab#1{#1}\fi

\bibitem[{{Arnaud}(1996)}]{Xspec1996}
{Arnaud}, K.~A. 1996, in Astronomical Society of the Pacific Conference Series,
  Vol. 101, Astronomical Data Analysis Software and Systems V, ed. G.~H.
  {Jacoby} \& J.~{Barnes}, 17

\bibitem[{{Astropy Collaboration} {et~al.}(2018){Astropy Collaboration},
  {Price-Whelan}, {Sip{H{o}}cz}, {G{"u}nther}, {Lim}, {Crawford}, {Conseil},
  {Shupe}, {Craig}, {Dencheva}, {Ginsburg}, {Vand erPlas}, {Bradley},
  {P{'e}rez-Su{'a}rez}, {de Val-Borro}, {Aldcroft}, {Cruz}, {Robitaille},
  {Tollerud}, {Ardelean}, {Babej}, {Bach}, {Bachetti}, {Bakanov}, {Bamford},
  {Barentsen}, {Barmby}, {Baumbach}, {Berry}, {Biscani}, {Boquien}, {Bostroem},
  {Bouma}, {Brammer}, {Bray}, {Breytenbach}, {Buddelmeijer}, {Burke},
  {Calderone}, {Cano Rodr{'i}guez}, {Cara}, {Cardoso}, {Cheedella}, {Copin},
  {Corrales}, {Crichton}, {D'Avella}, {Deil}, {Depagne}, {Dietrich}, {Donath},
  {Droettboom}, {Earl}, {Erben}, {Fabbro}, {Ferreira}, {Finethy}, {Fox},
  {Garrison}, {Gibbons}, {Goldstein}, {Gommers}, {Greco}, {Greenfield},
  {Groener}, {Grollier}, {Hagen}, {Hirst}, {Homeier}, {Horton}, {Hosseinzadeh},
  {Hu}, {Hunkeler}, {Ivezi{'c}}, {Jain}, {Jenness}, {Kanarek}, {Kendrew},
  {Kern}, {Kerzendorf}, {Khvalko}, {King}, {Kirkby}, {Kulkarni}, {Kumar},
  {Lee}, {Lenz}, {Littlefair}, {Ma}, {Macleod}, {Mastropietro}, {McCully},
  {Montagnac}, {Morris}, {Mueller}, {Mumford}, {Muna}, {Murphy}, {Nelson},
  {Nguyen}, {Ninan}, {N{"o}the}, {Ogaz}, {Oh}, {Parejko}, {Parley}, {Pascual},
  {Patil}, {Patil}, {Plunkett}, {Prochaska}, {Rastogi}, {Reddy Janga},
  {Sabater}, {Sakurikar}, {Seifert}, {Sherbert}, {Sherwood-Taylor}, {Shih},
  {Sick}, {Silbiger}, {Singanamalla}, {Singer}, {Sladen}, {Sooley},
  {Sornarajah}, {Streicher}, {Teuben}, {Thomas}, {Tremblay}, {Turner},
  {Terr{'o}n}, {van Kerkwijk}, {de la Vega}, {Watkins}, {Weaver}, {Whitmore},
  {Woillez}, {Zabalza}, \& {Astropy Contributors}}]{Astropy2018}
{Astropy Collaboration} \etal . 2018, \aj, 156, 123

\bibitem[{{Astropy Collaboration} {et~al.}(2013){Astropy Collaboration},
  {Robitaille}, {Tollerud}, {Greenfield}, {Droettboom}, {Bray}, {Aldcroft},
  {Davis}, {Ginsburg}, {Price-Whelan}, {Kerzendorf}, {Conley}, {Crighton},
  {Barbary}, {Muna}, {Ferguson}, {Grollier}, {Parikh}, {Nair}, {Unther},
  {Deil}, {Woillez}, {Conseil}, {Kramer}, {Turner}, {Singer}, {Fox}, {Weaver},
  {Zabalza}, {Edwards}, {Azalee Bostroem}, {Burke}, {Casey}, {Crawford},
  {Dencheva}, {Ely}, {Jenness}, {Labrie}, {Lim}, {Pierfederici}, {Pontzen},
  {Ptak}, {Refsdal}, {Servillat}, \& {Streicher}}]{Astropy2013}
{Astropy Collaboration} \etal . 2013, \aap, 558, A33

\bibitem[{{Berkhuijsen} {et~al.}(2006){Berkhuijsen}, {Mitra}, \&
  {Mueller}}]{Berkhuijsen2006}
{Berkhuijsen}, E.~M., {Mitra}, D., \& {Mueller}, P. 2006, AN, 327, 82

\bibitem[{{Blagrave} {et~al.}(2017){Blagrave}, {Martin}, {Joncas}, {Kothes},
  {Stil}, {Miville-Desch{\^e}nes}, {Lockman}, \& {Taylor}}]{Blagrave2017}
{Blagrave}, K., {Martin}, P.~G., {Joncas}, G., {Kothes}, R., {Stil}, J.~M.,
  {Miville-Desch{\^e}nes}, M.~A., {Lockman}, F.~J., \& {Taylor}, A.~R. 2017,
  \apj, 834, 126

\bibitem[{{Bracco} {et~al.}(2020){Bracco}, {Jeli{\'c}}, {Marchal}, {Turi{\'c}},
  {Erceg}, {Miville-Desch{\^e}nes}, \& {Bellomi}}]{Bracco2020}
{Bracco}, A., {Jeli{\'c}}, V., {Marchal}, A., {Turi{\'c}}, L., {Erceg}, A.,
  {Miville-Desch{\^e}nes}, M.~A., \& {Bellomi}, E. 2020, \aap, 644, L3

\bibitem[{{Brentjens} \& {de Bruyn}(2005)}]{Brentjens2005}
{Brentjens}, M.~A. \& {de Bruyn}, A.~G. 2005, \aap, 441, 1217

\bibitem[{{Burkhart} {et~al.}(2012){Burkhart}, {Lazarian}, \&
  {Gaensler}}]{Burkhart2012}
{Burkhart}, B., {Lazarian}, A., \& {Gaensler}, B.~M. 2012, \apj, 749, 145

\bibitem[{{Burn}(1966)}]{Burn1966}
{Burn}, B.~J. 1966, \mnras, 133, 67

\bibitem[{Cabral \& Leedom(1993)}]{Cabral1993}
Cabral, B. \& Leedom, L.~C. 1993, in Proceedings of the 20th Annual Conference
  on Computer Graphics and Interactive Techniques, SIGGRAPH '93 (New York, NY,
  USA: Association for Computing Machinery), 263–270

\bibitem[{{Calabretta} {et~al.}(2014){Calabretta}, {Staveley-Smith}, \&
  {Barnes}}]{Calabretta2014}
{Calabretta}, M.~R., {Staveley-Smith}, L., \& {Barnes}, D.~G. 2014, \pasa, 31,
  e007

\bibitem[{{Chambers} {et~al.}(2016){Chambers}, {Magnier}, {Metcalfe},
  {Flewelling}, {Huber}, {Waters}, {Denneau}, {Draper}, {Farrow}, {Finkbeiner},
  {Holmberg}, {Koppenhoefer}, {Price}, {Rest}, {Saglia}, {Schlafly}, {Smartt},
  {Sweeney}, {Wainscoat}, {Burgett}, {Chastel}, {Grav}, {Heasley}, {Hodapp},
  {Jedicke}, {Kaiser}, {Kudritzki}, {Luppino}, {Lupton}, {Monet}, {Morgan},
  {Onaka}, {Shiao}, {Stubbs}, {Tonry}, {White}, {Ba{\~n}ados}, {Bell},
  {Bender}, {Bernard}, {Boegner}, {Boffi}, {Botticella}, {Calamida},
  {Casertano}, {Chen}, {Chen}, {Cole}, {Deacon}, {Frenk}, {Fitzsimmons},
  {Gezari}, {Gibbs}, {Goessl}, {Goggia}, {Gourgue}, {Goldman}, {Grant},
  {Grebel}, {Hambly}, {Hasinger}, {Heavens}, {Heckman}, {Henderson}, {Henning},
  {Holman}, {Hopp}, {Ip}, {Isani}, {Jackson}, {Keyes}, {Koekemoer}, {Kotak},
  {Le}, {Liska}, {Long}, {Lucey}, {Liu}, {Martin}, {Masci}, {McLean}, {Mindel},
  {Misra}, {Morganson}, {Murphy}, {Obaika}, {Narayan}, {Nieto-Santisteban},
  {Norberg}, {Peacock}, {Pier}, {Postman}, {Primak}, {Rae}, {Rai}, {Riess},
  {Riffeser}, {Rix}, {R{\"o}ser}, {Russel}, {Rutz}, {Schilbach}, {Schultz},
  {Scolnic}, {Strolger}, {Szalay}, {Seitz}, {Small}, {Smith}, {Soderblom},
  {Taylor}, {Thomson}, {Taylor}, {Thakar}, {Thiel}, {Thilker}, {Unger},
  {Urata}, {Valenti}, {Wagner}, {Walder}, {Walter}, {Watters}, {Werner},
  {Wood-Vasey}, \& {Wyse}}]{Chambers2016}
{Chambers}, K.~C. \etal . 2016, arXiv:1612.05560

\bibitem[{{Chandrasekhar} \& {Fermi}(1953)}]{Chandrasekhar1953}
{Chandrasekhar}, S. \& {Fermi}, E. 1953, \apj, 118, 113

\bibitem[{{Cho} \& {Yoo}(2016)}]{Cho2016}
{Cho}, J. \& {Yoo}, H. 2016, \apj, 821, 21

\bibitem[{{Clark} {et~al.}(2015){Clark}, {Hill}, {Peek}, {Putman}, \&
  {Babler}}]{Clark2015}
{Clark}, S.~E., {Hill}, J.~C., {Peek}, J.~E.~G., {Putman}, M.~E., \& {Babler},
  B.~L. 2015, \prl, 115, 241302

\bibitem[{{Clark} {et~al.}(2019){Clark}, {Peek}, \&
  {Miville-Desch{\^e}nes}}]{Clark2019a}
{Clark}, S.~E., {Peek}, J.~E.~G., \& {Miville-Desch{\^e}nes}, M.~A. 2019, \apj,
  874, 171

\bibitem[{{Clark} {et~al.}(2014){Clark}, {Peek}, \& {Putman}}]{Clark2014}
{Clark}, S.~E., {Peek}, J.~E.~G., \& {Putman}, M.~E. 2014, \apj, 789, 82

\bibitem[{{Comrie} {et~al.}(2019){Comrie}, {Wang}, {Ford}, {Moraghan}, {Hsu},
  {Pińska}, {Chiang}, {Jan}, \& {Simmonds}}]{CARTA2019}
{Comrie}, A., {Wang}, K.-S., {Ford}, P., {Moraghan}, A., {Hsu}, S.-C.,
  {Pińska}, A., {Chiang}, C.-C., {Jan}, H., \& {Simmonds}, R. 2019, {CARTA:
  The Cube Analysis and Rendering Tool for Astronomy, v1.2.0 Zenodo,
  doi:10.5281/zenodo.3403491}

\bibitem[{{Crutcher} {et~al.}(2010){Crutcher}, {Wandelt}, {Heiles},
  {Falgarone}, \& {Troland}}]{Crutcher2010}
{Crutcher}, R.~M., {Wandelt}, B., {Heiles}, C., {Falgarone}, E., \& {Troland},
  T.~H. 2010, \apj, 725, 466

\bibitem[{{Davis}(1951)}]{Davis1951}
{Davis}, L. 1951, Physical Review, 890

\bibitem[{{Dawson} {et~al.}(2011){Dawson}, {McClure-Griffiths}, {Kawamura},
  {Mizuno}, {Onishi}, {Mizuno}, \& {Fukui}}]{Dawson2011}
{Dawson}, J.~R., {McClure-Griffiths}, N.~M., {Kawamura}, A., {Mizuno}, N.,
  {Onishi}, T., {Mizuno}, A., \& {Fukui}, Y. 2011, \apj, 728, 127

\bibitem[{{Dennison} {et~al.}(1998){Dennison}, {Simonetti}, \&
  {Topasna}}]{Dennison1998}
{Dennison}, B., {Simonetti}, J.~H., \& {Topasna}, G.~A. 1998, \pasa, 15, 147

\bibitem[{{Ferri{\`e}re}(2016)}]{Ferriere2016}
{Ferri{\`e}re}, K. 2016, J. Phys. Conf. Ser., 767, 012006

\bibitem[{{Ferri{\`e}re}(2020)}]{Ferriere2020}
{Ferri{\`e}re}, K. 2020, PPCF, 62, 014014

\bibitem[{{Ferri{\`e}re} {et~al.}(2021){Ferri{\`e}re}, {West}, \&
  {Jaffe}}]{Ferriere2021}
{Ferri{\`e}re}, K., {West}, J.~L., \& {Jaffe}, T.~R. 2021, \mnras, 507, 4968

\bibitem[{{Ferri{\`e}re}(2001)}]{Ferriere2001}
{Ferri{\`e}re}, K.~M. 2001, Reviews of Modern Physics, 73, 1031

\bibitem[{{Field} {et~al.}(1969){Field}, {Goldsmith}, \& {Habing}}]{Field1969}
{Field}, G.~B., {Goldsmith}, D.~W., \& {Habing}, H.~J. 1969, \apjl, 155, L149

\bibitem[{{Finkbeiner}(2003)}]{Finkbeiner2003}
{Finkbeiner}, D.~P. 2003, \apjs, 146, 407

\bibitem[{{Gaensler} {et~al.}(2001){Gaensler}, {Dickey}, {McClure-Griffiths},
  {Green}, {Wieringa}, \& {Haynes}}]{Gaensler2001}
{Gaensler}, B.~M., {Dickey}, J.~M., {McClure-Griffiths}, N.~M., {Green}, A.~J.,
  {Wieringa}, M.~H., \& {Haynes}, R.~F. 2001, \apj, 549, 959

\bibitem[{{Gaensler} {et~al.}(2011){Gaensler}, {Haverkorn}, {Burkhart},
  {Newton-McGee}, {Ekers}, {Lazarian}, {McClure-Griffiths}, {Robishaw},
  {Dickey}, \& {Green}}]{Gaensler2011}
{Gaensler}, B.~M. \etal . 2011, \nat, 478, 214

\bibitem[{{Gaensler} {et~al.}(2008){Gaensler}, {Madsen}, {Chatterjee}, \&
  {Mao}}]{Gaensler2008}
{Gaensler}, B.~M., {Madsen}, G.~J., {Chatterjee}, S., \& {Mao}, S.~A. 2008,
  \pasa, 25, 184

\bibitem[{{Gaustad} {et~al.}(2001){Gaustad}, {McCullough}, {Rosing}, \& {Van
  Buren}}]{Gaustad2001}
{Gaustad}, J.~E., {McCullough}, P.~R., {Rosing}, W., \& {Van Buren}, D. 2001,
  \pasp, 113, 1326

\bibitem[{{George} {et~al.}(2012){George}, {Stil}, \& {Keller}}]{George2012}
{George}, S.~J., {Stil}, J.~M., \& {Keller}, B.~W. 2012, \pasa, 29, 214

\bibitem[{{Goldsmith} {et~al.}(1969){Goldsmith}, {Habing}, \&
  {Field}}]{Goldsmith1969}
{Goldsmith}, D.~W., {Habing}, H.~J., \& {Field}, G.~B. 1969, \apj, 158, 173

\bibitem[{{Gray} {et~al.}(1998){Gray}, {Landecker}, {Dewdney}, \&
  {Taylor}}]{Gray1998}
{Gray}, A.~D., {Landecker}, T.~L., {Dewdney}, P.~E., \& {Taylor}, A.~R. 1998,
  \nat, 393, 660

\bibitem[{{Green}(2018)}]{Green2018}
{Green}, G. 2018, JOSS, 3, 695

\bibitem[{{Green} {et~al.}(2019){Green}, {Schlafly}, {Zucker}, {Speagle}, \&
  {Finkbeiner}}]{Green2019}
{Green}, G.~M., {Schlafly}, E., {Zucker}, C., {Speagle}, J.~S., \&
  {Finkbeiner}, D. 2019, \apj, 887, 93

\bibitem[{{Green} {et~al.}(2012){Green}, {McClure-Griffiths}, {Caswell},
  {Robishaw}, \& {Harvey-Smith}}]{Green2012}
{Green}, J.~A., {McClure-Griffiths}, N.~M., {Caswell}, J.~L., {Robishaw}, T.,
  \& {Harvey-Smith}, L. 2012, \mnras, 425, 2530

\bibitem[{{Guram} {et~al.}(2011){Guram}, {Andrecut}, {George}, \&
  {Taylor}}]{Guram2011}
{Guram}, S.~S., {Andrecut}, M., {George}, S.~J., \& {Taylor}, A.~R. 2011, in
  Astronomical Society of the Pacific Conference Series, Vol. 442, Astronomical
  Data Analysis Software and Systems XX, ed. I.~N. {Evans}, A.~{Accomazzi},
  D.~J. {Mink}, \& A.~H. {Rots}, 317

\bibitem[{{Guram} \& {Taylor}(2009)}]{Guram2009}
{Guram}, S.~S. \& {Taylor}, A.~R. 2009, Astronomical Society of the Pacific
  Conference Series, Vol. 407, {The Galactic ALFA Continuum Transit Survey:
  GALFACTS}, ed. D.~J. {Saikia}, D.~A. {Green}, Y.~{Gupta}, \& T.~{Venturi},
  282

\bibitem[{{Haffner} {et~al.}(2009){Haffner}, {Dettmar}, {Beckman}, {Wood},
  {Slavin}, {Giammanco}, {Madsen}, {Zurita}, \& {Reynolds}}]{Haffner2009}
{Haffner}, L.~M., {Dettmar}, R.~J., {Beckman}, J.~E., {Wood}, K., {Slavin},
  J.~D., {Giammanco}, C., {Madsen}, G.~J., {Zurita}, A., \& {Reynolds}, R.~J.
  2009, RvMP, 81, 969

\bibitem[{{Haffner} {et~al.}(1998){Haffner}, {Reynolds}, \&
  {Tufte}}]{Haffner1998}
{Haffner}, L.~M., {Reynolds}, R.~J., \& {Tufte}, S.~L. 1998, \apjl, 501, L83

\bibitem[{{Haffner} {et~al.}(2003){Haffner}, {Reynolds}, {Tufte}, {Madsen},
  {Jaehnig}, \& {Percival}}]{Haffner2003}
{Haffner}, L.~M., {Reynolds}, R.~J., {Tufte}, S.~L., {Madsen}, G.~J.,
  {Jaehnig}, K.~P., \& {Percival}, J.~W. 2003, \apjs, 149, 405

\bibitem[{{Harris} {et~al.}(2020){Harris}, {Millman}, {van der Walt},
  {Gommers}, {Virtanen}, {Cournapeau}, {Wieser}, {Taylor}, {Berg}, {Smith},
  {Kern}, {Picus}, {Hoyer}, {van Kerkwijk}, {Brett}, {Haldane}, {del R{\'\i}o},
  {Wiebe}, {Peterson}, {G{\'e}rard-Marchant}, {Sheppard}, {Reddy}, {Weckesser},
  {Abbasi}, {Gohlke}, \& {Oliphant}}]{Harris2020}
{Harris}, C.~R. \etal . 2020, \nat, 585, 357

\bibitem[{{Haverkorn} {et~al.}(2019){Haverkorn}, {Boulanger}, {En{\ss}lin},
  {H{\"o}randel}, {Jaffe}, {Jasche}, {Rachen}, \& {Shukurov}}]{Haverkorn2019}
{Haverkorn}, M., {Boulanger}, F., {En{\ss}lin}, T., {H{\"o}randel}, J.,
  {Jaffe}, T., {Jasche}, J., {Rachen}, J., \& {Shukurov}, A. 2019, Galaxies, 7,
  17

\bibitem[{{Haverkorn} \& {Heitsch}(2004)}]{Haverkorn2004a}
{Haverkorn}, M. \& {Heitsch}, F. 2004, \aap, 421, 1011

\bibitem[{{Haverkorn} {et~al.}(2003{\natexlab{a}}){Haverkorn}, {Katgert}, \&
  {de Bruyn}}]{Haverkorn2003a}
{Haverkorn}, M., {Katgert}, P., \& {de Bruyn}, A.~G. 2003{\natexlab{a}}, \aap,
  403, 1031

\bibitem[{{Haverkorn} {et~al.}(2003{\natexlab{b}}){Haverkorn}, {Katgert}, \&
  {de Bruyn}}]{Haverkorn2003b}
{Haverkorn}, M., {Katgert}, P., \& {de Bruyn}, A.~G. 2003{\natexlab{b}}, \aap,
  404, 233

\bibitem[{{Haverkorn} {et~al.}(2004{\natexlab{a}}){Haverkorn}, {Katgert}, \&
  {de Bruyn}}]{Haverkorn2004c}
{Haverkorn}, M., {Katgert}, P., \& {de Bruyn}, A.~G. 2004{\natexlab{a}}, \aap,
  427, 169

\bibitem[{{Haverkorn} {et~al.}(2004{\natexlab{b}}){Haverkorn}, {Katgert}, \&
  {de Bruyn}}]{Haverkorn2004b}
{Haverkorn}, M., {Katgert}, P., \& {de Bruyn}, A.~G. 2004{\natexlab{b}}, \aap,
  427, 549

\bibitem[{{Heald} {et~al.}(2009){Heald}, {Braun}, \& {Edmonds}}]{Heald2009}
{Heald}, G., {Braun}, R., \& {Edmonds}, R. 2009, \aap, 503, 409

\bibitem[{{Heiles}(1967)}]{Heiles1967}
{Heiles}, C. 1967, \apjs, 15, 97

\bibitem[{{Heiles}(1979)}]{Heiles1979}
{Heiles}, C. 1979, \apj, 229, 533

\bibitem[{{Heiles}(1984)}]{Heiles1984}
{Heiles}, C. 1984, \apjs, 55, 585

\bibitem[{{Heiles} \& {Haverkorn}(2012)}]{Heiles2012}
{Heiles}, C. \& {Haverkorn}, M. 2012, \ssr, 166, 293

\bibitem[{{Heiles} \& {Troland}(2003)}]{Heiles2003}
{Heiles}, C. \& {Troland}, T.~H. 2003, \apj, 586, 1067

\bibitem[{{Heiles} \& {Troland}(2004)}]{Heiles2004}
{Heiles}, C. \& {Troland}, T.~H. 2004, \apjs, 151, 271

\bibitem[{{Heitsch} {et~al.}(2001){Heitsch}, {Zweibel}, {Mac Low}, {Li}, \&
  {Norman}}]{Heitsch2001}
{Heitsch}, F., {Zweibel}, E.~G., {Mac Low}, M.-M., {Li}, P., \& {Norman}, M.~L.
  2001, \apj, 561, 800

\bibitem[{{Herron} {et~al.}(2018){Herron}, {Gaensler}, {Lewis}, \&
  {McClure-Griffiths}}]{Herron2018I}
{Herron}, C.~A., {Gaensler}, B.~M., {Lewis}, G.~F., \& {McClure-Griffiths},
  N.~M. 2018, \apj, 853, 9

\bibitem[{{HI4PI Collaboration} {et~al.}(2016){HI4PI Collaboration}, {Ben
  Bekhti}, {Fl{\"o}er}, {Keller}, {Kerp}, {Lenz}, {Winkel}, {Bailin},
  {Calabretta}, {Dedes}, {Ford}, {Gibson}, {Haud}, {Janowiecki}, {Kalberla},
  {Lockman}, {McClure-Griffiths}, {Murphy}, {Nakanishi}, {Pisano}, \&
  {Staveley-Smith}}]{HI4PI2016}
{HI4PI Collaboration} \etal . 2016, \aap, 594, A116

\bibitem[{{Hill}(2018)}]{Hill2018}
{Hill}, A. 2018, Galaxies, 6, 129

\bibitem[{{Hill} {et~al.}(2008){Hill}, {Benjamin}, {Kowal}, {Reynolds},
  {Haffner}, \& {Lazarian}}]{Hill2008}
{Hill}, A.~S., {Benjamin}, R.~A., {Kowal}, G., {Reynolds}, R.~J., {Haffner},
  L.~M., \& {Lazarian}, A. 2008, \apj, 686, 363

\bibitem[{{Hill} {et~al.}(2017){Hill}, {Landecker}, {Carretti}, {Douglas},
  {Sun}, {Gaensler}, {Mao}, {McClure-Griffiths}, {Reich}, {Wolleben}, {Dickey},
  {Gray}, {Haverkorn}, {Leahy}, \& {Schnitzeler}}]{Hill2017}
{Hill}, A.~S. \etal . 2017, \mnras, 467, 4631

\bibitem[{{Hu}(1981)}]{Hu1981}
{Hu}, E.~M. 1981, \apj, 248, 119

\bibitem[{{Hunter}(2007)}]{Hunter2007}
{Hunter}, J.~D. 2007, CSE, 9, 90

\bibitem[{{Iacobelli} {et~al.}(2014){Iacobelli}, {Burkhart}, {Haverkorn},
  {Lazarian}, {Carretti}, {Staveley-Smith}, {Gaensler}, {Bernardi}, {Kesteven},
  \& {Poppi}}]{Iacobelli2014}
{Iacobelli}, M. \etal . 2014, \aap, 566, A5

\bibitem[{{Jaffe}(2019)}]{Jaffe2019}
{Jaffe}, T.~R. 2019, Galaxies, 7, 52

\bibitem[{{Jaffe} {et~al.}(2010){Jaffe}, {Leahy}, {Banday}, {Leach}, {Lowe}, \&
  {Wilkinson}}]{Jaffe2010}
{Jaffe}, T.~R., {Leahy}, J.~P., {Banday}, A.~J., {Leach}, S.~M., {Lowe}, S.~R.,
  \& {Wilkinson}, A. 2010, \mnras, 401, 1013

\bibitem[{{Jeli{\'c}} {et~al.}(2015){Jeli{\'c}}, {de Bruyn}, {Pandey},
  {Mevius}, {Haverkorn}, {Brentjens}, {Koopmans}, {Zaroubi}, {Abdalla}, {Asad},
  {Bus}, {Chapman}, {Ciardi}, {Fernandez}, {Ghosh}, {Harker}, {Iliev},
  {Jensen}, {Kazemi}, {Mellema}, {Offringa}, {Patil}, {Vedantham}, \&
  {Yatawatta}}]{Jelic2015}
{Jeli{\'c}}, V. \etal . 2015, \aap, 583, A137

\bibitem[{{Jeli{\'c}} {et~al.}(2018){Jeli{\'c}}, {Prelogovi{\'c}}, {Haverkorn},
  {Remeijn}, \& {Klind{\v{z}}i{\'c}}}]{Jelic2018}
{Jeli{\'c}}, V., {Prelogovi{\'c}}, D., {Haverkorn}, M., {Remeijn}, J., \&
  {Klind{\v{z}}i{\'c}}, D. 2018, \aap, 615, L3

\bibitem[{{Kado-Fong} {et~al.}(2020){Kado-Fong}, {Kim}, {Ostriker}, \&
  {Kim}}]{Kado-Fong2020}
{Kado-Fong}, E., {Kim}, J.-G., {Ostriker}, E.~C., \& {Kim}, C.-G. 2020, \apj,
  897, 143

\bibitem[{{Kalberla} \& {Haud}(2015)}]{Kalberla2015}
{Kalberla}, P.~M.~W. \& {Haud}, U. 2015, \aap, 578, A78

\bibitem[{{Kalberla} \& {Haud}(2020)}]{Kalberla2020}
{Kalberla}, P.~M.~W. \& {Haud}, U. 2020, \aap, in press (arXiv:2003.01454)

\bibitem[{{Kalberla} \& {Kerp}(2016)}]{Kalberla2016a}
{Kalberla}, P.~M.~W. \& {Kerp}, J. 2016, \aap, 595, A37

\bibitem[{{Kalberla} {et~al.}(2017){Kalberla}, {Kerp}, {Haud}, \&
  {Haverkorn}}]{Kalberla2017}
{Kalberla}, P.~M.~W., {Kerp}, J., {Haud}, U., \& {Haverkorn}, M. 2017, \aap,
  607, A15

\bibitem[{{Kalberla} {et~al.}(2016){Kalberla}, {Kerp}, {Haud}, {Winkel}, {Ben
  Bekhti}, {Fl{\"o}er}, \& {Lenz}}]{Kalberla2016}
{Kalberla}, P.~M.~W., {Kerp}, J., {Haud}, U., {Winkel}, B., {Ben Bekhti}, N.,
  {Fl{\"o}er}, L., \& {Lenz}, D. 2016, \apj, 821, 117

\bibitem[{{Kalberla} {et~al.}(2010){Kalberla}, {McClure-Griffiths}, {Pisano},
  {Calabretta}, {Ford}, {Lockman}, {Staveley-Smith}, {Kerp}, {Winkel},
  {Murphy}, \& {Newton-McGee}}]{Kalberla2010}
{Kalberla}, P.~M.~W. \etal . 2010, \aap, 521, A17

\bibitem[{{Kanekar} {et~al.}(2003){Kanekar}, {Subrahmanyan}, {Chengalur}, \&
  {Safouris}}]{Kanekar2003}
{Kanekar}, N., {Subrahmanyan}, R., {Chengalur}, J.~N., \& {Safouris}, V. 2003,
  \mnras, 346, L57

\bibitem[{{Kim} \& {Ostriker}(2018)}]{Kim2018}
{Kim}, C.-G. \& {Ostriker}, E.~C. 2018, \apj, 853, 173

\bibitem[{{Lallement} {et~al.}(2019){Lallement}, {Babusiaux}, {Vergely},
  {Katz}, {Arenou}, {Valette}, {Hottier}, \& {Capitanio}}]{Lallement2019}
{Lallement}, R., {Babusiaux}, C., {Vergely}, J.~L., {Katz}, D., {Arenou}, F.,
  {Valette}, B., {Hottier}, C., \& {Capitanio}, L. 2019, \aap, 625, A135

\bibitem[{{Lazarian} \& {Yuen}(2018)}]{Lazarian2018}
{Lazarian}, A. \& {Yuen}, K.~H. 2018, \apj, 853, 96

\bibitem[{{Leahy}(2018)}]{Leahy2018}
{Leahy}, P. 2018, in GALFACTS Internal Memo, 1--30

\bibitem[{{Marchal} \& {Miville-Desch{\^e}nes}(2021)}]{Marchal2021}
{Marchal}, A. \& {Miville-Desch{\^e}nes}, M.-A. 2021, \apj, 908, 186

\bibitem[{{Marchal} {et~al.}(2019){Marchal}, {Miville-Desch{\^e}nes}, {Orieux},
  {Gac}, {Soussen}, {Lesot}, {d'Allonnes}, \& {Salom{\'e}}}]{Marchal2019}
{Marchal}, A., {Miville-Desch{\^e}nes}, M.-A., {Orieux}, F., {Gac}, N.,
  {Soussen}, C., {Lesot}, M.-J., {d'Allonnes}, A.~R., \& {Salom{\'e}}, Q. 2019,
  \aap, 626, A101

\bibitem[{{Martin} {et~al.}(2015){Martin}, {Blagrave}, {Lockman}, {Pinheiro
  Gon{\c{c}}alves}, {Boothroyd}, {Joncas}, {Miville-Desch{\^e}nes}, \&
  {Stephan}}]{Martin2015}
{Martin}, P.~G., {Blagrave}, K.~P.~M., {Lockman}, F.~J., {Pinheiro
  Gon{\c{c}}alves}, D., {Boothroyd}, A.~I., {Joncas}, G.,
  {Miville-Desch{\^e}nes}, M.~A., \& {Stephan}, G. 2015, \apj, 809, 153

\bibitem[{{McClure-Griffiths} {et~al.}(2002){McClure-Griffiths}, {Dickey},
  {Gaensler}, \& {Green}}]{McClure-Griffiths2002}
{McClure-Griffiths}, N.~M., {Dickey}, J.~M., {Gaensler}, B.~M., \& {Green},
  A.~J. 2002, \apj, 578, 176

\bibitem[{{McClure-Griffiths} {et~al.}(2003){McClure-Griffiths}, {Dickey},
  {Gaensler}, \& {Green}}]{McClure-Griffiths2003}
{McClure-Griffiths}, N.~M., {Dickey}, J.~M., {Gaensler}, B.~M., \& {Green},
  A.~J. 2003, \apj, 594, 833

\bibitem[{{McClure-Griffiths} {et~al.}(2006){McClure-Griffiths}, {Dickey},
  {Gaensler}, {Green}, \& {Haverkorn}}]{McClure-Griffiths2006}
{McClure-Griffiths}, N.~M., {Dickey}, J.~M., {Gaensler}, B.~M., {Green}, A.~J.,
  \& {Haverkorn}, M. 2006, \apj, 652, 1339

\bibitem[{{McClure-Griffiths} {et~al.}(2009){McClure-Griffiths}, {Pisano},
  {Calabretta}, {Ford}, {Lockman}, {Staveley-Smith}, {Kalberla}, {Bailin},
  {Dedes}, {Janowiecki}, {Gibson}, {Murphy}, {Nakanishi}, \&
  {Newton-McGee}}]{McClure-Griffiths2009}
{McClure-Griffiths}, N.~M. \etal . 2009, \apjs, 181, 398

\bibitem[{{McKee} \& {Ostriker}(1977)}]{Mckee1977}
{McKee}, C.~F. \& {Ostriker}, J.~P. 1977, \apj, 218, 148

\bibitem[{{Murray} {et~al.}(2015){Murray}, {Stanimirovi{\'c}}, {Goss},
  {Dickey}, {Heiles}, {Lindner}, {Babler}, {Pingel}, {Lawrence}, {Jencson}, \&
  {Hennebelle}}]{Murray2015}
{Murray}, C.~E. \etal . 2015, \apj, 804, 89

\bibitem[{{Murray} {et~al.}(2018){Murray}, {Stanimirovi{\'c}}, {Goss},
  {Heiles}, {Dickey}, {Babler}, \& {Kim}}]{Murray2018}
{Murray}, C.~E., {Stanimirovi{\'c}}, S., {Goss}, W.~M., {Heiles}, C., {Dickey},
  J.~M., {Babler}, B., \& {Kim}, C.-G. 2018, \apjs, 238, 14

\bibitem[{{Nota} \& {Katgert}(2010)}]{Nota2010}
{Nota}, T. \& {Katgert}, P. 2010, \aap, 513, A65

\bibitem[{{Ogbodo} {et~al.}(2020){Ogbodo}, {Green}, {Dawson}, {Breen}, {Mao},
  {McClure-Griffiths}, {Robishaw}, \& {Harvey-Smith}}]{Ogbodo2020}
{Ogbodo}, C.~S., {Green}, J.~A., {Dawson}, J.~R., {Breen}, S.~L., {Mao}, S.~A.,
  {McClure-Griffiths}, N.~M., {Robishaw}, T., \& {Harvey-Smith}, L. 2020,
  \mnras, 493, 199

\bibitem[{{Peek} {et~al.}(2018){Peek}, {Babler}, {Zheng}, {Clark}, {Douglas},
  {Korpela}, {Putman}, {Stanimirovi{\'c}}, {Gibson}, \& {Heiles}}]{Peek2018}
{Peek}, J.~E.~G. \etal . 2018, \apjs, 234, 2

\bibitem[{{Peek} \& {Clark}(2019)}]{Peek2019}
{Peek}, J.~E.~G. \& {Clark}, S.~E. 2019, \apjl, 886, L13

\bibitem[{{Planck Collaboration} {et~al.}(2014){Planck Collaboration},
  {Abergel}, {Ade}, {Aghanim}, {Alves}, {Aniano}, {Armitage-Caplan}, {Arnaud},
  {Ashdown}, {Atrio-Barand ela}, {Aumont}, {Baccigalupi}, {Banday}, {Barreiro},
  {Bartlett}, {Battaner}, {Benabed}, {Beno{\^\i}t}, {Benoit-L{\'e}vy},
  {Bernard}, {Bersanelli}, {Bielewicz}, {Bobin}, {Bock}, {Bonaldi}, {Bond},
  {Borrill}, {Bouchet}, {Boulanger}, {Bridges}, {Bucher}, {Burigana}, {Butler},
  {Cardoso}, {Catalano}, {Chamballu}, {Chary}, {Chiang}, {Chiang},
  {Christensen}, {Church}, {Clemens}, {Clements}, {Colombi}, {Colombo},
  {Combet}, {Couchot}, {Coulais}, {Crill}, {Curto}, {Cuttaia}, {Danese},
  {Davies}, {Davis}, {de Bernardis}, {de Rosa}, {de Zotti}, {Delabrouille},
  {Delouis}, {D{\'e}sert}, {Dickinson}, {Diego}, {Dole}, {Donzelli},
  {Dor{\'e}}, {Douspis}, {Draine}, {Dupac}, {Efstathiou}, {En{\ss}lin},
  {Eriksen}, {Falgarone}, {Finelli}, {Forni}, {Frailis}, {Fraisse},
  {Franceschi}, {Galeotta}, {Ganga}, {Ghosh}, {Giard}, {Giardino},
  {Giraud-H{\'e}raud}, {Gonz{\'a}lez-Nuevo}, {G{\'o}rski}, {Gratton},
  {Gregorio}, {Grenier}, {Gruppuso}, {Guillet}, {Hansen}, {Hanson}, {Harrison},
  {Helou}, {Henrot-Versill{\'e}}, {Hern{\'a}ndez-Monteagudo}, {Herranz},
  {Hildebrand t}, {Hivon}, {Hobson}, {Holmes}, {Hornstrup}, {Hovest},
  {Huffenberger}, {Jaffe}, {Jaffe}, {Jewell}, {Joncas}, {Jones}, {Juvela},
  {Keih{\"a}nen}, {Keskitalo}, {Kisner}, {Knoche}, {Knox}, {Kunz},
  {Kurki-Suonio}, {Lagache}, {L{\"a}hteenm{\"a}ki}, {Lamarre}, {Lasenby},
  {Laureijs}, {Lawrence}, {Leonardi}, {Le{\'o}n-Tavares}, {Lesgourgues},
  {Levrier}, {Liguori}, {Lilje}, {Linden-V{\o}rnle}, {L{\'o}pez-Caniego},
  {Lubin}, {Mac{\'\i}as-P{\'e}rez}, {Maffei}, {Maino}, {Mand olesi}, {Maris},
  {Marshall}, {Martin}, {Mart{\'\i}nez-Gonz{\'a}lez}, {Masi}, {Massardi},
  {Matarrese}, {Matthai}, {Mazzotta}, {McGehee}, {Melchiorri}, {Mendes},
  {Mennella}, {Migliaccio}, {Mitra}, {Miville-Desch{\^e}nes}, {Moneti},
  {Montier}, {Morgante}, {Mortlock}, {Munshi}, {Murphy}, {Naselsky}, {Nati},
  {Natoli}, {Netterfield}, {N{\o}rgaard-Nielsen}, {Noviello}, {Novikov},
  {Novikov}, {Osborne}, {Oxborrow}, {Paci}, {Pagano}, {Pajot}, {Paladini},
  {Paoletti}, {Pasian}, {Patanchon}, {Perdereau}, {Perotto}, {Perrotta},
  {Piacentini}, {Piat}, {Pierpaoli}, {Pietrobon}, {Plaszczynski},
  {Pointecouteau}, {Polenta}, {Ponthieu}, {Popa}, {Poutanen}, {Pratt},
  {Pr{\'e}zeau}, {Prunet}, {Puget}, {Rachen}, {Reach}, {Rebolo}, {Reinecke},
  {Remazeilles}, {Renault}, {Ricciardi}, {Riller}, {Ristorcelli}, {Rocha},
  {Rosset}, {Roudier}, {Rowan-Robinson}, {Rubi{\~n}o-Mart{\'\i}n}, {Rusholme},
  {Sandri}, {Santos}, {Savini}, {Scott}, {Seiffert}, {Shellard}, {Spencer},
  {Starck}, {Stolyarov}, {Stompor}, {Sudiwala}, {Sunyaev}, {Sureau}, {Sutton},
  {Suur-Uski}, {Sygnet}, {Tauber}, {Tavagnacco}, {Terenzi}, {Toffolatti},
  {Tomasi}, {Tristram}, {Tucci}, {Tuovinen}, {T{\"u}rler}, {Umana},
  {Valenziano}, {Valiviita}, {Van Tent}, {Verstraete}, {Vielva}, {Villa},
  {Vittorio}, {Wade}, {Wandelt}, {Welikala}, {Ysard}, {Yvon}, {Zacchei}, \&
  {Zonca}}]{Planck2013XI}
{Planck Collaboration} \etal . 2014, \aap, 571, A11

\bibitem[{{Planck Collaboration} {et~al.}(2016){Planck Collaboration}, {Adam},
  {Ade}, {Aghanim}, {Arnaud}, {Ashdown}, {Aumont}, {Baccigalupi}, {Banday},
  {Barreiro}, {Bartolo}, {Battaner}, {Benabed}, {Beno{\^\i}t},
  {Benoit-L{\'e}vy}, {Bernard}, {Bersanelli}, {Bertincourt}, {Bielewicz},
  {Bock}, {Bonavera}, {Bond}, {Borrill}, {Bouchet}, {Boulanger}, {Bucher},
  {Burigana}, {Calabrese}, {Cardoso}, {Catalano}, {Challinor}, {Chamballu},
  {Chiang}, {Christensen}, {Clements}, {Colombi}, {Colombo}, {Combet},
  {Couchot}, {Coulais}, {Crill}, {Curto}, {Cuttaia}, {Danese}, {Davies},
  {Davis}, {de Bernardis}, {de Rosa}, {de Zotti}, {Delabrouille}, {Delouis},
  {D{\'e}sert}, {Diego}, {Dole}, {Donzelli}, {Dor{\'e}}, {Douspis}, {Ducout},
  {Dupac}, {Efstathiou}, {Elsner}, {En{\ss}lin}, {Eriksen}, {Falgarone},
  {Fergusson}, {Finelli}, {Forni}, {Frailis}, {Fraisse}, {Franceschi},
  {Frejsel}, {Galeotta}, {Galli}, {Ganga}, {Ghosh}, {Giard},
  {Giraud-H{\'e}raud}, {Gjerl{\o}w}, {Gonz{\'a}lez-Nuevo}, {G{\'o}rski},
  {Gratton}, {Gruppuso}, {Gudmundsson}, {Hansen}, {Hanson}, {Harrison},
  {Henrot-Versill{\'e}}, {Herranz}, {Hildebrandt}, {Hivon}, {Hobson}, {Holmes},
  {Hornstrup}, {Hovest}, {Huffenberger}, {Hurier}, {Jaffe}, {Jaffe}, {Jones},
  {Juvela}, {Keih{\"a}nen}, {Keskitalo}, {Kisner}, {Kneissl}, {Knoche}, {Kunz},
  {Kurki-Suonio}, {Lagache}, {Lamarre}, {Lasenby}, {Lattanzi}, {Lawrence}, {Le
  Jeune}, {Leahy}, {Lellouch}, {Leonardi}, {Lesgourgues}, {Levrier}, {Liguori},
  {Lilje}, {Linden-V{\o}rnle}, {L{\'o}pez-Caniego}, {Lubin},
  {Mac{\'\i}as-P{\'e}rez}, {Maggio}, {Maino}, {Mandolesi}, {Mangilli}, {Maris},
  {Martin}, {Mart{\'\i}nez-Gonz{\'a}lez}, {Masi}, {Matarrese}, {McGehee},
  {Melchiorri}, {Mendes}, {Mennella}, {Migliaccio}, {Mitra},
  {Miville-Desch{\^e}nes}, {Moneti}, {Montier}, {Moreno}, {Morgante},
  {Mortlock}, {Moss}, {Mottet}, {Munshi}, {Murphy}, {Naselsky}, {Nati},
  {Natoli}, {Netterfield}, {N{\o}rgaard-Nielsen}, {Noviello}, {Novikov},
  {Novikov}, {Oxborrow}, {Paci}, {Pagano}, {Pajot}, {Paoletti}, {Pasian},
  {Patanchon}, {Pearson}, {Perdereau}, {Perotto}, {Perrotta}, {Pettorino},
  {Piacentini}, {Piat}, {Pierpaoli}, {Pietrobon}, {Plaszczynski},
  {Pointecouteau}, {Polenta}, {Pratt}, {Pr{\'e}zeau}, {Prunet}, {Puget},
  {Rachen}, {Reinecke}, {Remazeilles}, {Renault}, {Renzi}, {Ristorcelli},
  {Rocha}, {Rosset}, {Rossetti}, {Roudier}, {Rusholme}, {Sandri}, {Santos},
  {Sauv{\'e}}, {Savelainen}, {Savini}, {Scott}, {Seiffert}, {Shellard},
  {Spencer}, {Stolyarov}, {Stompor}, {Sudiwala}, {Sutton}, {Suur-Uski},
  {Sygnet}, {Tauber}, {Terenzi}, {Toffolatti}, {Tomasi}, {Tristram}, {Tucci},
  {Tuovinen}, {Valenziano}, {Valiviita}, {Van Tent}, {Vibert}, {Vielva},
  {Villa}, {Wade}, {Wandelt}, {Watson}, {Wehus}, {Yvon}, {Zacchei}, \&
  {Zonca}}]{Planck2016VIII}
{Planck Collaboration} \etal . 2016, \aap, 594, A8

\bibitem[{{Planck Collaboration} {et~al.}(2020{\natexlab{a}}){Planck
  Collaboration}, {Aghanim}, {Akrami}, {Alves}, {Ashdown}, {Aumont},
  {Baccigalupi}, {Ballardini}, {Banday}, {Barreiro}, {Bartolo}, {Basak},
  {Benabed}, {Bernard}, {Bersanelli}, {Bielewicz}, {Bock}, {Bond}, {Borrill},
  {Bouchet}, {Boulanger}, {Bracco}, {Bucher}, {Burigana}, {Calabrese},
  {Cardoso}, {Carron}, {Chary}, {Chiang}, {Colombo}, {Combet}, {Crill},
  {Cuttaia}, {de Bernardis}, {de Zotti}, {Delabrouille}, {Delouis}, {Di
  Valentino}, {Dickinson}, {Diego}, {Dor{\'e}}, {Douspis}, {Ducout}, {Dupac},
  {Efstathiou}, {Elsner}, {En{\ss}lin}, {Eriksen}, {Falgarone}, {Fantaye},
  {Fernandez-Cobos}, {Ferri{\`e}re}, {Finelli}, {Forastieri}, {Frailis},
  {Fraisse}, {Franceschi}, {Frolov}, {Galeotta}, {Galli}, {Ganga},
  {G{\'e}nova-Santos}, {Gerbino}, {Ghosh}, {Gonz{\'a}lez-Nuevo}, {G{\'o}rski},
  {Gratton}, {Green}, {Gruppuso}, {Gudmundsson}, {Guillet}, {Handley},
  {Hansen}, {Helou}, {Herranz}, {Hivon}, {Huang}, {Jaffe}, {Jones},
  {Keih{\"a}nen}, {Keskitalo}, {Kiiveri}, {Kim}, {Krachmalnicoff}, {Kunz},
  {Kurki-Suonio}, {Lagache}, {Lamarre}, {Lasenby}, {Lattanzi}, {Lawrence}, {Le
  Jeune}, {Levrier}, {Liguori}, {Lilje}, {Lindholm}, {L{\'o}pez-Caniego},
  {Lubin}, {Ma}, {Mac{\'\i}as-P{\'e}rez}, {Maggio}, {Maino}, {Mandolesi},
  {Mangilli}, {Marcos-Caballero}, {Maris}, {Martin},
  {Mart{\'\i}nez-Gonz{\'a}lez}, {Matarrese}, {Mauri}, {McEwen}, {Melchiorri},
  {Mennella}, {Migliaccio}, {Miville-Desch{\^e}nes}, {Molinari}, {Moneti},
  {Montier}, {Morgante}, {Moss}, {Natoli}, {Pagano}, {Paoletti}, {Patanchon},
  {Perrotta}, {Pettorino}, {Piacentini}, {Polastri}, {Polenta}, {Puget},
  {Rachen}, {Reinecke}, {Remazeilles}, {Renzi}, {Ristorcelli}, {Rocha},
  {Rosset}, {Roudier}, {Rubi{\~n}o-Mart{\'\i}n}, {Ruiz-Granados}, {Salvati},
  {Sandri}, {Savelainen}, {Scott}, {Sirignano}, {Sunyaev}, {Suur-Uski},
  {Tauber}, {Tavagnacco}, {Tenti}, {Toffolatti}, {Tomasi}, {Trombetti},
  {Valiviita}, {Vansyngel}, {Van Tent}, {Vielva}, {Villa}, {Vittorio},
  {Wandelt}, {Wehus}, {Zacchei}, \& {Zonca}}]{Planck2018XII}
{Planck Collaboration} \etal . 2020{\natexlab{a}}, \aap, 641, A12

\bibitem[{{Planck Collaboration} {et~al.}(2020{\natexlab{b}}){Planck
  Collaboration}, {Aghanim}, {Akrami}, {Ashdown}, {Aumont}, {Baccigalupi},
  {Ballardini}, {Banday}, {Barreiro}, {Bartolo}, {Basak}, {Benabed}, {Bernard},
  {Bersanelli}, {Bielewicz}, {Bond}, {Borrill}, {Bouchet}, {Boulanger},
  {Bucher}, {Burigana}, {Calabrese}, {Cardoso}, {Carron}, {Challinor},
  {Chiang}, {Colombo}, {Combet}, {Couchot}, {Crill}, {Cuttaia}, {de Bernardis},
  {de Rosa}, {de Zotti}, {Delabrouille}, {Delouis}, {Di Valentino}, {Diego},
  {Dor{\'e}}, {Douspis}, {Ducout}, {Dupac}, {Efstathiou}, {Elsner},
  {En{\ss}lin}, {Eriksen}, {Falgarone}, {Fantaye}, {Finelli}, {Frailis},
  {Fraisse}, {Franceschi}, {Frolov}, {Galeotta}, {Galli}, {Ganga},
  {G{\'e}nova-Santos}, {Gerbino}, {Ghosh}, {Gonz{\'a}lez-Nuevo}, {G{\'o}rski},
  {Gratton}, {Gruppuso}, {Gudmundsson}, {Handley}, {Hansen},
  {Henrot-Versill{\'e}}, {Herranz}, {Hivon}, {Huang}, {Jaffe}, {Jones},
  {Karakci}, {Keih{\"a}nen}, {Keskitalo}, {Kiiveri}, {Kim}, {Kisner},
  {Krachmalnicoff}, {Kunz}, {Kurki-Suonio}, {Lagache}, {Lamarre}, {Lasenby},
  {Lattanzi}, {Lawrence}, {Levrier}, {Liguori}, {Lilje}, {Lindholm},
  {L{\'o}pez-Caniego}, {Ma}, {Mac{\'\i}as-P{\'e}rez}, {Maggio}, {Maino},
  {Mandolesi}, {Mangilli}, {Martin}, {Mart{\'\i}nez-Gonz{\'a}lez}, {Matarrese},
  {Mauri}, {McEwen}, {Melchiorri}, {Mennella}, {Migliaccio},
  {Miville-Desch{\^e}nes}, {Molinari}, {Moneti}, {Montier}, {Morgante}, {Moss},
  {Mottet}, {Natoli}, {Pagano}, {Paoletti}, {Partridge}, {Patanchon},
  {Patrizii}, {Perdereau}, {Perrotta}, {Pettorino}, {Piacentini}, {Puget},
  {Rachen}, {Reinecke}, {Remazeilles}, {Renzi}, {Rocha}, {Roudier}, {Salvati},
  {Sandri}, {Savelainen}, {Scott}, {Sirignano}, {Sirri}, {Spencer}, {Sunyaev},
  {Suur-Uski}, {Tauber}, {Tavagnacco}, {Tenti}, {Toffolatti}, {Tomasi},
  {Tristram}, {Trombetti}, {Valiviita}, {Vansyngel}, {Van Tent}, {Vibert},
  {Vielva}, {Villa}, {Vittorio}, {Wandelt}, {Wehus}, \&
  {Zonca}}]{Planck2018III}
{Planck Collaboration} \etal . 2020{\natexlab{b}}, \aap, 641, A3

\bibitem[{{Purcell} {et~al.}(2020){Purcell}, {Van Eck}, {West}, {Sun}, \&
  {Gaensler}}]{Purcell2020}
{Purcell}, C.~R., {Van Eck}, C.~L., {West}, J., {Sun}, X.~H., \& {Gaensler},
  B.~M. 2020, {RM-Tools: Rotation measure (RM) synthesis and Stokes QU-fitting,
  ASCL, ascl:2005.003}

\bibitem[{{Rand} \& {Kulkarni}(1989)}]{Rand1989}
{Rand}, R.~J. \& {Kulkarni}, S.~R. 1989, \apj, 343, 760

\bibitem[{{Reid} {et~al.}(2014){Reid}, {Menten}, {Brunthaler}, {Zheng}, {Dame},
  {Xu}, {Wu}, {Zhang}, {Sanna}, {Sato}, {Hachisuka}, {Choi}, {Immer},
  {Moscadelli}, {Rygl}, \& {Bartkiewicz}}]{Reid2014}
{Reid}, M.~J. \etal . 2014, \apj, 783, 130

\bibitem[{{Reynolds}(1990{\natexlab{a}})}]{Reynolds1990a}
{Reynolds}, R. 1990{\natexlab{a}}, in IAU Symposium, Vol. 139, The Galactic and
  Extragalactic Background Radiation, ed. L.~C. {Bowyer}, S.

\bibitem[{{Reynolds}(1984)}]{Reynolds1984}
{Reynolds}, R.~J. 1984, \apj, 282, 191

\bibitem[{{Reynolds}(1988)}]{Reynolds1988}
{Reynolds}, R.~J. 1988, \apj, 333, 341

\bibitem[{{Reynolds}(1990{\natexlab{b}})}]{Reynolds1990b}
{Reynolds}, R.~J. 1990{\natexlab{b}}, \apjl, 349, L17

\bibitem[{{Reynolds}(1990{\natexlab{c}})}]{Reynolds1990c}
{Reynolds}, R.~J. 1990{\natexlab{c}}, \apj, 348, 153

\bibitem[{{Robishaw} \& {Heiles}(2021)}]{Robishaw2018}
{Robishaw}, T. \& {Heiles}, C. 2021, {The Measurement of Polarization in Radio
  Astronomy}, ed. A.~{Wolszczan}, 127--158

\bibitem[{{Robitaille} \& {Scaife}(2015)}]{Robitaille2015}
{Robitaille}, J.~F. \& {Scaife}, A.~M.~M. 2015, \mnras, 451, 372

\bibitem[{{Roy} {et~al.}(2013){Roy}, {Kanekar}, \& {Chengalur}}]{Roy2013}
{Roy}, N., {Kanekar}, N., \& {Chengalur}, J.~N. 2013, \mnras, 436, 2366

\bibitem[{{Saury} {et~al.}(2014){Saury}, {Miville-Desch{\^e}nes}, {Hennebelle},
  {Audit}, \& {Schmidt}}]{Saury2014}
{Saury}, E., {Miville-Desch{\^e}nes}, M.~A., {Hennebelle}, P., {Audit}, E., \&
  {Schmidt}, W. 2014, \aap, 567, A16

\bibitem[{{Schlafly} {et~al.}(2016){Schlafly}, {Meisner}, {Stutz},
  {Kainulainen}, {Peek}, {Tchernyshyov}, {Rix}, {Finkbeiner}, {Covey}, {Green},
  {Bell}, {Burgett}, {Chambers}, {Draper}, {Flewelling}, {Hodapp}, {Kaiser},
  {Magnier}, {Martin}, {Metcalfe}, {Wainscoat}, \& {Waters}}]{Schlafly2016}
{Schlafly}, E.~F. \etal . 2016, \apj, 821, 78

\bibitem[{{Schlegel} {et~al.}(1998){Schlegel}, {Finkbeiner}, \&
  {Davis}}]{Schlegel1998}
{Schlegel}, D.~J., {Finkbeiner}, D.~P., \& {Davis}, M. 1998, \apj, 500, 525

\bibitem[{{Schnitzeler} {et~al.}(2009){Schnitzeler}, {Katgert}, \& {de
  Bruyn}}]{Schnitzeler2009}
{Schnitzeler}, D.~H.~F.~M., {Katgert}, P., \& {de Bruyn}, A.~G. 2009, \aap,
  494, 611

\bibitem[{{Shajn}(1958)}]{Shajn1958}
{Shajn}, G.~A. 1958, in Electromagnetic Phenomena in Cosmical Physics, ed.
  B.~{Lehnert}, Vol.~6, 182

\bibitem[{{Shimwell} {et~al.}(2017){Shimwell}, {R{\"o}ttgering}, {Best},
  {Williams}, {Dijkema}, {de Gasperin}, {Hardcastle}, {Heald}, {Hoang},
  {Horneffer}, {Intema}, {Mahony}, {Mandal}, {Mechev}, {Morabito}, {Oonk},
  {Rafferty}, {Retana-Montenegro}, {Sabater}, {Tasse}, {van Weeren},
  {Br{\"u}ggen}, {Brunetti}, {Chy{\.z}y}, {Conway}, {Haverkorn}, {Jackson},
  {Jarvis}, {McKean}, {Miley}, {Morganti}, {White}, {Wise}, {van Bemmel},
  {Beck}, {Brienza}, {Bonafede}, {Calistro Rivera}, {Cassano}, {Clarke},
  {Cseh}, {Deller}, {Drabent}, {van Driel}, {Engels}, {Falcke}, {Ferrari},
  {Fr{\"o}hlich}, {Garrett}, {Harwood}, {Heesen}, {Hoeft}, {Horellou},
  {Israel}, {Kapi{\'n}ska}, {Kunert-Bajraszewska}, {McKay}, {Mohan},
  {Orr{\'u}}, {Pizzo}, {Prandoni}, {Schwarz}, {Shulevski}, {Sipior}, {Smith},
  {Sridhar}, {Steinmetz}, {Stroe}, {Varenius}, {van der Werf}, {Zensus}, \&
  {Zwart}}]{Shimwell2017}
{Shimwell}, T.~W. \etal . 2017, \aap, 598, A104

\bibitem[{{Skalidis} {et~al.}(2021){Skalidis}, {Sternberg}, {Beattie},
  {Pavlidou}, \& {Tassis}}]{Skalidis2021}
{Skalidis}, R., {Sternberg}, J., {Beattie}, J.~R., {Pavlidou}, V., \& {Tassis},
  K. 2021, \aap, in press (arXiv:2109.10925)

\bibitem[{{Skrutskie} {et~al.}(2006){Skrutskie}, {Cutri}, {Stiening},
  {Weinberg}, {Schneider}, {Carpenter}, {Beichman}, {Capps}, {Chester},
  {Elias}, {Huchra}, {Liebert}, {Lonsdale}, {Monet}, {Price}, {Seitzer},
  {Jarrett}, {Kirkpatrick}, {Gizis}, {Howard}, {Evans}, {Fowler}, {Fullmer},
  {Hurt}, {Light}, {Kopan}, {Marsh}, {McCallon}, {Tam}, {Van Dyk}, \&
  {Wheelock}}]{Skrutskie2006}
{Skrutskie}, M.~F. \etal . 2006, \aj, 131, 1163

\bibitem[{{Snowden} {et~al.}(1997){Snowden}, {Egger}, {Freyberg}, {McCammon},
  {Plucinsky}, {Sanders}, {Schmitt}, {Tr{\"u}mper}, \& {Voges}}]{Snowden1997}
{Snowden}, S.~L., {Egger}, R., {Freyberg}, M.~J., {McCammon}, D., {Plucinsky},
  P.~P., {Sanders}, W.~T., {Schmitt}, J.~H.~M.~M., {Tr{\"u}mper}, J., \&
  {Voges}, W. 1997, \apj, 485, 125

\bibitem[{{Soler} {et~al.}(2020){Soler}, {Beuther}, {Syed}, {Wang}, {Anderson},
  {Glover}, {Hennebelle}, {Heyer}, {Henning}, {Izquierdo}, {Klessen}, {Linz},
  {McClure-Griffiths}, {Ott}, {Ragan}, {Rugel}, {Schneider}, {Smith},
  {Sormani}, {Stil}, {Tre{\ss}}, \& {Urquhart}}]{Soler2020}
{Soler}, J.~D. \etal . 2020, \aap, 642, A163

\bibitem[{{Soler} {et~al.}(2018){Soler}, {Bracco}, \& {Pon}}]{Soler2018}
{Soler}, J.~D., {Bracco}, A., \& {Pon}, A. 2018, \aap, 609, L3

\bibitem[{{Stil} \& {Hryhoriw}(2016)}]{Stil2016}
{Stil}, J.~M. \& {Hryhoriw}, A. 2016, \apj, 826, 202

\bibitem[{{Sun} {et~al.}(2008){Sun}, {Reich}, {Waelkens}, \&
  {En{\ss}lin}}]{Sun2008}
{Sun}, X.~H., {Reich}, W., {Waelkens}, A., \& {En{\ss}lin}, T.~A. 2008, \aap,
  477, 573

\bibitem[{{Tahani} {et~al.}(2018){Tahani}, {Plume}, {Brown}, \&
  {Kainulainen}}]{Tahani2018}
{Tahani}, M., {Plume}, R., {Brown}, J.~C., \& {Kainulainen}, J. 2018, \aap,
  614, A100

\bibitem[{{Taylor}(2012)}]{Taylor2012}
{Taylor}, A.~R. 2012, in {GALFACTS Internal Note}, 1--9

\bibitem[{Taylor(2013)}]{Taylor2013}
Taylor, A.~R. 2013, {IOP} Conference Series: Materials Science and Engineering,
  44, 012019

\bibitem[{{Taylor} \& {Salter}(2010)}]{Taylor2010}
{Taylor}, A.~R. \& {Salter}, C.~J. 2010, Astronomical Society of the Pacific
  Conference Series, Vol. 438, {GALFACTS: The G-ALFA Continuum Transit Survey},
  ed. R.~{Kothes}, T.~L. {Landecker}, \& A.~G. {Willis}, 402

\bibitem[{{Taylor} {et~al.}(2009){Taylor}, {Stil}, \& {Sunstrum}}]{Taylor2009}
{Taylor}, A.~R., {Stil}, J.~M., \& {Sunstrum}, C. 2009, \apj, 702, 1230

\bibitem[{{Thomson} {et~al.}(2019){Thomson}, {Landecker}, {Dickey},
  {McClure-Griffiths}, {Wolleben}, {Carretti}, {Fletcher}, {Federrath}, {Hill},
  {Mao}, {Gaensler}, {Haverkorn}, {Clark}, {Van Eck}, \& {West}}]{Thomson2019}
{Thomson}, A. J.~M. \etal . 2019, \mnras, 487, 4751

\bibitem[{{Tritsis} {et~al.}(2019){Tritsis}, {Federrath}, \&
  {Pavlidou}}]{Tritsis2019}
{Tritsis}, A., {Federrath}, C., \& {Pavlidou}, V. 2019, \apj, 873, 38

\bibitem[{{Turi{\'c}} {et~al.}(2021){Turi{\'c}}, {Jeli{\'c}}, {Jaspers},
  {Haverkorn}, {Bracco}, {Erceg}, {Ceraj}, {van Eck}, \& {Zaroubi}}]{Turic2021}
{Turi{\'c}}, L., {Jeli{\'c}}, V., {Jaspers}, R., {Haverkorn}, M., {Bracco}, A.,
  {Erceg}, A., {Ceraj}, L., {van Eck}, C., \& {Zaroubi}, S. 2021, \aap, 654, A5

\bibitem[{{Uyaniker} {et~al.}(2003){Uyaniker}, {Landecker}, {Gray}, \&
  {Kothes}}]{Uyaniker2003}
{Uyaniker}, B., {Landecker}, T.~L., {Gray}, A.~D., \& {Kothes}, R. 2003, \apj,
  585, 785

\bibitem[{{van de Hulst}(1967)}]{vandeHulst1967}
{van de Hulst}, H.~C. 1967, \araa, 5, 167

\bibitem[{{Van Eck} {et~al.}(2019){Van Eck}, {Haverkorn}, {Alves}, {Beck},
  {Best}, {Carretti}, {Chy{\.z}y}, {En{\ss}lin}, {Farnes}, {Ferri{\`e}re},
  {Heald}, {Iacobelli}, {Jeli{\'c}}, {Reich}, {R{\"o}ttgering}, \&
  {Schnitzeler}}]{VanEck2019}
{Van Eck}, C.~L. \etal . 2019, \aap, 623, A71

\bibitem[{{Van Eck} {et~al.}(2017){Van Eck}, {Haverkorn}, {Alves}, {Beck}, {de
  Bruyn}, {En{\ss}lin}, {Farnes}, {Ferri{\`e}re}, {Heald}, {Horellou},
  {Horneffer}, {Iacobelli}, {Jeli{\'c}}, {Mart{\'\i}-Vidal}, {Mulcahy},
  {Reich}, {R{\"o}ttgering}, {Scaife}, {Schnitzeler}, {Sobey}, \&
  {Sridhar}}]{VanEck2017}
{Van Eck}, C.~L. \etal . 2017, \aap, 597, A98

\bibitem[{{van Haarlem} {et~al.}(2013){van Haarlem}, {Wise}, {Gunst}, {Heald},
  {McKean}, {Hessels}, {de Bruyn}, {Nijboer}, {Swinbank}, {Fallows},
  {Brentjens}, {Nelles}, {Beck}, {Falcke}, {Fender}, {H{\"o}randel},
  {Koopmans}, {Mann}, {Miley}, {R{\"o}ttgering}, {Stappers}, {Wijers},
  {Zaroubi}, {van den Akker}, {Alexov}, {Anderson}, {Anderson}, {van Ardenne},
  {Arts}, {Asgekar}, {Avruch}, {Batejat}, {B{\"a}hren}, {Bell}, {Bell}, {van
  Bemmel}, {Bennema}, {Bentum}, {Bernardi}, {Best}, {B{\^\i}rzan}, {Bonafede},
  {Boonstra}, {Braun}, {Bregman}, {Breitling}, {van de Brink}, {Broderick},
  {Broekema}, {Brouw}, {Br{\"u}ggen}, {Butcher}, {van Cappellen}, {Ciardi},
  {Coenen}, {Conway}, {Coolen}, {Corstanje}, {Damstra}, {Davies}, {Deller},
  {Dettmar}, {van Diepen}, {Dijkstra}, {Donker}, {Doorduin}, {Dromer}, {Drost},
  {van Duin}, {Eisl{\"o}ffel}, {van Enst}, {Ferrari}, {Frieswijk}, {Gankema},
  {Garrett}, {de Gasperin}, {Gerbers}, {de Geus}, {Grie{\ss}meier}, {Grit},
  {Gruppen}, {Hamaker}, {Hassall}, {Hoeft}, {Holties}, {Horneffer}, {van der
  Horst}, {van Houwelingen}, {Huijgen}, {Iacobelli}, {Intema}, {Jackson},
  {Jelic}, {de Jong}, {Juette}, {Kant}, {Karastergiou}, {Koers}, {Kollen},
  {Kondratiev}, {Kooistra}, {Koopman}, {Koster}, {Kuniyoshi}, {Kramer},
  {Kuper}, {Lambropoulos}, {Law}, {van Leeuwen}, {Lemaitre}, {Loose}, {Maat},
  {Macario}, {Markoff}, {Masters}, {McFadden}, {McKay-Bukowski}, {Meijering},
  {Meulman}, {Mevius}, {Middelberg}, {Millenaar}, {Miller-Jones}, {Mohan},
  {Mol}, {Morawietz}, {Morganti}, {Mulcahy}, {Mulder}, {Munk}, {Nieuwenhuis},
  {van Nieuwpoort}, {Noordam}, {Norden}, {Noutsos}, {Offringa}, {Olofsson},
  {Omar}, {Orr{\'u}}, {Overeem}, {Paas}, {Pand ey-Pommier}, {Pandey}, {Pizzo},
  {Polatidis}, {Rafferty}, {Rawlings}, {Reich}, {de Reijer}, {Reitsma},
  {Renting}, {Riemers}, {Rol}, {Romein}, {Roosjen}, {Ruiter}, {Scaife}, {van
  der Schaaf}, {Scheers}, {Schellart}, {Schoenmakers}, {Schoonderbeek},
  {Serylak}, {Shulevski}, {Sluman}, {Smirnov}, {Sobey}, {Spreeuw}, {Steinmetz},
  {Sterks}, {Stiepel}, {Stuurwold}, {Tagger}, {Tang}, {Tasse}, {Thomas},
  {Thoudam}, {Toribio}, {van der Tol}, {Usov}, {van Veelen}, {van der Veen},
  {ter Veen}, {Verbiest}, {Vermeulen}, {Vermaas}, {Vocks}, {Vogt}, {de Vos},
  {van der Wal}, {van Weeren}, {Weggemans}, {Weltevrede}, {White}, {Wijnholds},
  {Wilhelmsson}, {Wucknitz}, {Yatawatta}, {Zarka}, {Zensus}, \& {van
  Zwieten}}]{vanHaarlem2013}
{van Haarlem}, M.~P. \etal . 2013, \aap, 556, A2

\bibitem[{{Verschuur}(1970)}]{Verschuur1970}
{Verschuur}, G.~L. 1970, \aj, 75, 687

\bibitem[{{Virtanen} {et~al.}(2020){Virtanen}, {Gommers}, {Burovski},
  {Oliphant}, {Weckesser}, {Cournapeau}, {Alexbrc}, {Peterson}, {Reddy},
  {Wilson}, {Haberland}, {Mayorov}, {Endolith}, {Nelson}, {Der Van Walt},
  {Laxalde}, {Brett}, {Polat}, {Larson}, {Millman}, {Lars}, {Van Mulbregt},
  {Eric-Jones}, {Carey}, {Moore}, {Kern}, {Leslie}, {Perktold}, {Striega}, \&
  {Feng}}]{Virtanen2020}
{Virtanen}, P. \etal . 2020, {scipy/scipy: SciPy 1.7.0, v1.5.3 Zenodo,
  doi:10.5281/zenodo.595738}

\bibitem[{{Weisberg} {et~al.}(2008){Weisberg}, {Stanimirovi{\'c}}, {Xilouris},
  {Hedden}, {de la Fuente}, {Anderson}, \& {Jenet}}]{Weisberg2008}
{Weisberg}, J.~M., {Stanimirovi{\'c}}, S., {Xilouris}, K., {Hedden}, A., {de la
  Fuente}, A., {Anderson}, S.~B., \& {Jenet}, F.~A. 2008, \apj, 674, 286

\bibitem[{{Wenger} {et~al.}(2018){Wenger}, {Balser}, {Anderson}, \&
  {Bania}}]{Wenger2018}
{Wenger}, T.~V., {Balser}, D.~S., {Anderson}, L.~D., \& {Bania}, T.~M. 2018,
  \apj, 856, 52

\bibitem[{{Wieringa} {et~al.}(1993){Wieringa}, {de Bruyn}, {Jansen}, {Brouw},
  \& {Katgert}}]{Wieringa1993}
{Wieringa}, M.~H., {de Bruyn}, A.~G., {Jansen}, D., {Brouw}, W.~N., \&
  {Katgert}, P. 1993, \aap, 268, 215

\bibitem[{{Winkel} {et~al.}(2016){Winkel}, {Kerp}, {Fl{\"o}er}, {Kalberla},
  {Ben Bekhti}, {Keller}, \& {Lenz}}]{Winkel2016}
{Winkel}, B., {Kerp}, J., {Fl{\"o}er}, L., {Kalberla}, P.~M.~W., {Ben Bekhti},
  N., {Keller}, R., \& {Lenz}, D. 2016, \aap, 585, A41

\bibitem[{{Wolfire} {et~al.}(2003){Wolfire}, {McKee}, {Hollenbach}, \&
  {Tielens}}]{Wolfire2003}
{Wolfire}, M.~G., {McKee}, C.~F., {Hollenbach}, D., \& {Tielens}, A.~G.~G.~M.
  2003, \apj, 587, 278

\bibitem[{{Wolleben} {et~al.}(2009){Wolleben}, {Landecker}, {Carretti},
  {Dickey}, {Fletcher}, {Gaensler}, {Han}, {Haverkorn}, {Leahy},
  {McClure-Griffiths}, {McConnell}, {Reich}, \& {Taylor}}]{Wolleben2009}
{Wolleben}, M. \etal . 2009, in Cosmic Magnetic Fields: From Planets, to Stars
  and Galaxies, ed. K.~G. {Strassmeier}, A.~G. {Kosovichev}, \& J.~E.
  {Beckman}, Vol. 259, 89--90

\bibitem[{{Wolleben} {et~al.}(2006){Wolleben}, {Landecker}, {Reich}, \&
  {Wielebinski}}]{Wolleben2006}
{Wolleben}, M., {Landecker}, T.~L., {Reich}, W., \& {Wielebinski}, R. 2006,
  \aap, 448, 411

\bibitem[{{Wood} {et~al.}(2010){Wood}, {Hill}, {Joung}, {Mac Low}, {Benjamin},
  {Haffner}, {Reynolds}, \& {Madsen}}]{Wood2010}
{Wood}, K., {Hill}, A.~S., {Joung}, M.~R., {Mac Low}, M.-M., {Benjamin}, R.~A.,
  {Haffner}, L.~M., {Reynolds}, R.~J., \& {Madsen}, G.~J. 2010, \apj, 721, 1397

\bibitem[{{Yoon} \& {Cho}(2019)}]{Yoon2019}
{Yoon}, H. \& {Cho}, J. 2019, \apj, 880, 137

\bibitem[{{Zaroubi} {et~al.}(2015){Zaroubi}, {Jelic}, {de Bruyn}, {Boulanger},
  {Bracco}, {Kooistra}, {Alves}, {Brentjens}, {Ferriere}, {Ghosh}, {Koopmans},
  {Levrier}, {Miville-Deschenes}, {Montier}, {Pandey}, \&
  {Soler}}]{Zaroubi2015}
{Zaroubi}, S. \etal . 2015, \mnras, 454, L46

\bibitem[{{Zonca} {et~al.}(2019){Zonca}, {Singer}, {Lenz}, {Reinecke},
  {Rosset}, {Hivon}, \& {Gorski}}]{Zonca2019}
{Zonca}, A., {Singer}, L., {Lenz}, D., {Reinecke}, M., {Rosset}, C., {Hivon},
  E., \& {Gorski}, K. 2019, JOSS, 4, 1298

\end{thebibliography}

\end{document}